\documentclass[12pt]{article}
\usepackage{epsfig,cite}
\usepackage {amsmath}
\usepackage{amssymb}
\include{epsf}
\newcount\timecount
\newcount\hours \newcount\minutes  \newcount\temp \newcount\pmhours
\hours = \time
\divide\hours by 60
\temp = \hours
\multiply\temp by 60
\minutes = \time
\advance\minutes by -\temp
\def\hour{\the\hours}
\def\minute{\ifnum\minutes<10 0\the\minutes
            \else\the\minutes\fi}
\def\clock{
\ifnum\hours=0 12:\minute\ AM
\else\ifnum\hours<12 \hour:\minute\ AM
      \else\ifnum\hours=12 12:\minute\ PM
            \else\ifnum\hours>12
                 \pmhours=\hours
                 \advance\pmhours by -12
                 \the\pmhours:\minute\ PM
                 \fi
            \fi
      \fi
\fi
}

\def\monthname{\relax\ifcase\month 0/\or January\or February\or
   March\or April\or May\or June\or July\or August\or September\or
   October\or November\or December\else\number\month/\fi}

\def\bold#1{\setbox0=\hbox{$#1$}%
     \kern-.025em\copy0\kern-\wd0
     \kern.05em\copy0\kern-\wd0
     \kern-.025em\raise.0433em\box0 }

\def\beq{\begin{equation}}
\def\eeq{\end{equation}}

\def\ga{\mathrel{\raise.3ex\hbox{$>$\kern-.75em\lower1ex\hbox{$\sim$}}}}
\def\la{\mathrel{\raise.3ex\hbox{$<$\kern-.75em\lower1ex\hbox{$\sim$}}}}
\def\gev{{\rm \, Ge\kern-0.125em V}}
\def\tev{{\rm \, Te\kern-0.125em V}}
\def\gyr{{\rm \, G\kern-0.125em yr}}

\def\tbt{\tan \beta}

\def\gappeq{\mathrel{\rlap {\raise.5ex\hbox{$>$}}
{\lower.5ex\hbox{$\sim$}}}}
\def\lappeq{\mathrel{\rlap{\raise.5ex\hbox{$<$}}
{\lower.5ex\hbox{$\sim$}}}}
\def\Toprel#1\over#2{\mathrel{\mathop{#2}\limits^{#1}}}

\def\sel{{\widetilde e}}

\def\stau{\widetilde \tau}

\def\snu{\widetilde \nu}

\def\m12{m_{1\!/2}}

\def\PL{{Phys.~Lett.} }

\def\PRL{{Phys.~Rev.~Lett.} }

\def\stau{\tilde{\tau}}
\def\snu{\tilde{\nu}}
\def\sel{\tilde{e}}

\def\tanb{\tan \beta}

\def\bea{\begin{eqnarray}}
\def\eea{\end{eqnarray}}
\newcommand{\goto}{\rightarrow}

\begin{document}
\begin{titlepage}
\pagestyle{empty}
\baselineskip=21pt
\rightline{CERN-PH-TH/2008-106, UMN--TH--2645/08, FTPI--MINN--08/14}
\vskip 0.2in
\begin{center}
{\large{\bf Varying the Universality of Supersymmetry-Breaking Contributions to MSSM
Higgs Boson Masses}}
\end{center}
\begin{center}
\vskip 0.2in
{\bf John~Ellis}$^1$, {\bf Keith~A.~Olive}$^{2}$ and
{\bf Pearl Sandick}$^{2}$
\vskip 0.1in

{\it
$^1${TH Division, PH Department, CERN, CH-1211 Geneva 23, Switzerland}\\
$^2${William I. Fine Theoretical Physics Institute, \\
University of Minnesota, Minneapolis, MN 55455, USA}\\
}

{\bf Abstract}
\end{center}
\baselineskip=18pt \noindent

We consider the minimal supersymmetric extension of the Standard Model (MSSM)
with varying amounts of non-universality in the soft supersymmetry-breaking
contributions to the Higgs scalar masses. In addition to the constrained MSSM (CMSSM)
in which these are universal with the soft supersymmetry-breaking
contributions to the squark and slepton masses at the input GUT scale, we
consider scenarios in which both the Higgs scalar masses are non-universal by the
same amount (NUHM1), and scenarios in which they are independently non-universal
(NUHM2). We show how the NUHM1 scenarios generalize the $(m_{1/2}, m_0)$
planes of the CMSSM by allowing either $\mu$ or $m_A$ to take different (fixed)
values and we also show how the NUHM1 scenarios are embedded as special
cases of the more general NUHM2 scenarios.
Generalizing from the CMSSM, we find regions of the NUHM1 parameter space that
are excluded because the LSP is a selectron. We also find new regions where the
neutralino relic density falls within the range preferred by astrophysical and
cosmological measurements, thanks to rapid annihilation through direct-channel
Higgs poles, or coannihilation with selectrons, or because the LSP composition
crosses over from being mainly bino to mainly Higgsino. Generalizing further to the
NUHM2, we find regions of its parameter space where a sneutrino is the LSP,
and others where neutralino coannihilation with sneutrinos is important for the
relic density. In both the NUHM1 and the NUHM2, there are slivers of parameter
space where the LHC has fewer prospects for discovering sparticles than in the
CMSSM, because either $m_{1/2}$ and/or $m_0$ may be considerably larger than
in the CMSSM.

\vfill
\leftline{CERN-PH-TH/2008-106}
\leftline{May 2008}
\end{titlepage}

\section{Introduction}

The simplest supersymmetric model is the minimal supersymmetric extension of the 
Standard Model (MSSM), and it is commonly assumed that the soft supersymmetry-breaking
contributions to the squark, slepton and Higgs scalar masses are universal at
some GUT input scale (CMSSM) \cite{funnel,cmssm}. 
This is certainly the simplest assumption, but it is neither 
the only nor necessarily the most plausible version of the MSSM. For example,
universality might hold at some lower renormalization scale \cite{eosk}, as in some mirage unification
scenarios \cite{mixed}. Alternatively, the soft supersymmetry-breaking masses may not be
universal at any renormalization scale, as occurs in some string scenarios for
supersymmetry breaking \cite{string}. The suppression of flavour-changing supersymmetric
interactions suggests that the soft supersymmetry-breaking masses of all generations of
squarks and sleptons with the same electroweak quantum numbers
may be the same, i.e., $m_{\tilde e_L}^2 = m_{\tilde \mu_L}^2 = 
m_{\tilde \tau_L}^2$, $m_{\tilde e_R}^2 = m_{\tilde \mu_R}^2 = m_{\tilde \tau_R}^2$, 
and similarly for the ${\tilde q_{L,R}}$ of charges $+2/3$ and $-1/3$ \cite{flavor}. 
However, this argument does 
not motivate universality between sleptons and squarks, or even
between left- and right-handed sleptons
or squarks. Some degree of universality would be expected in supersymmetric GUTs. For
example, in supersymmetric SU(5) one would expect $m_{\tilde e_L}^2 = m_{\tilde d_R}^2$
and $m_{\tilde e_R}^2 =  m_{\tilde u_L}^2 = m_{\tilde u_R}^2$. Supersymmetric SO(10)
would further predict universality between all the soft supersymmetry-breaking squark and
slepton masses. However, supersymmetric GUTs do not give any reason to think
that the soft supersymmetry-breaking contributions to the Higgs scalar masses should be
universal with the squark and slepton masses. This full universality, postulated in the CMSSM,
would occur in minimal supergravity (mSUGRA) scenarios \cite{vcmssm}, but not in more general
effective no-scale supergravity theories such as those derived from string models \cite{noscale}.

On the basis of the above discussion, it is natural to consider models with non-universal 
soft supersymmetry-breaking contributions to the Higgs scalar masses \cite{nonu}. In general,
one may introduce two independent non-universality parameters, scenarios which can be termed
NUHM2 \cite{nuhm}, but one could also consider scenarios with equal amounts of non-universality
for the two Higgs doublets, scenarios which can be termed NUHM1 \cite{nuhm1}.
 Such scenarios would be natural in
a supersymmetric SO(10) GUT framework, since the two Higgs multiplets occupy a
common vectorial 10-dimensional representation, while each matter generation occupies a
common spinorial 16-dimensional representation of SO(10).

CMSSM scenarios have four continuous parameters, which may be taken as
$m_0, m_{1/2}, A_0,$ $\tan \beta$, with the values of $|\mu|$ and $m_A$ then 
being fixed by the electroweak vacuum conditions. Correspondingly, NUHM1
scenarios have one additional parameter, that may be taken as either $\mu$ or $m_A$,
whereas both $\mu$ and $m_A$ are free parameters in NUHM2 scenarios. The
full six-dimensional NUHM2 parameter space has been explored in a number of
studies \cite{nuhm}, but its higher dimensionality renders its complete characterization quite
complicated, and it is less amenable to a Markov Chain Monte Carlo analysis than
the NUHM1 and particularly CMSSM scenarios \cite{mcmc}. The main purpose of this paper is
to discuss how the CMSSM, NUHM1 and NUHM2 scenarios may be related by
processes of dimensional enhancement: CMSSM $\in$ NUHM1 $\in$ NUHM2 and
reduction: NUHM2 $\ni$ NUHM1 $\ni$ CMSSM, laying the basis for more complete
understanding of the NUHM1 and NUHM2 parameter spaces.
Accordingly, in the following sections we focus first on the relationship between the CMSSM
and NUHM1 scenarios, and subsequently on the relationship between the NUHM1 and NUHM2
scenarios.

The most important contributions to most sparticle masses are those due to $m_{1/2}$
and $m_0$, so studies of the phenomenological constraints on the CMSSM parameter
space \cite{eoss,cmssmwmap}
and the prospects for experimental searches at the LHC and elsewhere are
frequently displayed in $(m_{1/2}, m_0)$ planes for different values of $\tan \beta$, $A_0$
and the sign of $\mu$. The values of $|\mu|$ and $m_A$ then vary across these planes
according to the electroweak vacuum conditions. In our first exploration of the NUHM1
parameter space, we display and discuss $(m_{1/2}, m_0)$ planes for different choices
of fixed values of $m_A$ and positive $\mu$, seeking to understand, in particular, the
dependences on $m_A$ and $\mu$ of the strips of parameter space compatible with the 
cold dark matter density inferred from WMAP and other observations \cite{WMAP}.
A key question
here is whether the good (but not complete) LHC coverage of the CMSSM WMAP strips \cite{eoss}
is repeated also in NUHM1 scenarios.
We find that there are extensions of the preferred regions of the 
$(m_{1/2}, m_0)$ planes to larger values of these parameters that are
affected by the choices of $\mu$ or $m_A$, whereas the preferred
regions of these latter parameters are more sensitive to the choices of the other
NUHM1 parameters. In some of the extensions, the LHC would either have difficulty in
detecting supersymmetry at all, or would only provide access to a limited range of
sparticles. Since the interest of NUHM1 scenarios lies largely with the new possibilities for
varying $m_A$ and $\mu$, which have in turn important implications for the spectrum of
heavy MSSM Higgs bosons and gauginos, we also display explicitly the variations of the
various phenomenological constraints in planes correlating $m_{1/2}$ or $m_0$ with
$m_A$ or $\mu$. 

In our discussion of the relationship between the NUHM1 and NUHM2 scenarios, we
display the allowed regions of parameter space as explicit functions of the degrees of
non-universality of the soft supersymmetry-breaking scalar mass parameters of the
two MSSM Higgs multiplets. We find that the WMAP relic density constraint, in
particular, generally favours models with a relatively high degree of non-universality,
close to the boundaries of the NUHM2 parameter space imposed by other
theoretical and phenomenological constraints such as the breakdown of electroweak
symmetry breaking or the absence of charged dark matter. This reflects the fact,
known already from studies of the CMSSM with GUT-scale universality, that the
supersymmetric relic density is too large in generic domains of parameter space,
being brought down into the WMAP range in particular cases such as the coannihilation~\cite{coann}
and focus-point regions (close to the charged dark matter and electroweak
symmetry breaking boundaries, respectively)~\cite{fp}, 
or in rapid-annihilation funnel regions~\cite{funnel}.

\section{From the CMSSM to the NUHM1}
\label{sec:CMSSMtoNUHM1}

In the CMSSM, the weak-scale observables are determined by four continuous parameters and a sign; the
universal scalar mass $m_0$, the universal gaugino mass $m_{1/2}$, the
universal trilinear coupling $A_0$, the ratio of the Higgs vacuum
expectation values $\tanb$, and the sign of the Higgs mass parameter
$\mu$. We consider the values of the parameters $m_0$, $m_{1/2}$ and $A_0$ to be
specified at the SUSY GUT scale.  The effective Higgs masses-squared, $m_1^2$
and $m_2^2$ are responsible for generating electroweak symmetry
breaking through their running from the input scale down to low
energies. In the CMSSM, $m_1^2(M_{GUT}) = m_2^2(M_{GUT})=m_0^2$, and
$|\mu|$ and $m_A$ are calculated from the electroweak vacuum
conditions,
\beq
m_A^2(Q) = m_1^2(Q)+m_2^2(Q)+2\mu^2(Q)+\Delta_A(Q)
\label{eq:mA}
\eeq  
and
\beq
\mu^2=\frac{m_1^2-m_2^2\tan^2\beta+\frac{1}{2}m_Z^2(1-\tan^2\beta)+\Delta_{\mu}^{(1)}}{\tan^2\beta-1+\Delta_{\mu}^{(2)}},
\label{eq:mu}
\eeq
where $\Delta_A$ and $\Delta_\mu^{(1,2)}$ are loop
corrections~\cite{Barger:1993gh,deBoer:1994he,Carena:2001fw},
$Q=(m_{\stau_R}m_{\stau_L})^{1/2}$, and all quantities in
(\ref{eq:mu}) are defined at the electroweak scale, $m_Z$. Unless otherwise noted,
$m_A \equiv m_A(Q)$ and $\mu \equiv \mu(m_Z)$.  The values of the parameters in (\ref{eq:mA})
and (\ref{eq:mu}) are related through well-known radiative corrections \cite{Barger:1993gh,IL,Martin:1993zk}  $c_1$, $c_2$
and $c_{\mu}$ such that
\begin{eqnarray}
m_1^2(Q)=m_1^2+c_1, \nonumber\\
m_2^2(Q)=m_2^2+c_2, \label{eq:radcorr}\\
\mu^2(Q)=\mu^2+c_{\mu}. \nonumber
\end{eqnarray}

In the NUHM1 one still has $m_1^2(M_{GUT}) = m_2^2(M_{GUT})$, but these
are no longer identified with the universal scalar mass, $m_0$, so an
additional parameter is necessary to fix the common GUT-scale value of the
Higgs masses-squared.  This additional parameter may be taken to be either
$\mu$ or $m_A$, and the relationship between $m_1^2$ and $m_2^2$ at
the weak scale can be calculated from (\ref{eq:mA}) - (\ref{eq:radcorr}) so as to
respect the electroweak boundary conditions at $m_Z$ and the weakened
universality condition at $M_{GUT}$.  

If $m_A$ is taken to be the free parameter (input), then at
$m_Z$ we have
\begin{eqnarray}
m_1^2(\tan^2\beta+1+\Delta_{\mu}^{(2)})
=m_2^2(\tan^2\beta+1-\Delta_{\mu}^{(2)})+m_Z^2(\tan^2\beta-1)-2\Delta_{\mu}^{(1)}\nonumber\\
+\big(m_A^2-(\Delta_A(Q)+c_1+c_2+2c_{\mu})\big)
(\tan^2\beta-1+\Delta_{\mu}^{(2)}).
\end{eqnarray}
Alternatively, if $\mu$ is taken as the free parameter, then at $m_Z$
we have
\beq
m_1^2=m_2^2\tan^2\beta+\mu^2(\tan^2\beta-1+\Delta_{\mu}^{(2)})+\frac{1}{2}m_Z^2(\tan^2\beta-1)-\Delta_{\mu}^{(1)}.
\eeq
In each case, the boundary condition at $M_{GUT}$ is $m_1^2 =
m_2^2$.  Clearly, for some specific input values of $\mu$ and $m_A$, one
finds $m_1^2(M_{GUT}) = m_2^2(M_{GUT}) = m_0^2$, thereby recovering the
CMSSM. The characteristics of the parameter space as one deviates
from this scenario are the subjects of the following subsections.


\subsection{The NUHM1 with $m_A$ as a Free Parameter}
\label{sec:mA}

We begin our characterization of the relationship between
the CMSSM and NUHM1 scenarios by taking $m_A$ as the additional free
parameter, and assume positive $\mu$, as suggested by $g_\mu - 2$ and
$b \goto s \gamma$, at least within the CMSSM.

\begin{figure}[ht!]
\begin{center}
\mbox{\epsfig{file=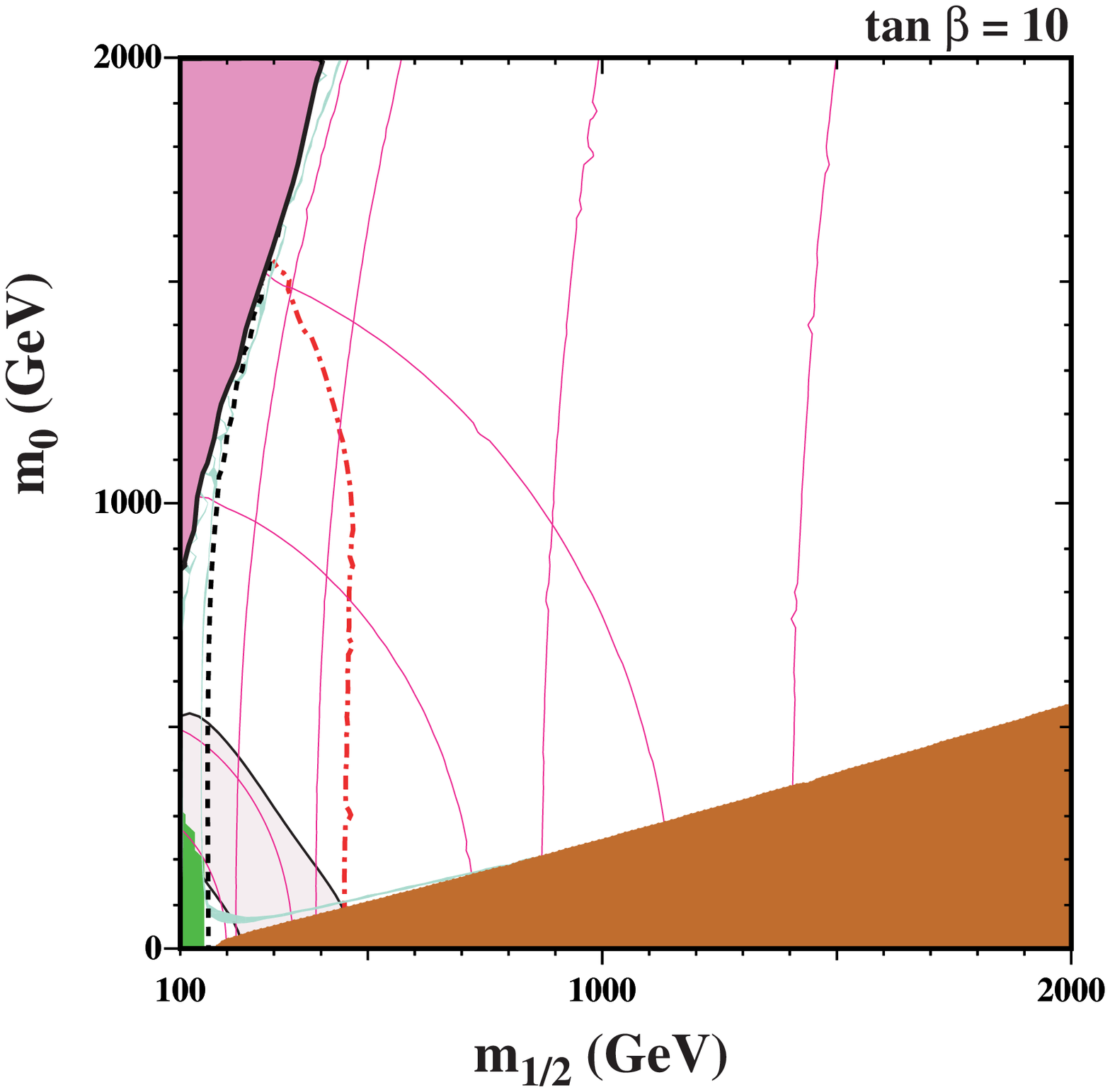,height=6.8cm}}
\mbox{\epsfig{file=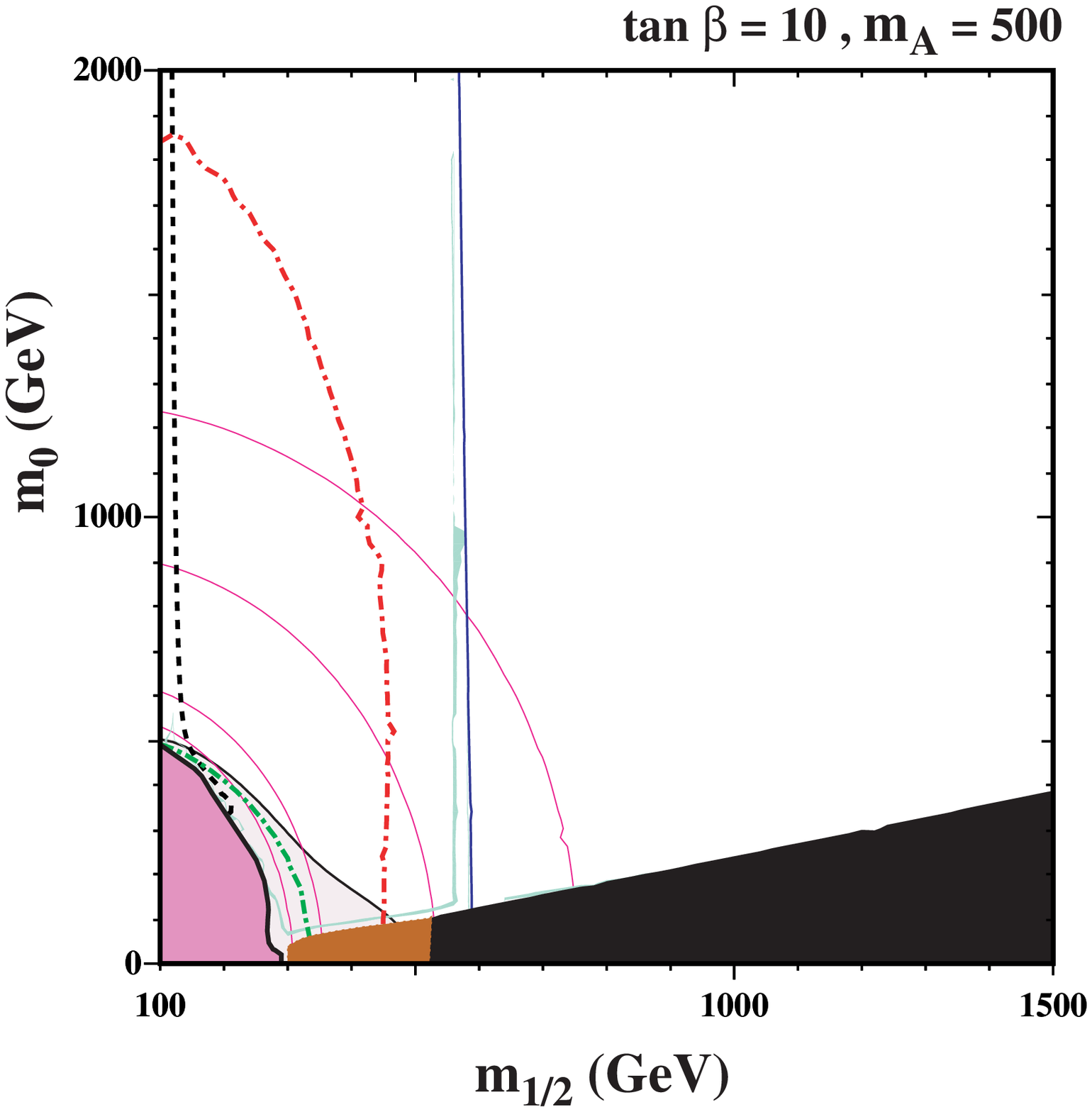,height=7cm}}
\end{center}
\begin{center}
\mbox{\epsfig{file=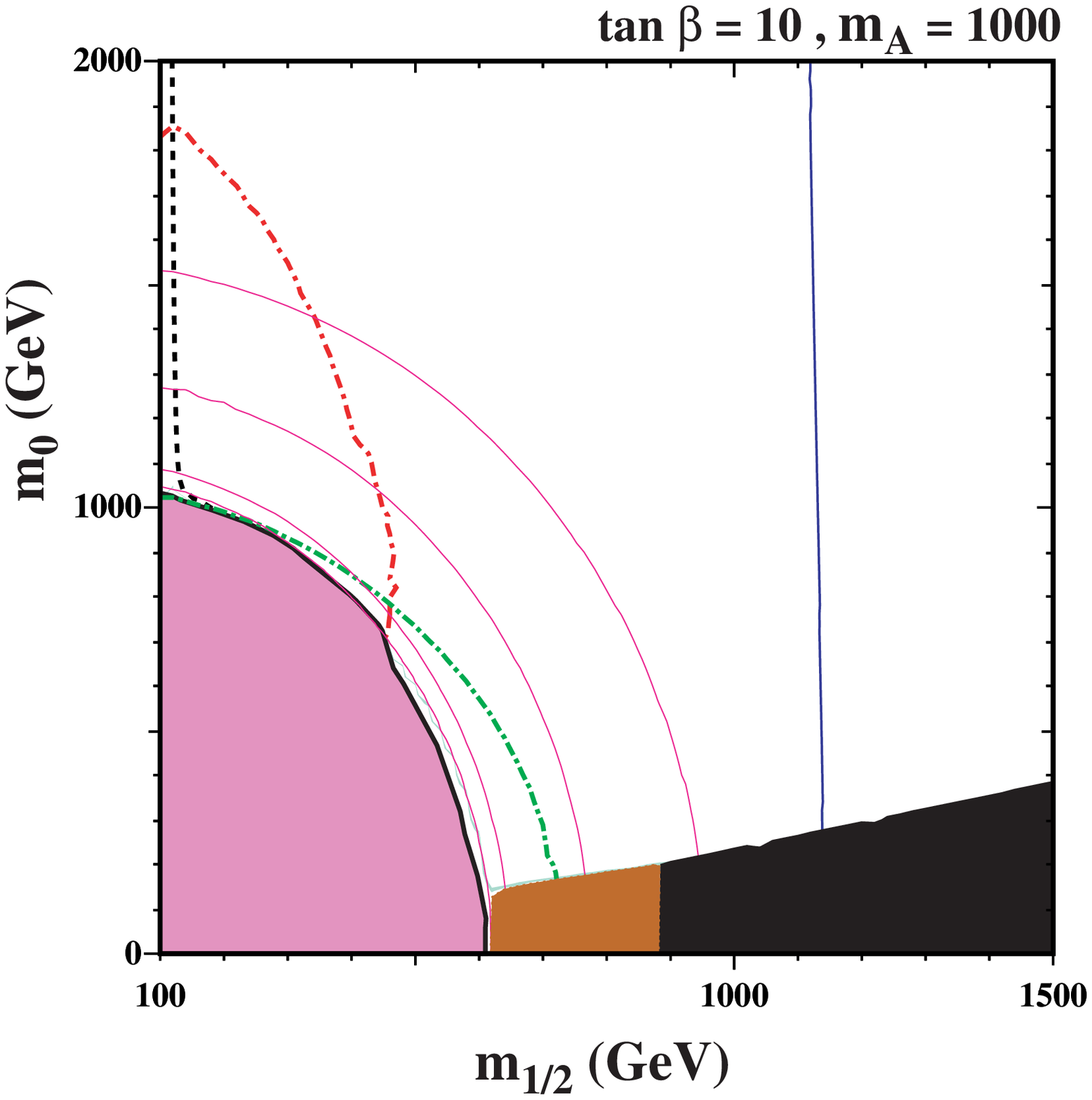,height=7cm}}
\mbox{\epsfig{file=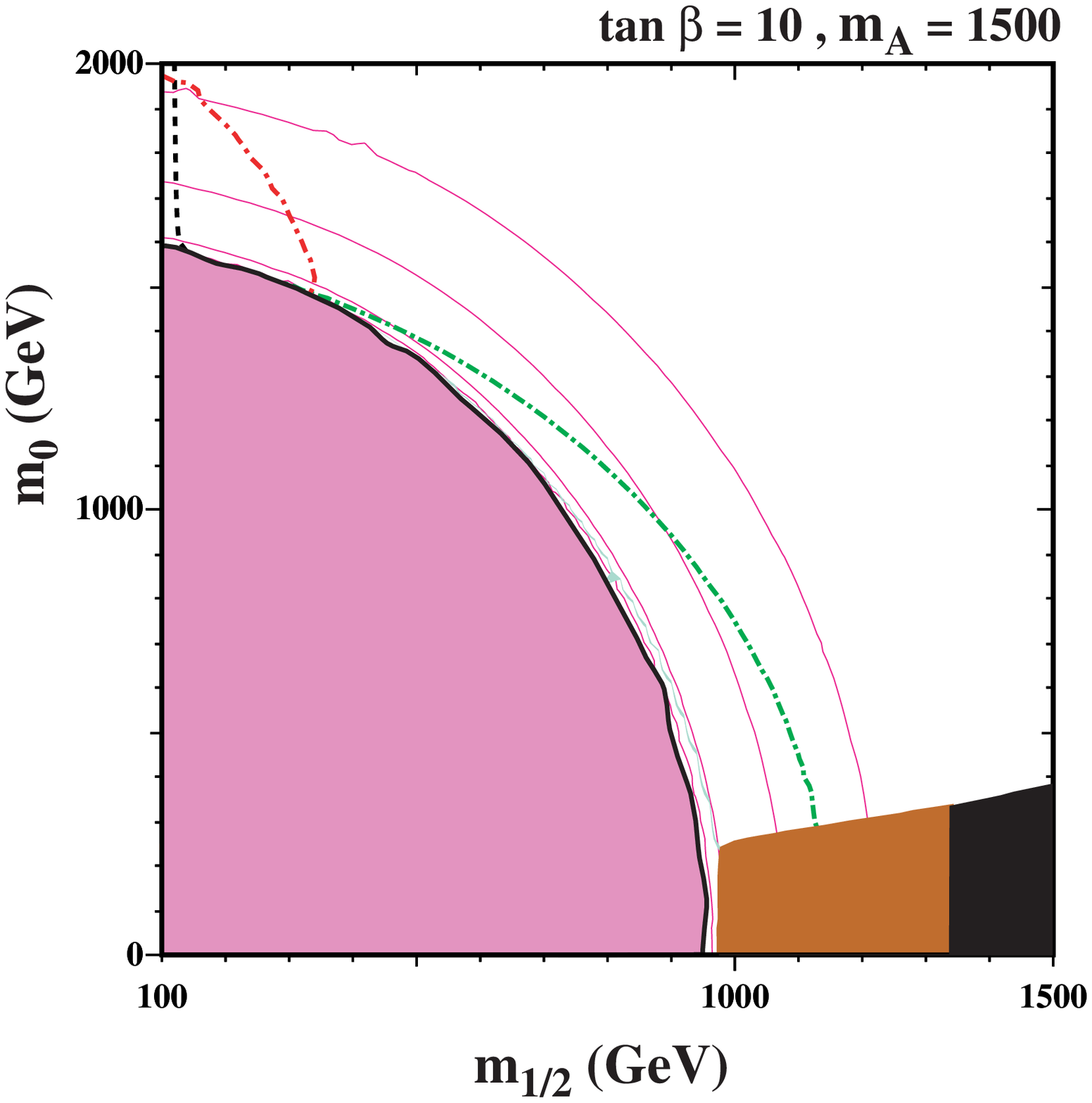,height=7cm}}
\end{center}
\caption{\it Panel (a) shows the $(m_{1/2},m_0)$ plane for the CMSSM for $\tanb = 10$, 
with contours of $m_A$ and
$\mu$ of 300, 500, 1000, and 1500 GeV as described in the
text. Panels~(b), (c), and (d) show the NUHM1 $(m_{1/2},m_0)$ planes for $\tanb = 10$
with $m_A = 500$, 1000, and 1500 GeV, respectively. Constraints and
contours are as
described in the text.
}
\label{fig:m12m0fixmA}
\end{figure}

As a basis for the comparison, in Fig.~\ref{fig:m12m0fixmA} we show
in panel~(a) a CMSSM $(m_{1/2},m_0)$ plane with $\tanb=10$ and
$A_0=0$.  We have plotted (pink) contours of constant $\mu$ and $m_A$ of 300,
500, 1000, and 1500 GeV, with $\mu$ contours appearing roughly
vertical and $m_A$ contours appearing as quarter-ellipses centered at the
origin.  There are also several phenomenological constraints shown in
panel~(a)~\footnote{We use the same notations for these constraints in this
and the following figures.}.  In the region at low $m_{1/2}$ and large $m_0$ there 
is a (dark pink) shaded region where there are no
consistent solutions to the electroweak vacuum conditions, since they would require
$\mu^2 < 0$.  An additional unphysical region is found along the
bottom of the plane at larger $m_{1/2}$ and low $m_0$, where the 
lightest supersymmetric particle (LSP) is
a charged stau $\stau$ (brown shading). Contours of $m_h = 114$~GeV (red
dot-dashed) and $m_{\chi^{\pm}}=104$~GeV (black dashed) mark, 
approximately, the edges of the regions excluded by unsuccessful searches
at LEP \cite{LEPsusy}. Both $m_h$ and $m_{\chi^{\pm}}$ increase with $m_{1/2}$, so
portions to the right of these contours are allowed.  The region
favored by the measurement of the muon anomalous magnetic moment \cite{g-2},
$g_\mu-2$, at the two-$\sigma$ level (light pink shading bounded by solid black lines) 
is also visible at very low
$(m_{1/2},m_0)$, and the region disfavoured by $b \goto s \gamma$ \cite{bsgex} is shaded green. 

Finally, the regions of the plane where the relic
density of neutralino LSPs falls in the range favoured by WMAP and other
observations for the dark matter abundance appear
as thin turquoise strips.  For the chosen value of $\tanb=10$, the relic
density of neutralinos is too large over the bulk of the plane, 
and falls within the WMAP range in two distinct regions.  
In the upper left corner, tracking the region excluded
by the electroweak vacuum conditions, lies the focus-point region \cite{fp}, where
the lightest neutralino is Higgsino-like and annihilations to gauge
bosons bring the relic density down into the WMAP range. Alongside
the forbidden $\stau$-LSP region lies the coannihilation strip \cite{coann},
where $\chi$-$\stau$ coannihilations reduce the relic density of
neutralinos.  At larger $\tanb$, a rapid-annihilation funnel \cite{funnel} may exist
where $2m_\chi \sim m_A$ and $s$-channel annihilations mediated by the
pseudoscalar Higgs decrease drastically the relic density of
neutralino LSPs, though not for $\tan \beta = 10$. 
We see that the CMSSM predicts values of $m_A$ between
$\sim 500$~GeV and $\sim 1500$~GeV and $\mu$ between 
$\sim 500$~GeV and $\sim 1200$~GeV in the parts of the coannihilation
strips compatible with the LEP constraints, while values of $m_A > 1500$~GeV
and $\mu < 500$~GeV are favoured in the focus-point region for $m_0 < 2$ TeV. 

Panels (b), (c), and (d) of Fig.~\ref{fig:m12m0fixmA} show NUHM1 $(m_{1/2},m_0)$ planes
for $\tan \beta = 10, A_0 = 0$ and $\mu > 0$
with $m_A=500$, 1000, and 1500 GeV, respectively, and $\mu$ calculated
using (\ref{eq:mu}). In addition to the constraints discussed above,
we also plot contours of $\mu = 300$, 500, 1000, and 1500 GeV (light pink). The
most prominent departure from the CMSSM is that the requirement of
electroweak symmetry breaking constrains the plane at low 
$m_0$ rather than at large $m_0$. In this region (below the CMSSM
contour), $m_A$ is fixed to be larger than its CMSSM value, resulting
in correspondingly larger $m_1^2$ and $m_2^2$. We see from
(\ref{eq:mu}) that, with $m_2^2 < 0$ and weighted by
$\tan^2\beta$, the effect is to drive $\mu^2$ smaller, and eventually negative. 
The excluded region grows with $m_A$ as $m_1^2$ and $m_2^2$ are pushed
farther from their CMSSM values, and is
flanked by concentric contours of constant $\mu$. The stau LSP exclusion regions are qualitatively similar to those in the CMSSM,
shown in panel (a), however for moderate values of $m_A$ there is a
(black shaded) region
of the plane where the lighter selectron is the LSP.
Also apparent in panel~(b) for $m_0 = 300$~GeV is a small region at low
$m_{1/2}$ and $m_0$ that is favored by $g_\mu - 2$, which disappears
for larger $m_A$ beneath the expanding region where electroweak symmetry
breaking is not possible. There is no region of this or the following panels
that is excluded by $b \goto s \gamma$.

The LSP mass
and composition are roughly the same as they are in the CMSSM at
large $\mu$: at all but the smallest values of $\mu$, the LSP is bino-like
in the CMSSM.  At moderate and large $\mu$, the masses of the sparticles are only minimally
affected by the fact that $m_A$ is fixed, causing several of the
constraints to appear similar to the CMSSM case. In particular, the LEP chargino and Higgs
constraints again exclude smaller values of $m_{1/2}$, though the shape of both the Higgs
and the chargino exclusions change with increasing $m_A$. 

The strip where the relic LSP density falls within the range preferred by WMAP
and other data stays, in general, close to the regions excluded by the requirement that
the LSP be neutral and by the electroweak vacuum conditions. However, one difference from the 
CMSSM for $\tan \beta = 10$ that is very prominent in panel~(b) is a rapid-annihilation funnel,
straddling the dark blue contour where $2 m_{\chi} = m_A$, that
rises out of the coannihilation strip at $m_{1/2}\approx 570$~GeV, reaching
$m_0 \gtrsim 2300$ GeV. Branches of
good relic density form the inner and outer funnel walls, between which the
relic density falls below the WMAP range.
At larger $m_A$, the dark matter strip changes somewhat.  For $m_A = 1000$ GeV, 
shown in panel~(c) of Fig.~\ref{fig:m12m0fixmA}, $2m_{\chi} = m_A$ at
$m_{1/2} \approx 1130$ GeV. However the coannihilation strip has essentially
terminated at lower $m_{1/2}$, so there is no prominent rapid-annihilation funnel.
Finally, at $m_A = 1500$ GeV, shown in panel (d), $2m_{\chi} = m_A$ at
$m_{1/2} \approx 1680$ GeV, well beyond the end of the coannihilation
strip.  The relic density still decreases in these regions,
but it remains above the WMAP range, so there is no visible funnel.  

We have already emphasized that the parameter space expands
by one dimension between the CMSSM and the NUHM1.  
In each plane (b)-(d) of Fig.~\ref{fig:m12m0fixmA}, there is a green dot-dashed contour
tracking the CMSSM parameters in the NUHM1 $(m_{1/2},m_0)$ plane.
The change in position of this contour as
$m_A$ is increased can be understood by comparison with the contours of
constant $m_A$ in the CMSSM panel~(a). As an example, we consider 
the variation in $\mu$ on the CMSSM contour and how its position
changes in the NUHM1 plane. Examining the contour of
$m_A = 1000$~GeV in the CMSSM plane, we find that in the
$\stau$-LSP region, the value of $\mu$ along the contour reaches a maximum of about 860~GeV. Following the curve to larger $m_0$, we see that it terminates
at the boundary of the region where $\mu^2 < 0$.  So we expect that
the CMSSM contour in the NUHM1 $(m_{1/2},m_0)$ plane with $m_A=1000$
GeV runs smoothly through the contours of constant $\mu$ from $\mu =
860$~GeV in the $\stau$-LSP region to the boundary of the electroweak
symmetry breaking region.
As $m_A$ increases, the CMSSM contour begins near the
coannihilation strip at
correspondingly larger values of $\mu$, but it always terminates at
$\mu=0$. The points of intersection of the CMSSM line with the electroweak
vacuum boundary move to larger values of $m_{1/2}$ and $m_0$ as
$m_A$ increases in panels (b), (c) and (d), tracking the focus-point region in panel (a).

It is clear from panels (b) to (d) that the NUHM1 shares
some small pieces of the cosmologically preferred regions of the parameter
space of the CMSSM for moderate and large values of $m_A$. Only for
500~GeV$\lesssim m_A \lesssim 1500$~GeV does the CMSSM contour
intersect a phenomenologically viable portion of the coannihilation
strip, and only for $m_A \gtrsim 1500$~GeV does it intersect the focus-point region.
Moving away from the CMSSM 
contours in the NUHM1 planes, we find that cosmologically preferred areas
in the focus-point regions are now available at lower $m_A$.
For example, at $m_A = 1000$ GeV in the CMSSM, the focus-point is found
at low values of $m_{1/2}$ where both the Higgs and chargino mass constraints are violated.
In the NUHM1, as seen in panel (c), we find
a viable focus-point strip at $m_{1/2} > 500$ GeV at values of $\mu$
lower than in the CMSSM. Furthermore, we find additional
coannihilation strip at both larger and smaller $\mu$ than what would
be expected in the CMSSM, and for a range of $m_A$ there is even a
rapid-annihilation funnel.

The funnel region is
interesting in that it passes all constraints and may have fairly heavy
scalars, as does the focus-point region in the CMSSM, but with a
bino-like neutralino LSP.  A key difference between the two cases is
illustrated by the following simple example.  If the LHC discovers a
gluino weighing 1.5~TeV, which is estimated to be possible with less than 1~fb$^{-1}$
of integrated luminosity~\cite{lhc,cmstdr,LHC}, then, in the CMSSM the lightest charged
sparticles are encouragingly light with $m_{\chi^{\pm}}=340$ GeV in
the focus-point region and $m_{\stau}=280$ GeV in the coannihilation
strip. However, in the NUHM1, although we
will discover charged staus easily if Nature has chosen the coannihilation strip,
at the peak of the funnel in panel (b) the
lighter chargino could be heavier than 900~GeV, and staus would be as
heavy as $m_{\stau} \approx 2300$~GeV.  In this case, the rapid-annihilation
funnel represents a continuum of viable sparticle masses between
the two extremes. Both the CMSSM points and the NUHM1 points have a
light LSP with 250~GeV $\lesssim m_{\chi} \lesssim$ 280~GeV, but the
pseudoscalar Higgs mass is quite large in the CMSSM and highly
dependent on the value of $m_0$, whereas in the NUHM1 $m_A=550$~GeV in this case.
According to previous studies in the CMSSM, detecting supersymmetry at the
LHC should be possible along the rapid-annihilation strip in panel (b)
for $m_0 < 2000$~GeV with roughly 10 fb$^{-1}$ of integrated
luminosity, though the number of sparticles accessible with dedicated
follow-up searches would
decrease as $m_0$ increases.

\begin{figure}[ht!]
\begin{center}
\mbox{\epsfig{file=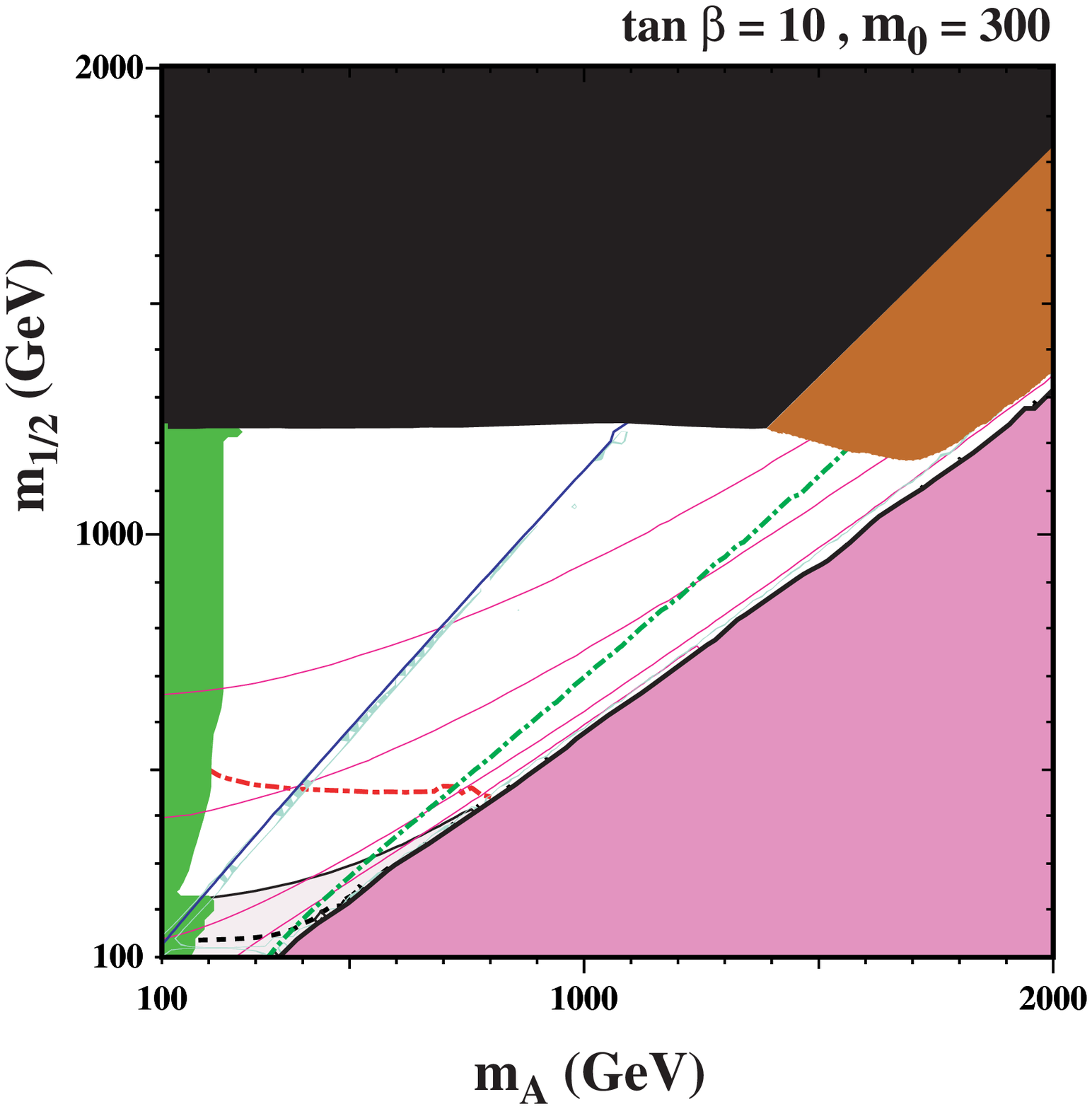,height=7cm}}
\mbox{\epsfig{file=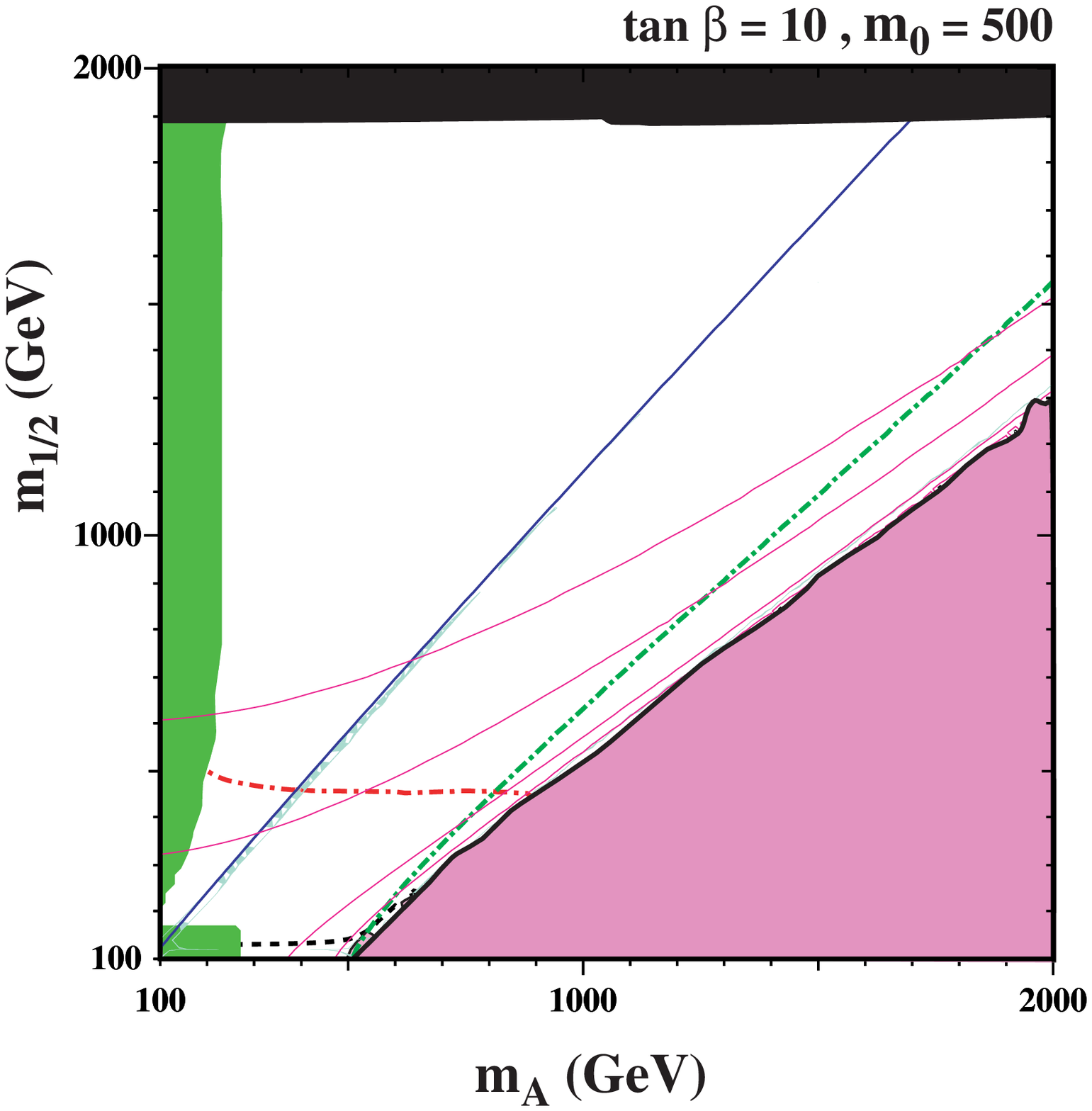,height=7cm}}
\end{center}
\begin{center}
\mbox{\epsfig{file=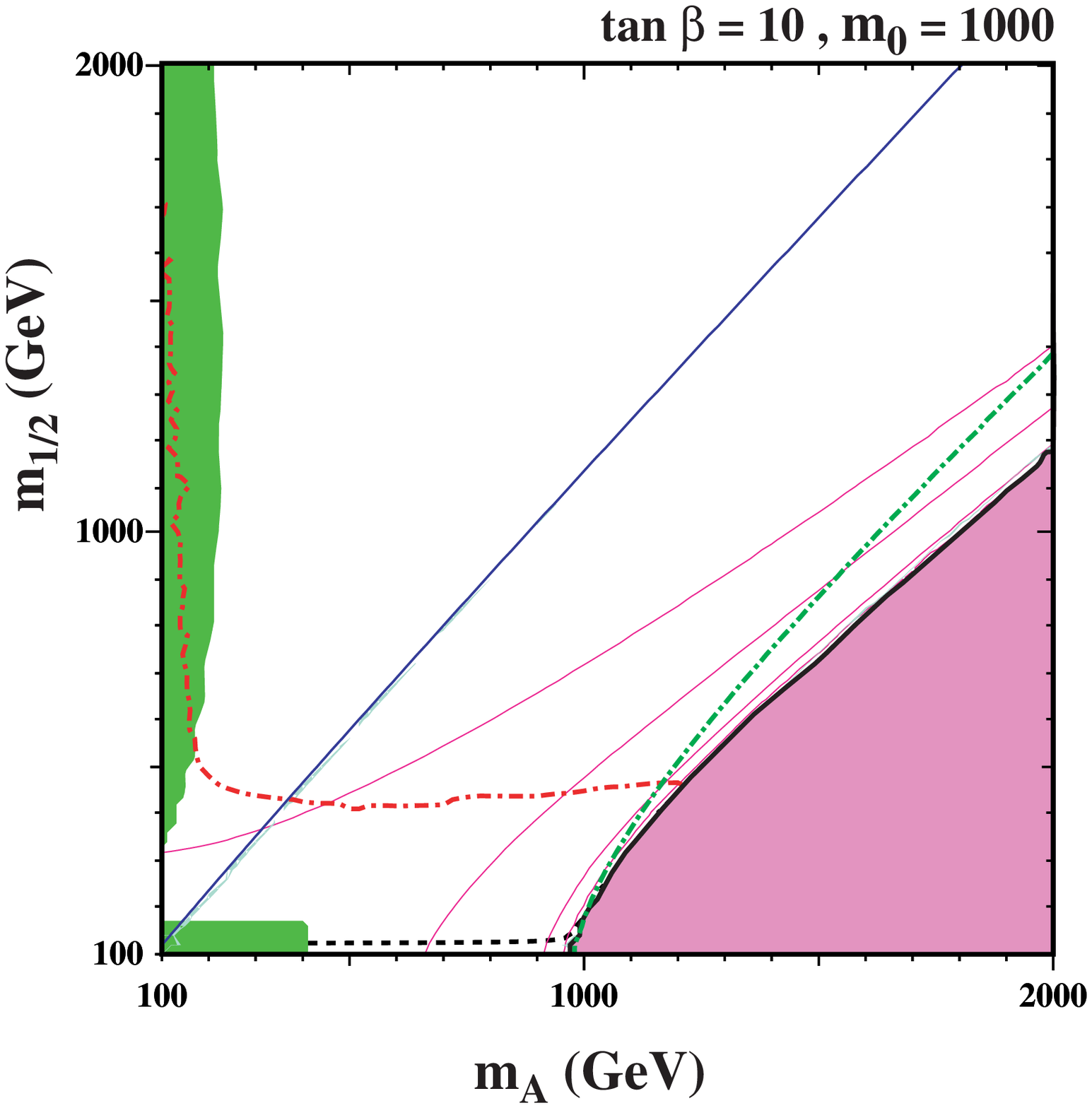,height=7cm}}
\mbox{\epsfig{file=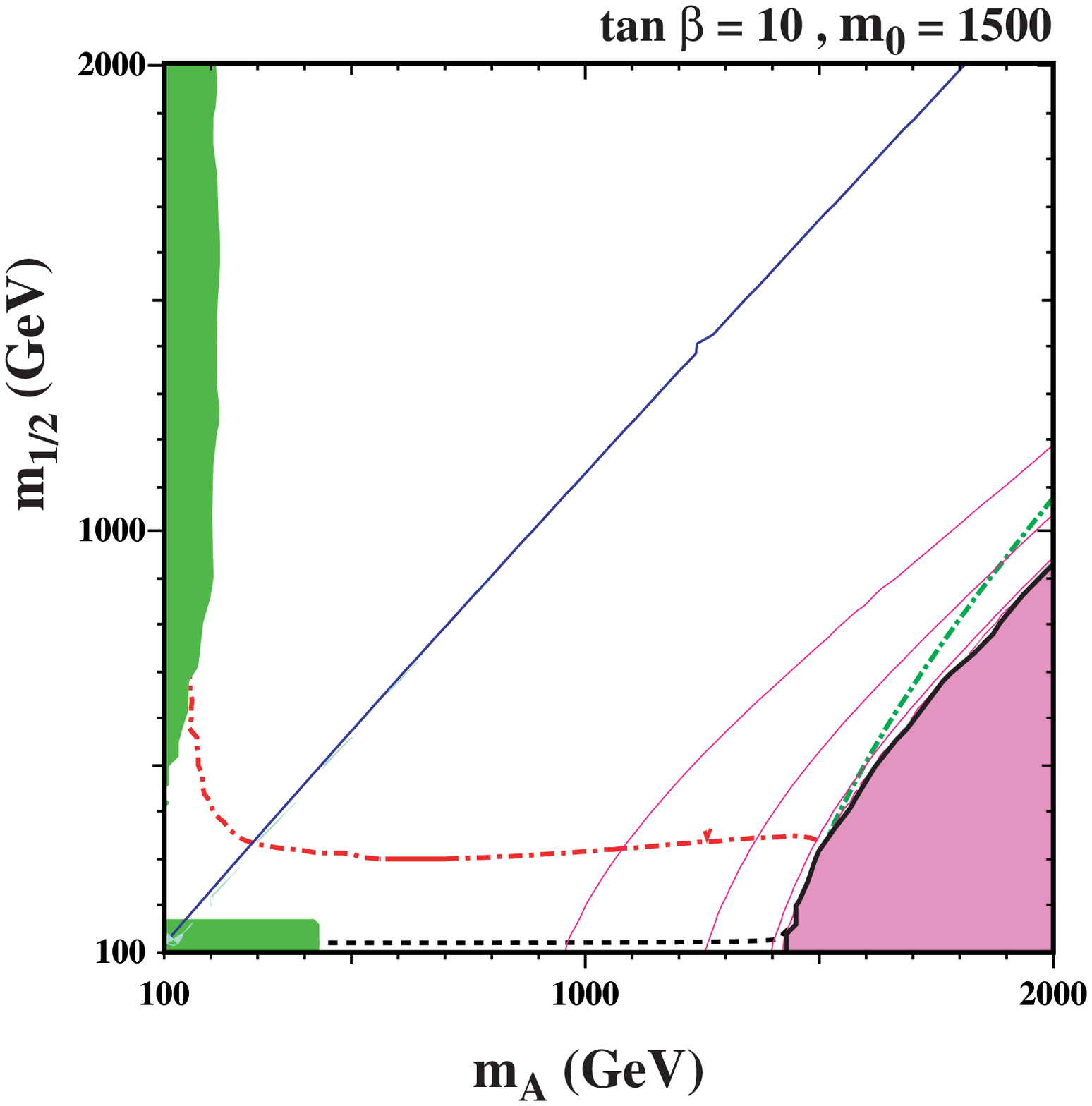,height=7cm}}
\end{center}
\caption{\it Examples of NUHM1 $(m_A,m_{1/2})$ planes with $\tanb = 10$,
$A_0=0$, $\mu>0$, and $m_0=300$, 500, 1000, and 1500 GeV in Panels
(a), (b), (c), and (d), respectively.  Constraints are displayed as in
Figure~\ref{fig:m12m0fixmA}.
}
\label{fig:mAm12fixm0}
\end{figure}

\subsubsection{Fixed $m_0$}
\label{sec:mAvm12fixedm0}

Alternative ways to view the NUHM1 parameter space include fixing
either $m_0$ or $m_{1/2}$ and scanning over $m_A$. We first examine
the former option.  

We show in Fig.~\ref{fig:mAm12fixm0}
examples of the $(m_A,m_{1/2})$ planes for $m_0 = 300$, 500, 1000, and
1500 GeV.  The unfamiliar appearances of
the constraints can once again be understood by comparison with
panel~(a) of Fig.~\ref{fig:m12m0fixmA}.  For example, for $m_0=300$~GeV,
as seen in panel~(a), we note that the upper third of the plane is
excluded due to a charged LSP. This reflects the fact that in the CMSSM plane, for fixed
$m_0$, $m_{\stau}$ increases more slowly than $m_\chi$ as $m_{1/2}$ increases, so that at
large $m_{1/2}$ the $\stau$ becomes the LSP.  Increasing $m_0$
postpones the $\stau$-LSP region to larger $m_{1/2}$, so that this
constraint almost disappears in panel (b) where $m_0 = 500$~GeV, and does not appear at all
in panels~(c) and (d), where $m_0 = 1000$
and 1500~GeV, respectively.  While there is no ${\tilde e}$-LSP region in
the CMSSM plane, as seen in panel (a) of Fig.~1,
the selectron mass renormalization is similar to that
of the stau, so the selectron-LSP regions in the NUHM1 planes shift
similarly to larger $m_{1/2}$.

The other unphysical regions in CMSSM planes occur in their
upper left corners, where there is no consistent
electroweak vacuum.  As seen in panel (a) of
Fig.~\ref{fig:m12m0fixmA}, this issue arises at low $m_{1/2}$ and large $m_0$.  As
$m_0$ is increased, the boundary of this region moves to larger $m_{1/2}$ and $m_A$. 
The positive correlation between $m_A$ and $m_{1/2}$ along this boundary is
seen clearly in all the panels of Fig.~\ref{fig:mAm12fixm0}. We also see that,
particularly at small $m_{1/2}$, this boundary also retreats to larger $m_A$ as
$m_0$ increases. Following the
boundary of this excluded region are the contours of constant
$\mu$, which converge slightly as $m_A$ and $m_{1/2}$ increase. 
Also apparent in panel~(a) for $m_0 = 300$~GeV is a small region at low
$m_{1/2}$ and $m_0$ that is favored by $g_\mu - 2$, which disappears
for larger $m_0$.
We also see at very low $m_{1/2}$ the LEP chargino bound. The dominant 
experimental constraints in these planes are the LEP limits on the
Higgs mass and the branching ratio of \mbox{$b \goto s \gamma$,} which exclude the areas
below the dot-dashed red contour and in the green shaded region, respectively. 

There are two viable
WMAP-compatible regions in these planes. One is the upper 
portion of the rapid-annihilation funnel, which is
oriented diagonally in the planes, close to the diagonal blue line where $m_\chi = m_A/2$.  
Since the position of the funnel
is defined by the LSP mass, which in this case depends primarily on
$m_{1/2}$ due to its bino-like character, and the pseudoscalar Higgs
mass, which forms the $x$-axis, the rapid-annihilation funnel is fixed
in the plane as $m_0$ is varied. The other viable WMAP-compatible region
(less immediately apparent in these plots) is the focus-point region which tracks the boundary
of the region where electroweak symmetry breaking is not possible. 

In each plane of Fig.~\ref{fig:mAm12fixm0}, the CMSSM contour runs diagonally 
through the contours of constant
$\mu$. For $m_0 = 300$ GeV, the CMSSM contour starts in the bulk region
at low $m_{1/2}$. Many of these points lie in the region favored by
$g_\mu - 2$, but this portion of the plane is excluded by the LEP bound
on the Higgs mass. As we follow the CMSSM contour to larger $m_{1/2}$
(larger $m_A$), we see that $\mu$ is increasing along the contour. This
corresponds to following a contour of constant $m_0$ horizontally
across the CMSSM $(m_{1/2},m_0)$ plane. Eventually, at large $m_{1/2}$
and any fixed value of $m_0$, the CMSSM contour
intersects the region where the $\stau$ is the LSP, but not the
${\tilde e}$-LSP region. As we increase
$m_0$, the $\stau$-LSP region is postponed to larger $m_{1/2}$. 
The CMSSM contours at large $m_0$ lie above the bulk region, but the
LEP constraint on the Higgs mass is still important, as it is only very weakly dependent on
$m_0$. The rapid-annihilation funnel 
region of WMAP-compatible neutralino relic density is bounded at low
$m_{1/2}$ by the LEP Higgs constraint and, for low $m_0$, at large
$m_{1/2}$ by the $\stau$-LSP region.  The funnel occurs at larger
$\mu$ than we expect in the CMSSM.

According to previous studies~\cite{lhc,cmstdr}, 
the LHC should find a signal of supersymmetry in the CMSSM scenario with 10~fb$^{-1}$ of
integrated luminosity if
$m_{1/2} \lesssim 900 (900) (800) (700)$~GeV for $m_0 = 300 (500)
(1000) (1500)$~GeV. In the NUHM1, for fixed $m_{1/2}$ and $m_0$,
 the spectrum of charged scalars and
gauginos is only affected through loop corrections to the RGEs, so we expect a
similar LHC reach for these values of $m_0$, shown in panels (a, b, c) and (d) of Fig.~\ref{fig:mAm12fixm0}. This means that
progressively shorter sections of the rapid-annihilation funnels and focus-point
strips are likely to be accessible to the LHC.

\begin{figure}[ht!]
\begin{center}
\mbox{\epsfig{file=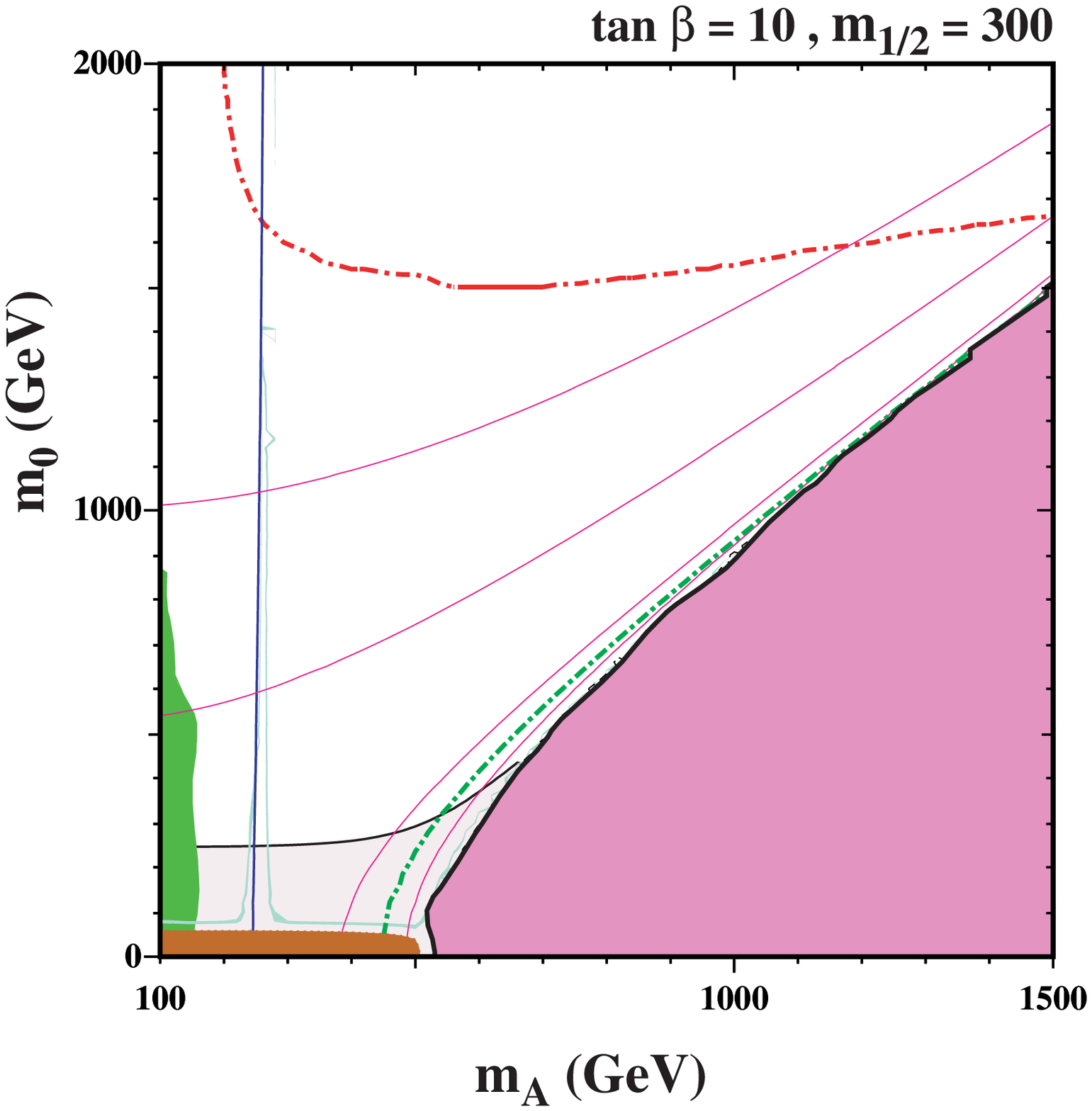,height=7cm}}
\mbox{\epsfig{file=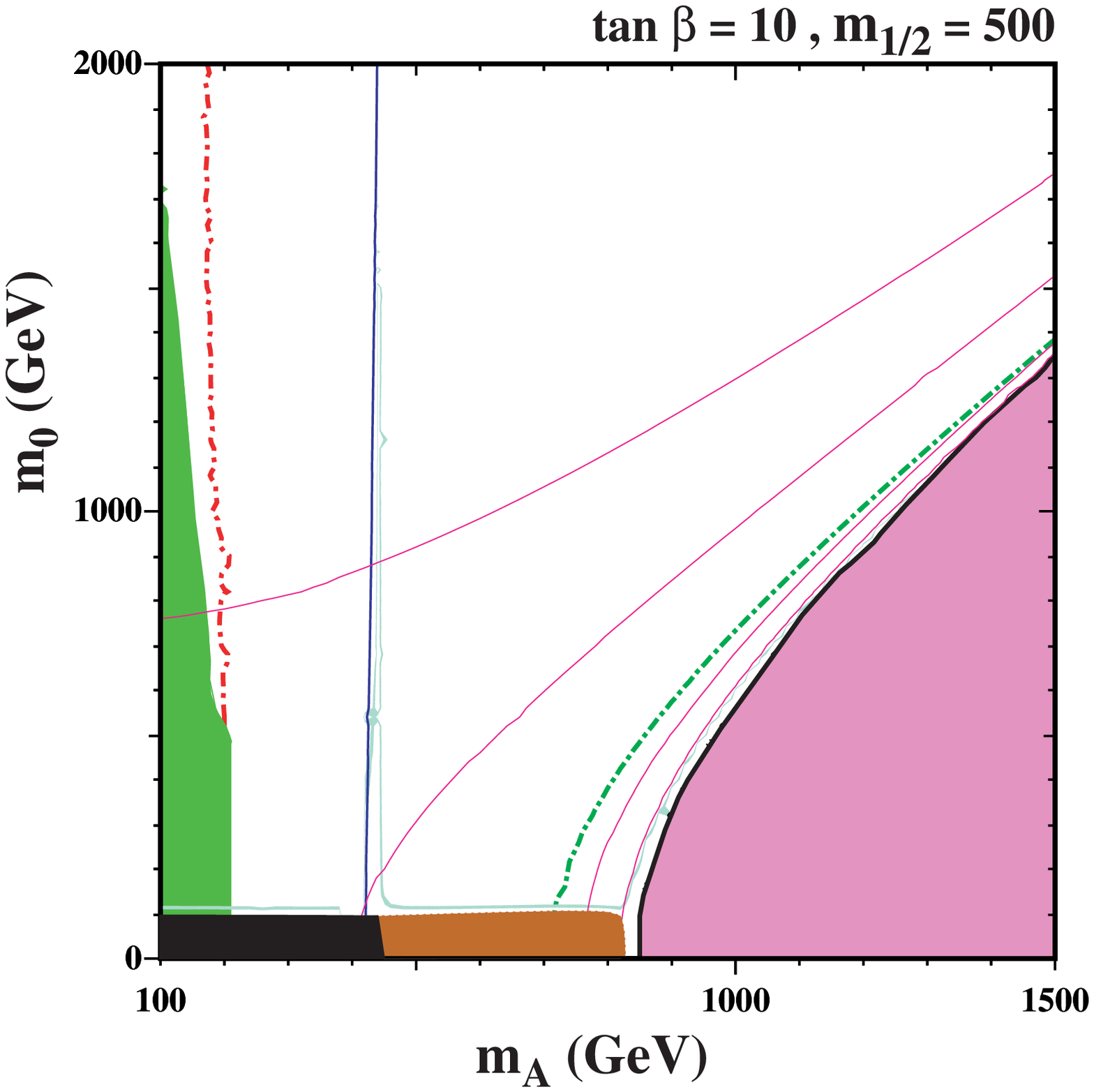,height=7cm}}
\end{center}
\begin{center}
\mbox{\epsfig{file=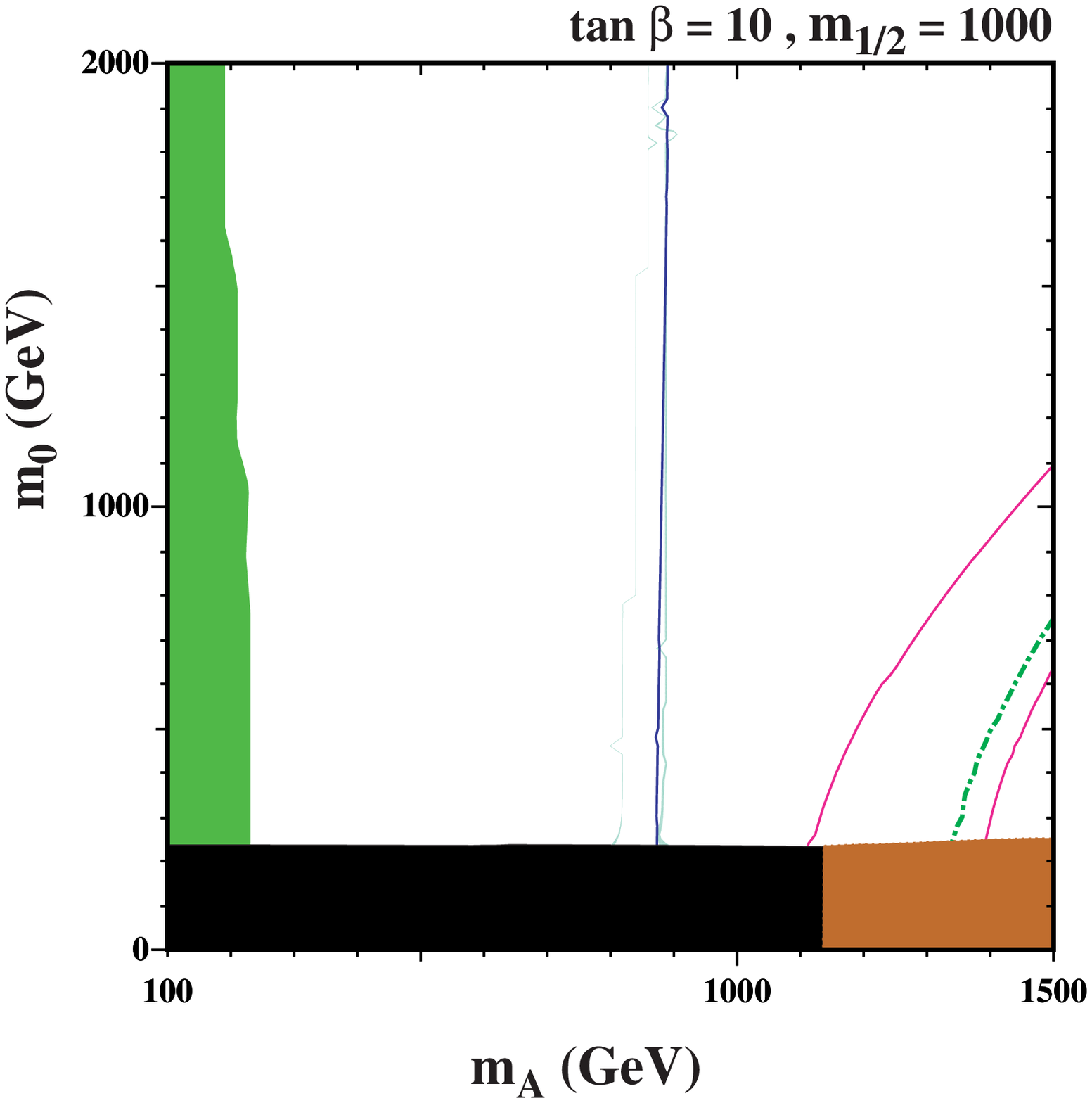,height=7cm}}
\mbox{\epsfig{file=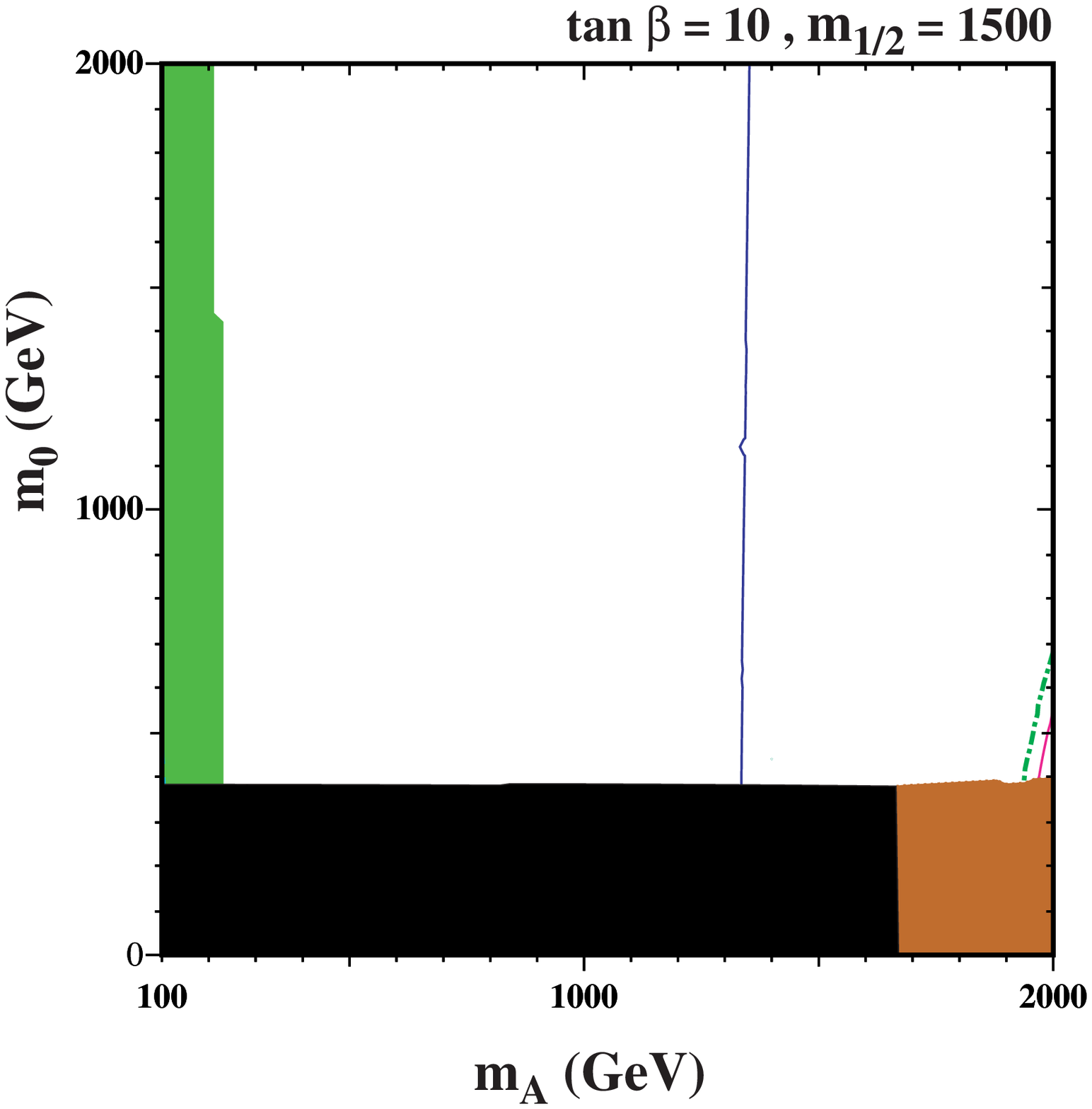,height=7cm}}
\end{center}
\caption{\it Examples of NUHM1 $(m_A,m_0)$ planes with $\tanb = 10$,
$A_0=0$, $\mu>0$, and $m_{1/2}=300$, 500, 1000, and 1500 GeV in Panels
(a), (b), (c), and (d), respectively.  Constraints are displayed as in Figure~\ref{fig:m12m0fixmA}.}
\label{fig:mAm0fixm12}
\end{figure}

\subsubsection{Fixed $m_{1/2}$}

We now discuss NUHM1 parameter space for various fixed values of $m_{1/2}$, as shown in the
$(m_A, m_0)$ planes in Fig.~\ref{fig:mAm0fixm12}. 
We note first that the forbidden stau LSP region is absent for low $m_{1/2} = 300$~GeV,
as seen in panel (a), puts in an appearance at low $m_0$ when $m_{1/2} = 500$~GeV, as seen in
panel (b), and reaches progressively to larger $m_0$ at larger $m_{1/2}$, as seen in
panels (c) and (d). This behaviour was to be expected from the analogous feature in the
CMSSM, shown in panel (a) of Fig.~\ref{fig:m12m0fixmA}, and reflects the fact that
$m_\chi$ increases more rapidly with $m_{1/2}$ than does $m_{\tilde
\tau_1}$. At larger $m_{1/2}$ we see the emergence of the selectron
LSP region at low $m_A$. We also note that the electroweak vacuum exclusion retreats to smaller $m_0$ and
larger $m_A$ as $m_{1/2}$ increases, disappearing altogether for
$m_{1/2} = 1000$ and 1500~GeV, again reflecting the CMSSM feature seen in
panel (a) of Fig.~\ref{fig:m12m0fixmA}.

One of the dominant experimental constraints on the parameter space
is that due to the LEP Higgs mass bound, 
which excludes most of the plane for $m_{1/2} = 300$~GeV and low $m_A$ for 
$m_{1/2} = 500$~GeV, as seen in panels (a) and (b), respectively. 
The Higgs mass is more sensitive to variations in $m_0$ at
lower $m_{1/2}$, whereas at large $m_{1/2}$ the Higgs mass is primarily
sensitive to $m_{1/2}$ and less dependent on $m_0$ (as in the
CMSSM). 
We also note that the branching ratio of $b \goto s \gamma$ excludes a strip
of parameter space that expands slowly with $m_A$.

There are three distinct regions of WMAP-compatible relic density in these
$(m_A,m_0)$ planes. The first is the vertical rapid-annihilation
funnel, where the relic density decreases drastically. This moves to larger $m_A$ as
$m_{1/2}$ increases, reflecting the movement of the blue line where $m_{\chi} = m_A/2$.
The second region of good relic density is the coannihilation strip, which is present when
$m_{1/2} \lesssim 900$~GeV. In fact, we see that the rapid-annihilation funnel rises
directly out of the coannihilation strip where the two coincide, as
also seen in Fig.~\ref{fig:m12m0fixmA}. Finally, the third is the focus-point strip,
which tracks the region excluded by the requirement of electroweak symmetry breaking.
As $m_{1/2}$ continues to increase, this strip
is pushed to values of $m_A$ beyond those plotted. 

The CMSSM contours in the $(m_A,m_0)$ planes correspond to following
a strip of constant $m_{1/2}$ in the $(m_{1/2}, m_0)$ plane shown in panel (a)
of Fig.~\ref{fig:m12m0fixmA} upwards from the coannihilation strip. 
Since $\mu$ depends strongly on $m_{1/2}$, but has little
sensitivity to the value of $m_0$, these contours appear to be roughly
contours of constant $\mu$ in each case. For low values of $m_{1/2}$,
the CMSSM contour begins in the bulk region at low $m_0$. This is a
region favoured by $g_\mu - 2$ but strongly excluded by the LEP Higgs bound. 
Eventually, we find the focus-point region at very large $m_0$.
In panel (b), the CMSSM line
arches up from the $\stau$-LSP region towards the region where there is no
electroweak symmetry breaking. In Panels~(c) and
(d), the CMSSM contour begins at low $m_0$ and large $m_A$ in the $\stau$-LSP
region, but there are no further visible features of interest. As already noted,
both the CMSSM contour and the rapid
annihilation funnel move to larger $m_A$ as
$m_{1/2}$ increases. However, since the CMSSM
contour moves more quickly than the funnel, there is no rapid
annihilation funnel in the CMSSM for $\tanb =10$, unlike the NUHM1 case.

According to previous studies~\cite{lhc,cmstdr}, 
the LHC should find a signal of supersymmetry in the CMSSM scenario with 10~fb$^{-1}$ of
integrated luminosity if
$m_{0} < 2000$~GeV for $m_{1/2} = 300 (500)$~GeV. As discussed in
section~\ref{sec:mAvm12fixedm0}, we expect a similar reach in the
NUHM1 for comparable values of $m_{1/2}$,
as shown in panels (a) and (b) of Fig.~\ref{fig:m12m0fixmA}. This means that
all of the visible parts of these planes should be accessible to the LHC. On the
other hand, previous analyses~\cite{lhc,cmstdr} suggest that in the CMSSM, the parameter
space with $m_{1/2} \gtrsim 1000$~GeV would be inaccessible without an increase in the
integrated luminosity. In the NUHM1 planes, due to the appearance of
the rapid-annihilation funnel, one may find fairly light
charged scalars even if $m_{1/2} > 1000$~GeV, as shown in panels (c) and (d).

\subsubsection{Varying $\tan \beta$}


Finally, we discuss the characteristics of the NUHM1 parameter space
as we vary $\tanb$.  We recall that in the CMSSM at large $\tanb$ a
rapid-annihilation funnel appears in the $(m_{1/2},m_0)$ plane
when $\tanb > 35$, extending from the coannihilation strip to larger $(m_{1/2},m_0)$. In addition, at large $\tanb$ the
excluded $\stau$-LSP region becomes more prominent in the $(m_{1/2},m_0)$
plane at low $m_0$, and the branching ratio of $b \goto s \gamma$
excludes more of the plane at low $m_{1/2}$~\footnote{{\it Modulo}
cancellations between different contributions, that sometimes
introduce an allowed corridor through the excluded region, even at low
$m_{1/2}$.}.  The effects of variations
in $\tanb$ on these constraints alter the appearance of the NUHM1
planes, as well.

\begin{figure}[ht!]
\begin{center}
\mbox{\epsfig{file=nuhm1_mAvM_10_500s.eps,height=7cm}}
\mbox{\epsfig{file=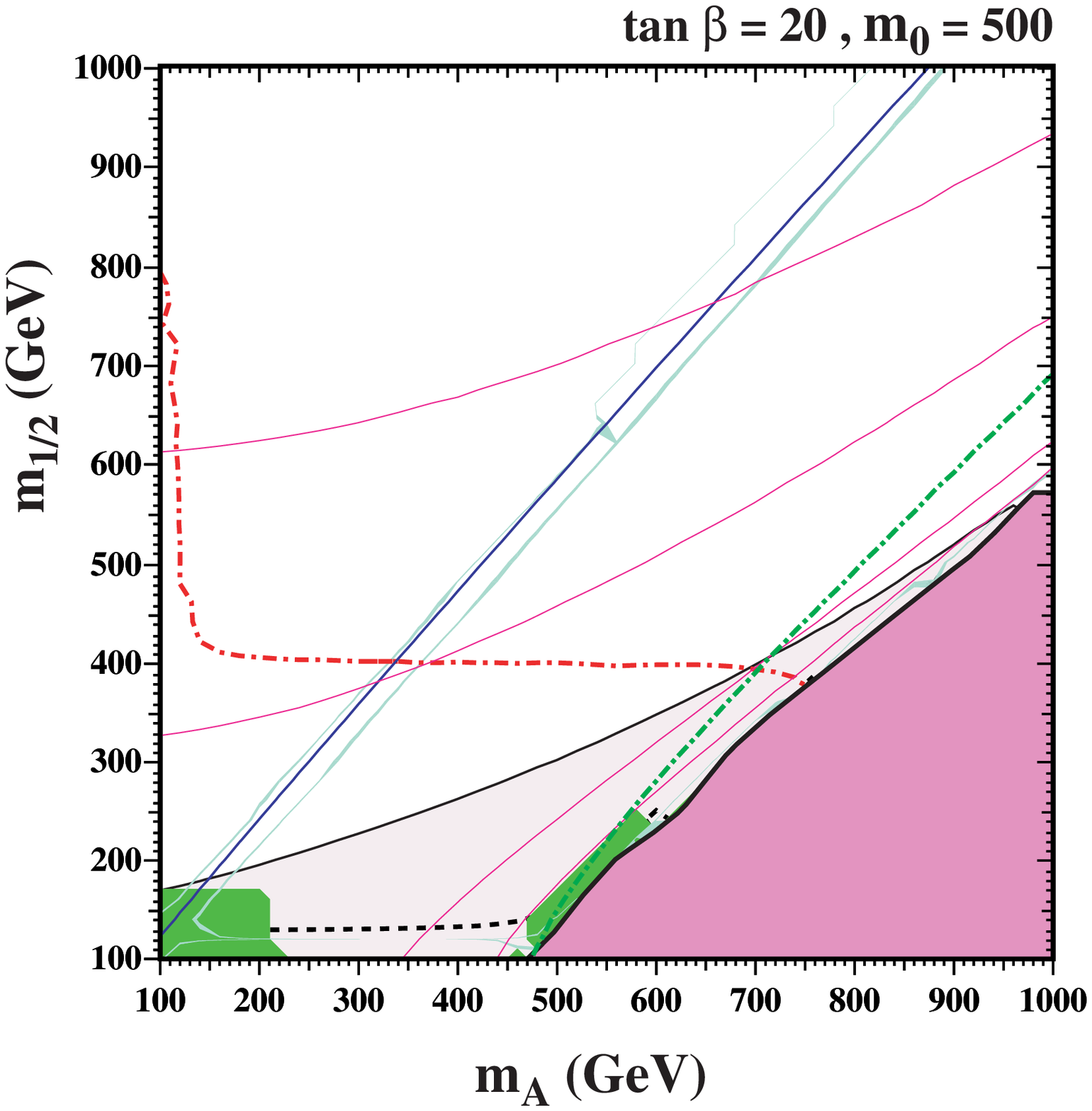,height=7cm}}
\end{center}
\begin{center}
\mbox{\epsfig{file=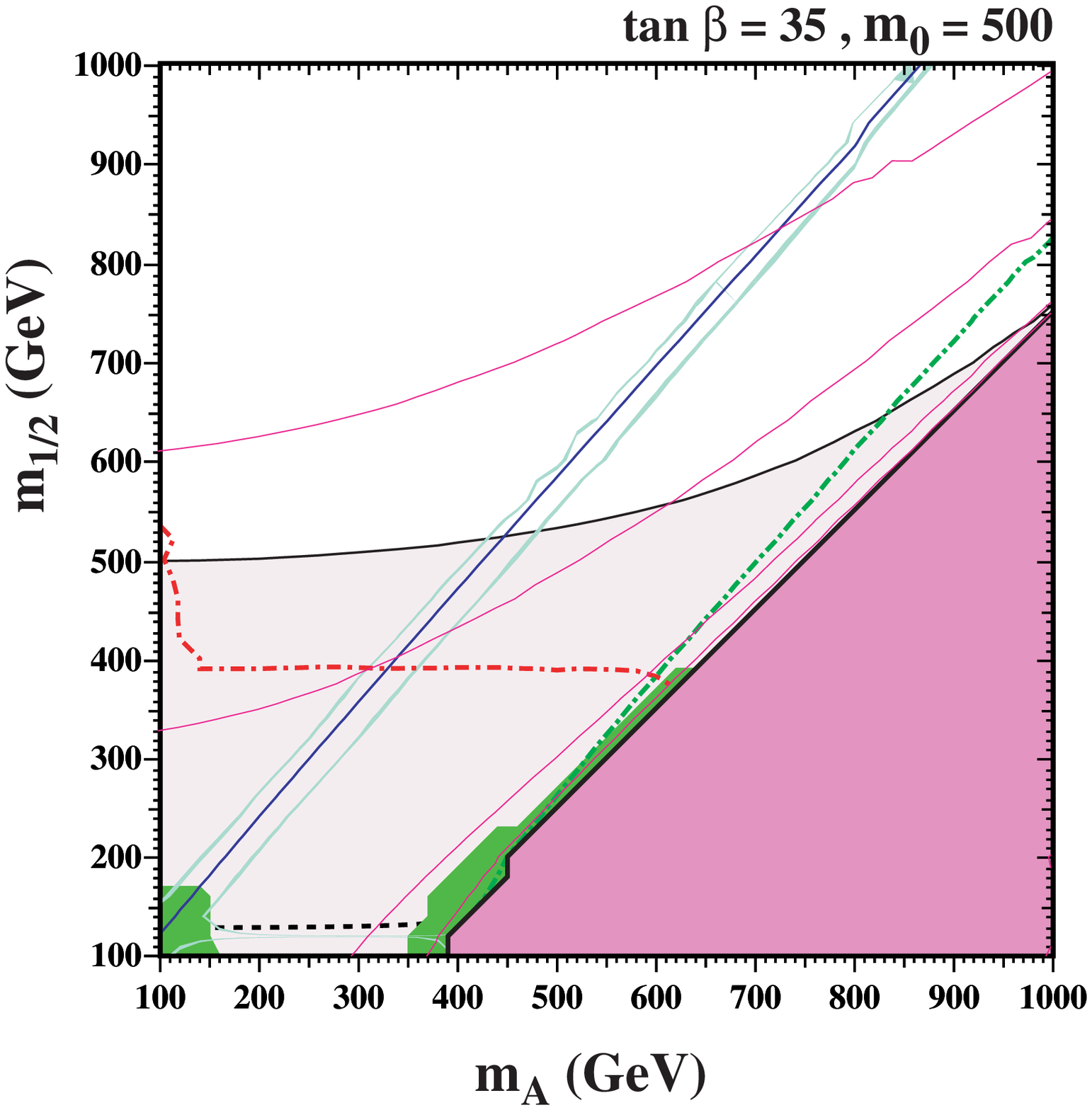,height=7cm}}
\mbox{\epsfig{file=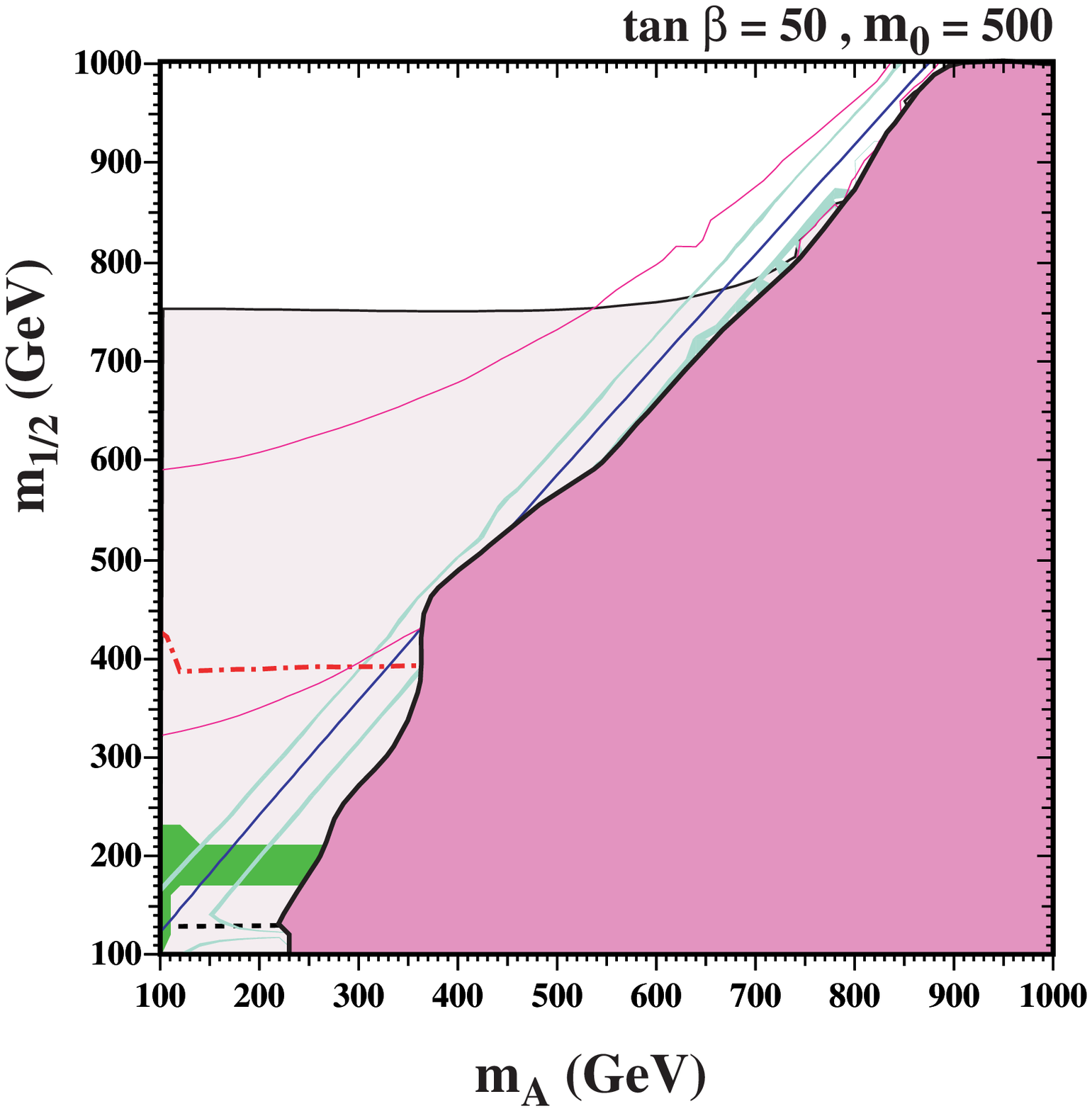,height=7cm}}
\end{center}
\caption{\it Examples of NUHM1 $(m_A,m_{1/2})$ planes with $m_0 = 500$ GeV,
$A_0=0$, $\mu>0$, and $\tanb=10$, 20, 35, and 50 in Panels
(a), (b), (c), and (d), respectively.  Constraints are displayed as in Figure~\ref{fig:m12m0fixmA}.
}
\label{fig:mAm12tanb}
\end{figure}

In Fig.~\ref{fig:mAm12tanb}, we show NUHM1 $(m_A,m_{1/2})$ planes
with $m_0 = 500$~GeV and $\tanb = 10$, 20, 35, and 50 in panels~(a),
(b), (c), and (d), respectively. Note that panel~(a) of
Fig.~\ref{fig:mAm12tanb} is the same as panel~(b) of
Fig.~\ref{fig:mAm12fixm0}. As $\tanb$ is increased, $\mu$ decreases,
as is evidenced by the movement of the contours of constant $\mu$ out into
the plane and the expansion of the region where there are no
consistent solutions to the electroweak vacuum conditions. As a
result, the CMSSM contour is pushed to lower $m_A$ for fixed
$m_{1/2}$, moving closer to the rapid-annihilation funnel. 
In the CMSSM, however, the rapid-annihilation
funnel begins at roughly $m_0 = 800$~GeV, so the
CMSSM contour does not cross the rapid-annihilation funnel even at $\tanb =
50$ in these planes with $m_0 = 500$~GeV. In these NUHM1 planes, the
location of the rapid-annihilation funnel is almost independent of $\tanb$.

In contrast to the CMSSM, in these particular NUHM1
planes the constraint due to the branching ratio of $b
\goto s \gamma$ becomes insignificant at large $\tanb$. On the other hand, the region
favored by $g_\mu - 2$ expands such that a significant portion of the 
rapid-annihilation funnel falls within it, as well as the LEP constraint
on the Higgs mass. In addition to the fixed rapid-annihilation funnel, in each panel of
Fig.~\ref{fig:mAm12tanb} there is a narrow WMAP strip close to the electroweak
symmetry-breaking boundary. For $\tanb = 10$, portions of the funnel and this boundary strip
are compatible with all these constraints, except $g_\mu - 2$, for $m_{1/2} > 500$~GeV.
When $\tanb = 20$, $m_{1/2} > 400$~GeV is allowed by the Higgs constraint, 
and part of this boundary strip is also compatible with $g_\mu - 2$.
When $\tanb = 35$, the region
allowed by $g_\mu -2$ extends to larger $m_{1/2}$, and parts of both the
rapid-annihilation funnel and the boundary strip are compatible with it and with $m_h$.

\begin{figure}[ht!]
\begin{center}
\mbox{\epsfig{file=nuhm1_mAvm0_10_500s.eps,height=7cm}}
\mbox{\epsfig{file=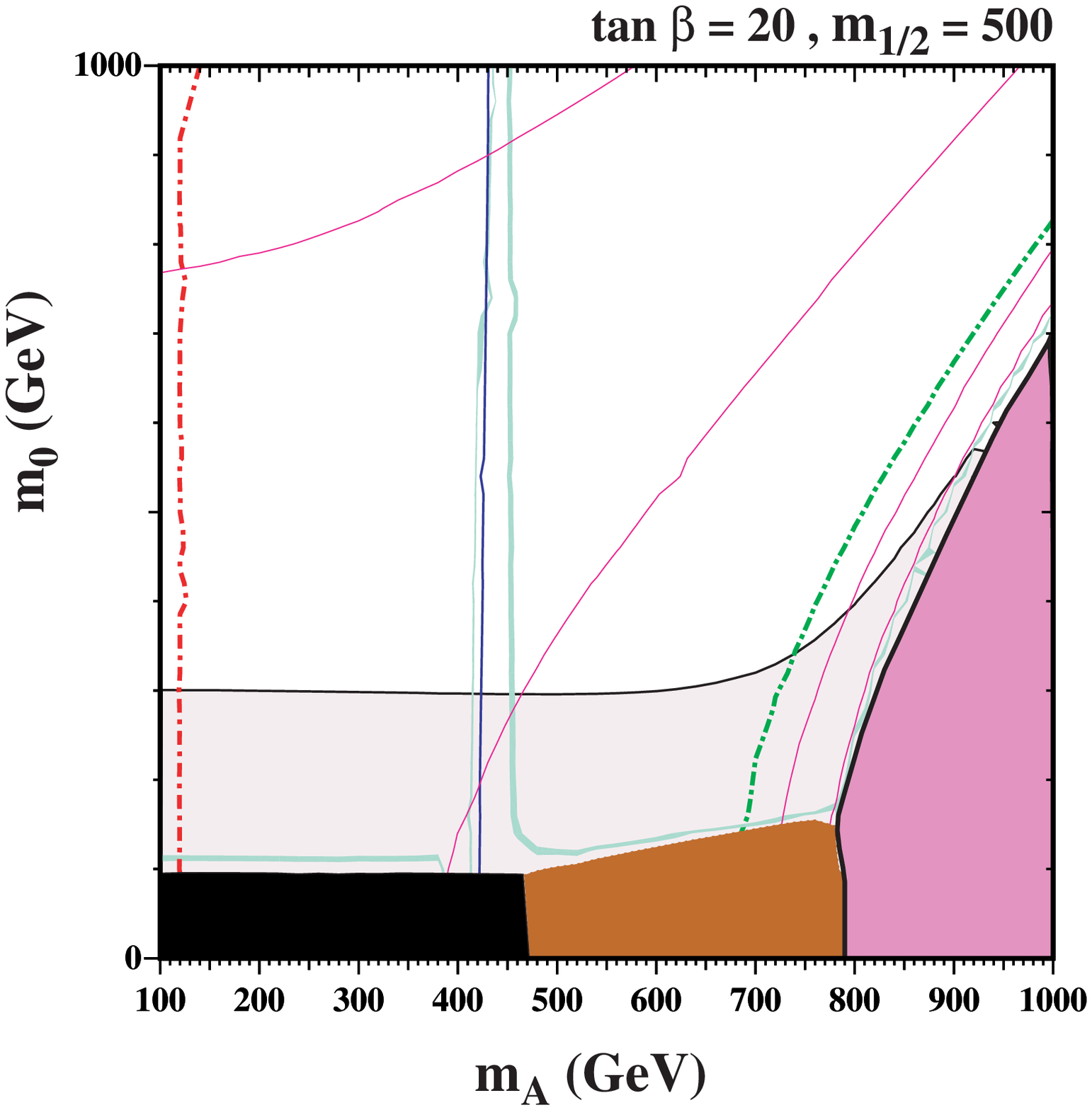,height=7cm}}
\end{center}
\begin{center}
\mbox{\epsfig{file=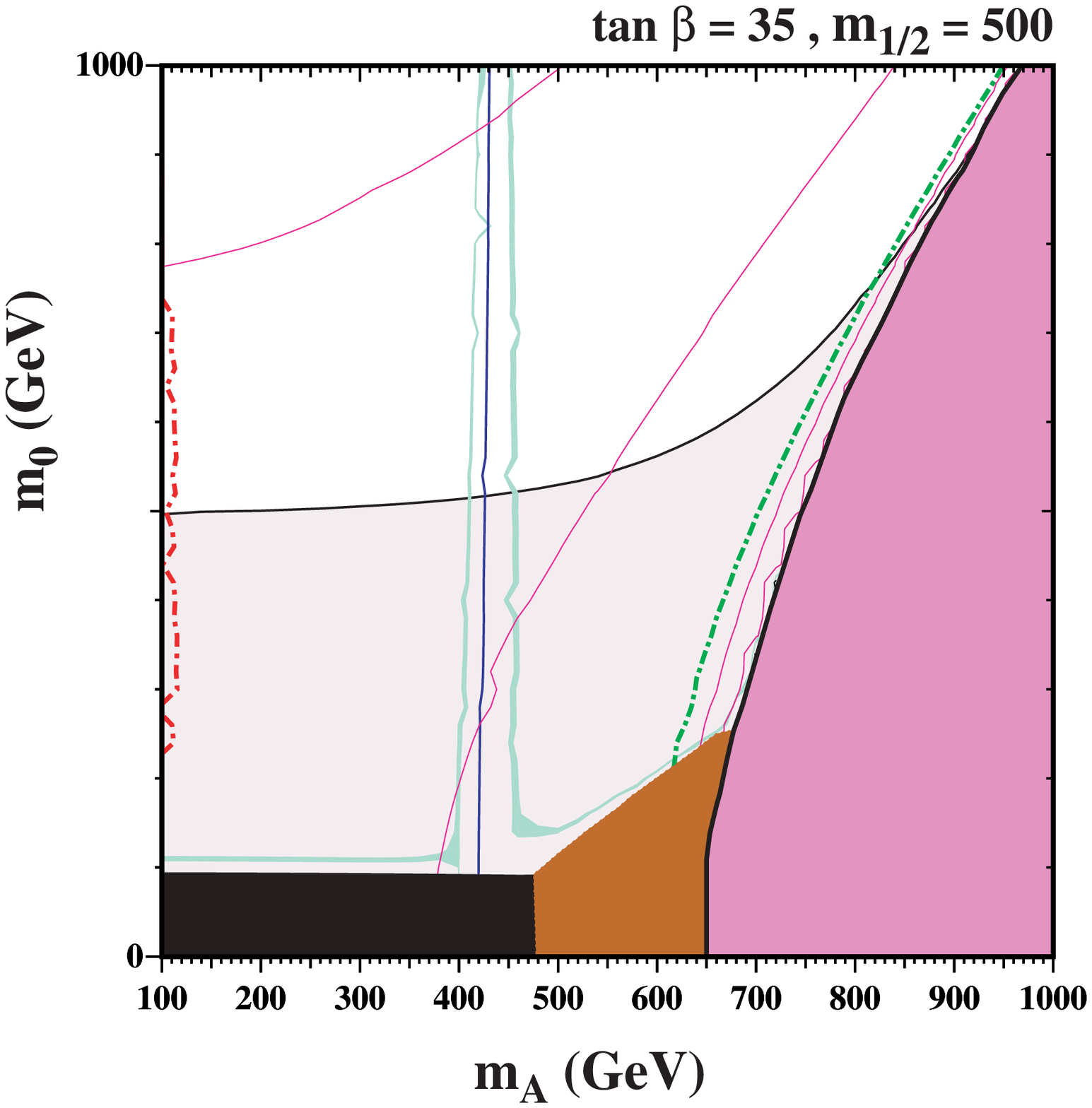,height=7cm}}
\mbox{\epsfig{file=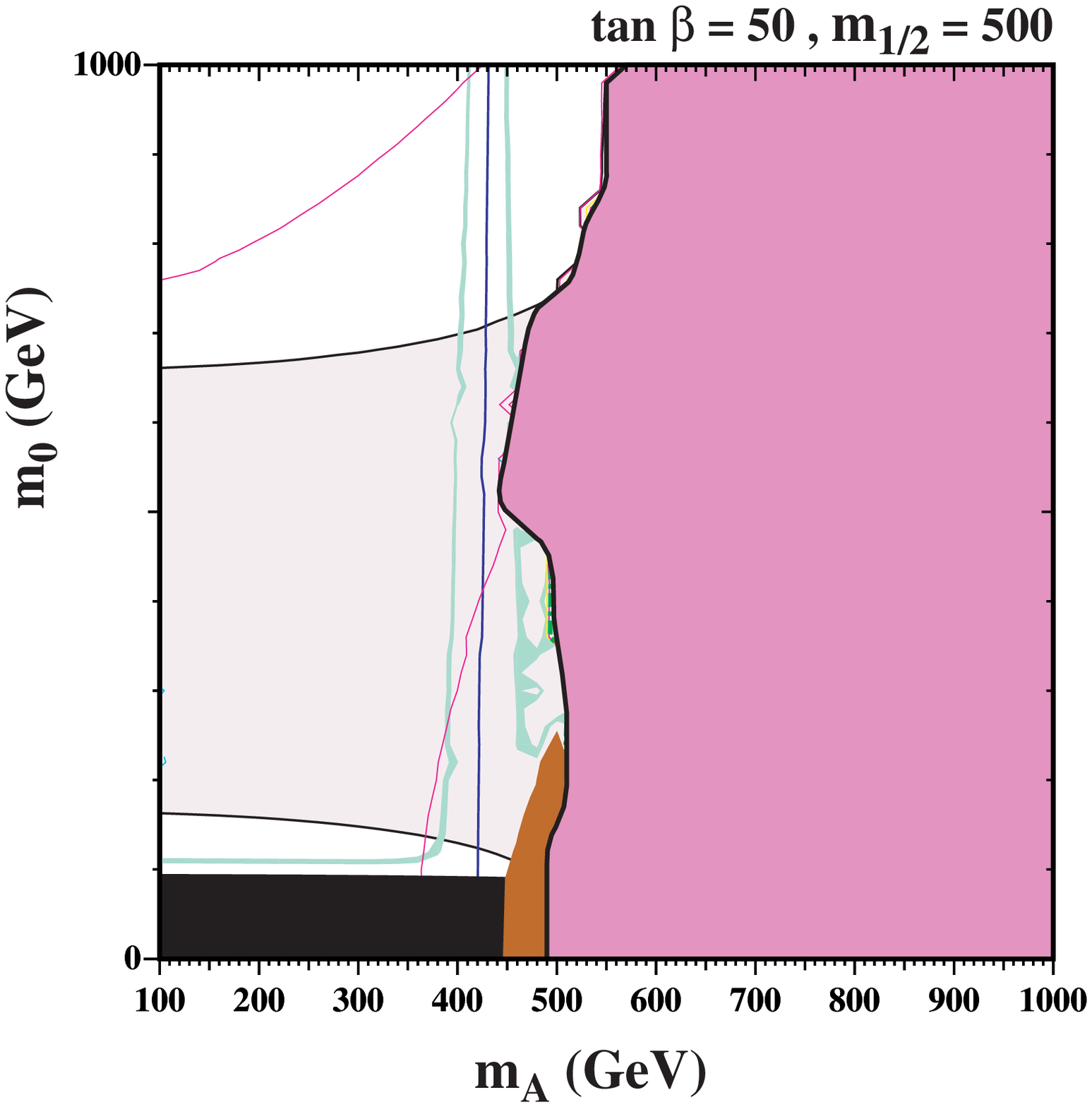,height=7cm}}
\end{center}
\caption{\it Examples of NUHM1 $(m_A,m_0)$ planes with $m_{1/2} = 500$ GeV,
$A_0=0$, $\mu>0$, and $\tanb=10$, 20, 35, and 50 in Panels
(a), (b), (c), and (d), respectively.  Constraints are displayed as in Figure~\ref{fig:m12m0fixmA}.
}
\label{fig:mAm0tanb}
\end{figure}

In Fig.~\ref{fig:mAm0tanb}, we show NUHM1 $(m_A,m_0)$ planes
with $m_{1/2} = 500$~GeV and $\tanb = 10$, 20, 35, and 50 in panels~(a),
(b), (c), and (d), respectively. Note that panel~(a) of
Fig.~\ref{fig:mAm0tanb} is the same as panel~(b) of
Fig.~\ref{fig:mAm12fixm0}. As $\tanb$ increases, we see that
the boundary of the electroweak symmetry-breaking region moves to
lower values of $m_A$, while the $\stau$-LSP region changes its shape, 
becoming less important at small $m_A$ but more important at larger
$m_A$. In contrast, the $\sel$-LSP region is fixed
at very low $m_0$ as $\tanb$ is increased, becoming visible as the
$\stau$-LSP region shifts, and it is bordered by a $\chi-\sel$
coannihilation strip. The LEP Higgs constraint excludes only a narrow strip at small $m_A$, almost
independent of $m_0$, that narrows as $\tanb$ increases. The $b \to s \gamma$
constraint is visible only for $\tanb = 10$, at small $m_A$. There is no region favoured
by $g_\mu - 2$ when $\tanb = 10$, but this appears and expands as $\tanb$ increases.
The CMSSM line arches up and outwards in each panel, following and gradually
approaching the boundary of electroweak symmetry breaking.

The strip where the dark matter density falls within the WMAP range exhibits
the familiar features of a rapid-annihilation funnel, which is near-vertical and 
straddles the blue line where $m_\chi = m_A/2$, a coannihilation strip near
the boundary of the charged LSP regions, and a strip near the boundary of the
region where there is no electroweak symmetry breaking. This region is
compatible with all the phenomenological constraints, including also
$g_\mu - 2$ when $\tanb = 20$ or more. There are in general two intersections
with the CMSSM line, corresponding to the coannihilation and fixed-point
strips in the $(m_{1/2}, m_0)$ planes for different values of $\tanb$.
The rapid-annihilation funnel is in general at lower $m_A$ than the CMSSM
line, except for $\tanb = 50$.
The analogous planes for larger $m_{1/2}$ would exhibit more intersections between
the CMSSM line and the rapid-annihilation funnel.

According to previous studies~\cite{lhc,cmstdr}, 
the LHC should find a signal of supersymmetry in the CMSSM scenario
for $\tanb =10$ with 10~fb$^{-1}$ of
integrated luminosity if
$m_{0} < 2000$~GeV for $m_{1/2} = 500$~GeV.  Given the sensitivity of
the sparticle spectrum to the value of $\tanb$, we estimate that the visible parts of the planes in 
Fig.~\ref{fig:mAm0tanb} should be accessible to the LHC.


\subsection{The NUHM1 with $\mu$ as a Free Parameter}
\label{sec:mu}

\begin{figure}[ht!]
\begin{center}
\mbox{\epsfig{file=mvMcontours_10s.eps,height=7cm}}
\mbox{\epsfig{file=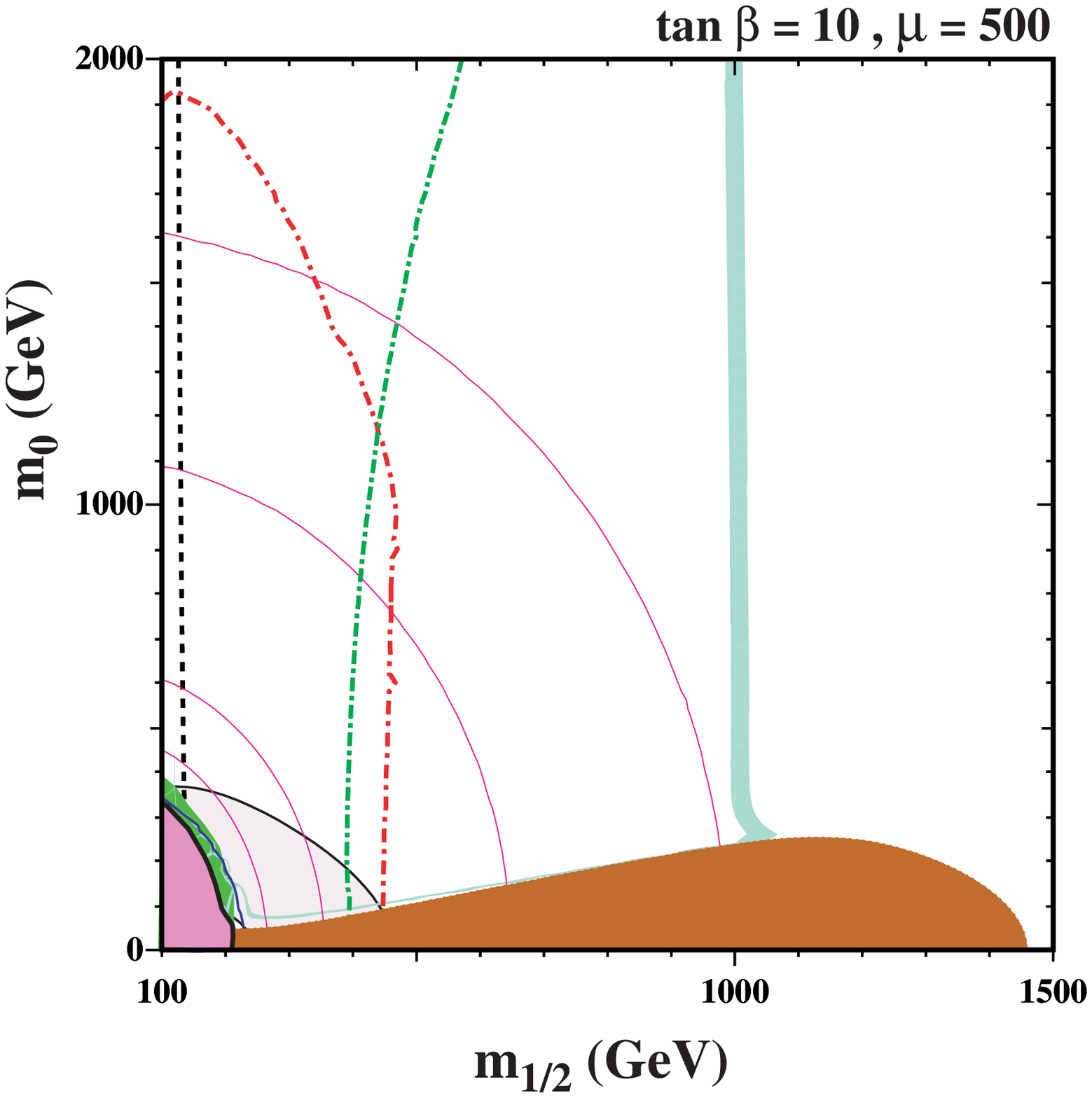,height=7cm}}
\end{center}
\begin{center}
\mbox{\epsfig{file=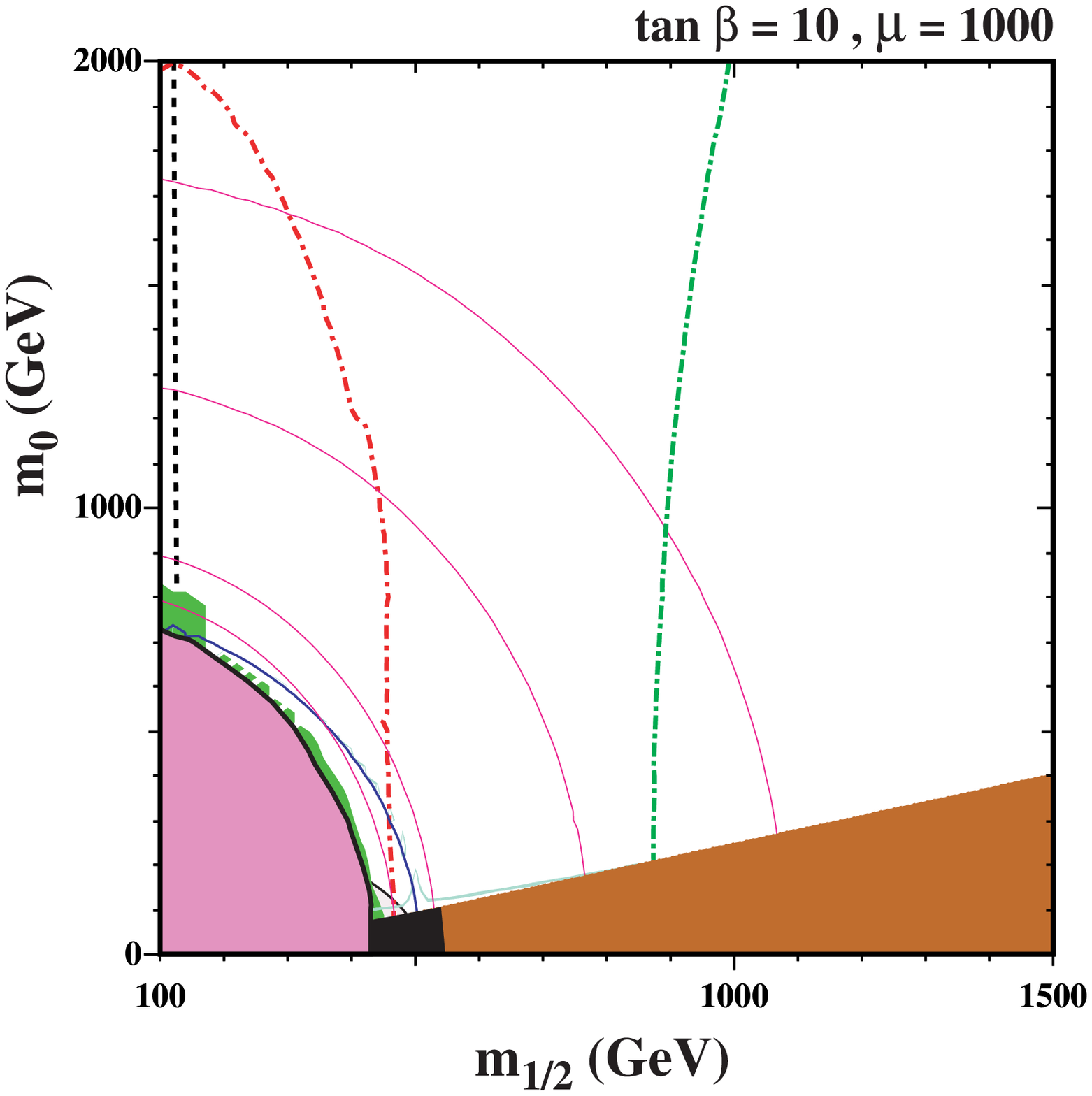,height=7cm}}
\mbox{\epsfig{file=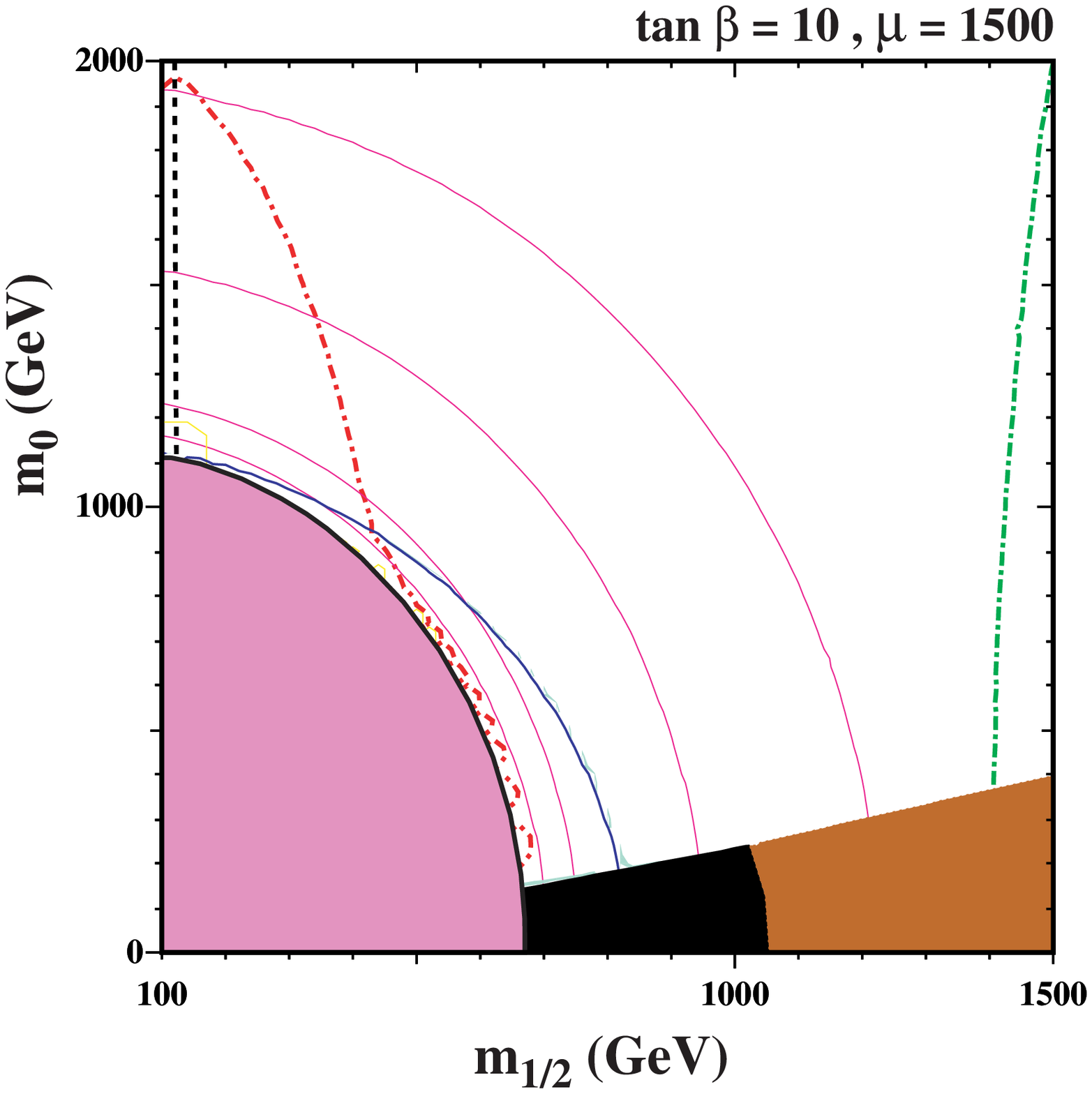,height=7cm}}
\end{center}
\caption{\it  Panel (a) shows the $(m_{1/2},m_0)$ plane for the CMSSM, with contours of $m_A$ and
$\mu$ of 300, 500, 1000, and 1500 GeV as described in the
text. Panels~(b), (c), and (d) show the NUHM1 $(m_{1/2},m_0)$ plane
with $m_u = 500$, 1000, and 1500 GeV, respectively.  Constraints are displayed as in Figure~\ref{fig:m12m0fixmA}.
}
\label{fig:m12m0fixmu}
\end{figure}

As discussed above, in the NUHM1, one may choose either $m_A$ or $\mu$
as the additional input to those of the CMSSM. In this subsection, we
re-examine the parameter space, this time choosing $\mu$ as a free parameter.  We
begin, as in Section~\ref{sec:mA}, with a comparison of the CMSSM
$(m_{1/2},m_0)$ planes with NUHM1 planes, now at fixed $\mu$.
In Fig.~\ref{fig:m12m0fixmu}, we show in panel~(a) the
CMSSM plane (identical to panel~(a) of Fig.~\ref{fig:m12m0fixmA}),
including the contours of constant $m_A$ and $\mu$ of 300, 500 1000,
and 1500~GeV.  Panels (b), (c), and (d) show the NUHM1 planes with
$\mu=500$, 1000, and 1500~GeV, respectively.

At first glance, the $(m_{1/2}, m_0)$ planes with fixed $\mu$ have
some similarities with those with fixed $m_A$. There are excluded regions at
very low $(m_{1/2}, m_0)$ where the pseudoscalar Higgs mass squared is
negative, corresponding to the absence of electroweak symmetry breaking,
surrounded by four contours of fixed $m_A = 300$, 500, 1000, and 1500~GeV. 
At small values of $m_0$, extending out to large $m_{1/2}$, there are
excluded $\stau$-LSP regions resembling those in the CMSSM.
As usual, the LEP chargino and Higgs constraints exclude regions at small $m_{1/2}$,
and $b \to s \gamma$ excludes strips near the electroweak symmetry-breaking boundaries
for $\mu = 500, 1000$~GeV, shown in panels (b) and (c), respectively. We also see in
these planes regions at low $m_{1/2}$ and $m_0$ that are favoured by $g_\mu - 2$.

There are three generic parts of the WMAP relic density strips in panels (b, c) and (d)
of Fig.~\ref{fig:m12m0fixmu}. There are
coannihilation strips close to the $\stau$-and ${\tilde e}$-LSP boundaries, and other strips
close to the electroweak symmetry-breaking boundaries. Arching between these
are curved rapid-annihilation funnels that appear at low $m_A$, with
strips of good relic density forming the funnel walls. 
For $\mu = 1000$ GeV, the rapid-annihilation funnel is
partially excluded by the branching ratio of $b \goto s \gamma$ and even more so
by $m_h$. 
Additionally, in panel (b) for $\mu = 500$~GeV, there is a fourth, near-vertical strip,
where the relic density is brought down into the WMAP range because of the large
mixing between the bino and Higgsino components in the LSP. For smaller $m_{1/2} < 500$~GeV,
the LSP is almost pure bino, and the relic density is too large except in the narrow
strips mentioned previously. This is the opposite of what
happens in the CMSSM, where the Higgsino fraction increases
at smaller $m_{1/2}$ at large $m_0$. On the other hand, for larger $m_{1/2} > 1000$~GeV, the LSP is
almost pure Higgsino, and the relic density falls below the WMAP 
range~\footnote{It is also this change in the nature of the LSP that causes the
boundary of the $\stau$-LSP region to drop. Since the
$\stau$ mass is affected only minimally by the value of $\mu$, we find
that $\stau$-LSP region terminates at some value of $m_{1/2}$ related
primarily to $\mu$.}. At large $m_0$ in panel (b) 
of Fig.~\ref{fig:m12m0fixmu}, 
it is only in the `crossover' strip  that the relic density falls within the WMAP range.
Analogous near-vertical crossover strips
are not visible in panels (c) and (d) of Fig.~\ref{fig:m12m0fixmu}, but would in principle
appear at larger $m_{1/2} \sim 2000, 3000$~GeV, respectively.

The CMSSM contour in each of panels (b, c) and (d) of Fig.~\ref{fig:m12m0fixmu}
is a roughly vertical line, the position of which
is determined by the value of $m_A$ that one would find from the
electroweak vacuum conditions in the
standard CMSSM. Since the contours of constant $m_A$ in these NUHM1
$(m_{1/2}, m_0)$ planes look very similar to the corresponding
contours in the CMSSM plane shown above, the CMSSM contours here in
turn look qualitatively similar to contours of constant $\mu$ in the CMSSM plane.
The CMSSM lines are compatible with WMAP only in infinitesimal cuts across
the coannihilation strips, missing all the excitement occurring elsewhere
in the planes, namely the focus-point, rapid-annihilation and
crossover strips.

In the NUHM1 $(m_{1/2},m_0)$ planes with fixed $\mu$, the crossover
strip and the rapid-annihilat- ion funnel comprise regions of interest
in addition to those commonly found in the CMSSM. Whereas the standard
CMSSM regions will be fairly well-covered by the LHC, there are
significant regions of the NUHM1 plane which may not be so easily
accessed.  For example, for $\mu=500$~GeV, 
as shown in panel (b) of Fig.~\ref{fig:m12m0fixmu}, the crossover strip runs at
$m_{1/2} \approx 1000$~GeV from $m_0 = 260$~GeV, where it is terminated
by the $\stau$-LSP region, to well beyond 10~TeV, crossing the CMSSM
contour at $m_0 =3400$~GeV.  Since the strip is roughly constant
in $m_{1/2}$, at any point along it one finds $m_{\chi} \approx
430$~GeV and $m_{\chi^{\pm}} \approx 510$~GeV.  
The gluino mass is 2.2 to 2.3~TeV along this strip, which is expected
to be within the LHC's reach with just over 10~fb$^{-1}$ of
integrated luminosity~\cite{lhc,cmstdr}. If $m_0$ is low, then
charged scalar particles may be accessible, with masses as low as
450~GeV. Above the CMSSM contour, however, all scalar particles have
masses well above 3~TeV.

Turning to panel~(d), when $\mu =
1500$~GeV, we find a different situation.  The rapid-annihilation
funnel represents a cosmologically preferred region that occurs at
moderate values of both $m_{1/2}$ and $m_0$, in contrast to the CMSSM,
where cosmologically-preferred regions generally occur at either small
$m_{1/2}$ or small $m_0$.  Taking as an example the point
$(m_{1/2},m_0) = (640,700)$~GeV, we find a rather light neutralino with
$m_{\chi} = 275$~GeV.  The chargino and psuedoscalar Higgs are
somewhat heavier at 545 and 570~GeV, respectively, and charged scalars
have masses of 735~GeV.  This point is particularly interesting in
that $m_{\widetilde{g}} = 1480$~GeV, which should be accessible at the LHC with only
1~fb$^{-1}$ of integrated luminosity.  In the CMSSM, a gluino of 1480~GeV would imply either the coannihilation strip, where $m_{\stau} =
280$~GeV and $m_{\sel} = 285$~GeV, or the focus-point region,
where charged scalars are much heavier. 
In the NUHM1,
several sparticles may have masses below 1~TeV, and points on the
rapid-annihilation strip should be distinguishable from points on the CMSSM
coannihilation strip.

\subsubsection{Fixed $m_0$}

\begin{figure}[ht!]
\begin{center}
\mbox{\epsfig{file=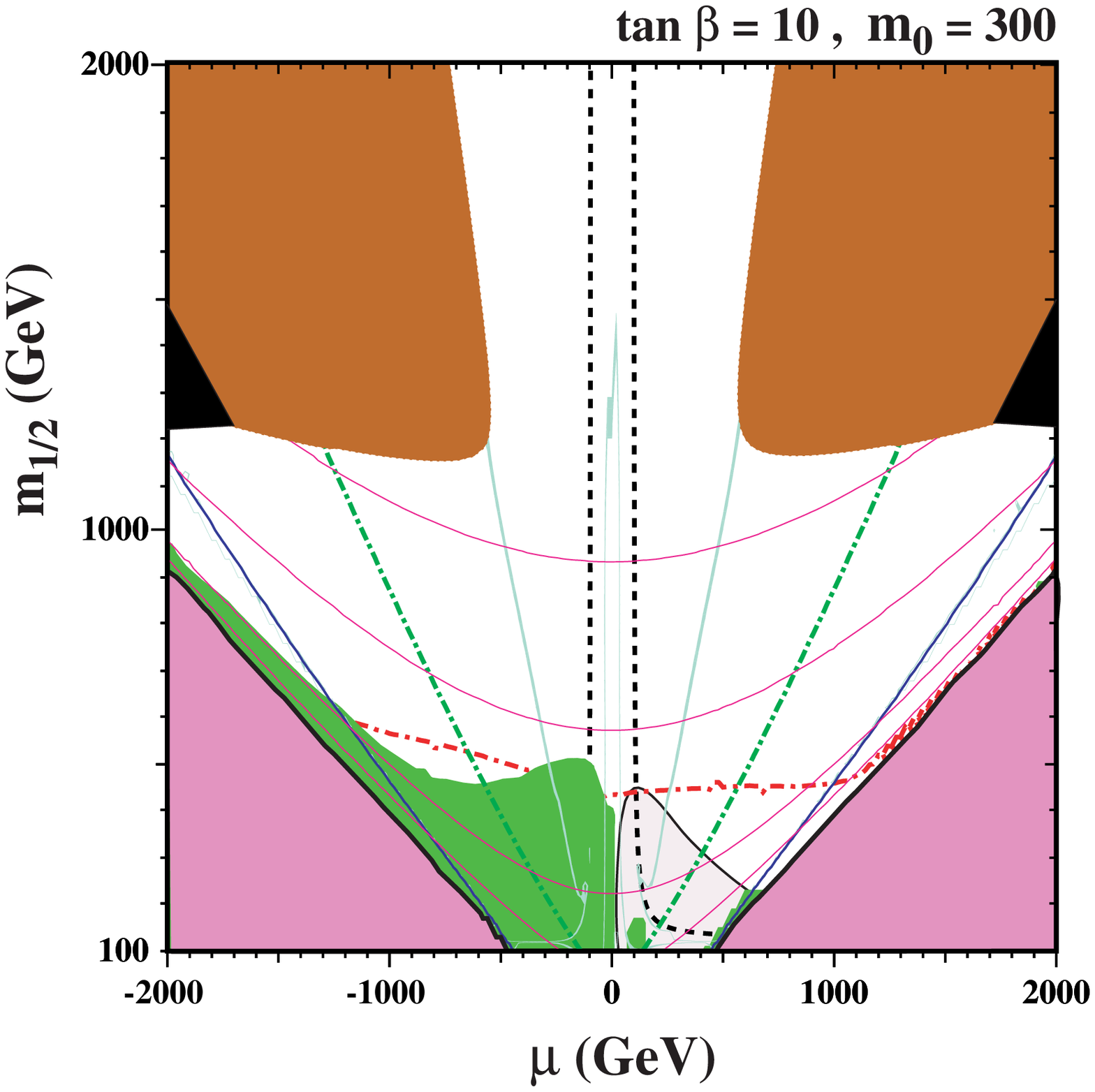,height=7cm}}
\mbox{\epsfig{file=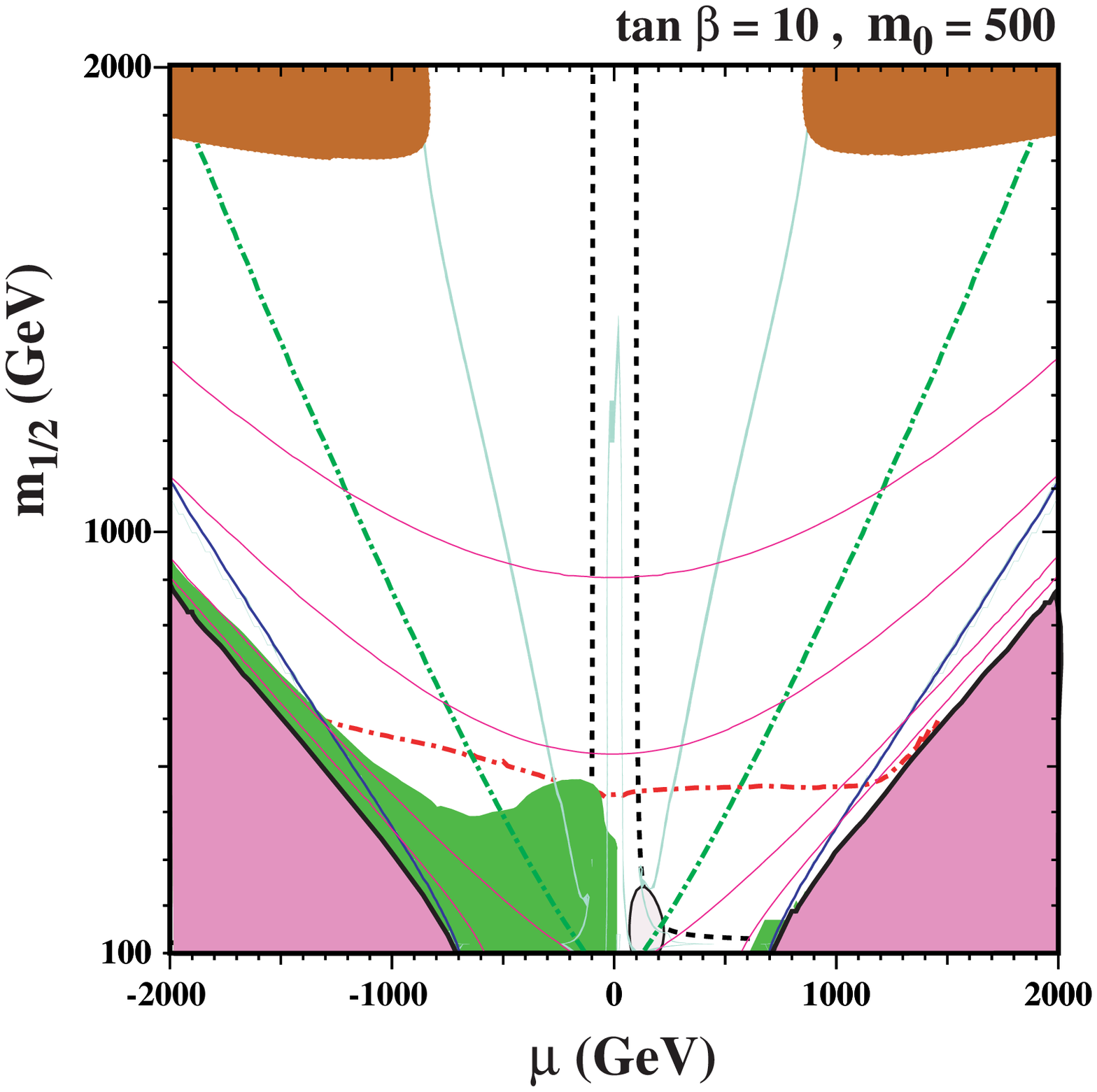,height=7cm}}
\end{center}
\begin{center}
\mbox{\epsfig{file=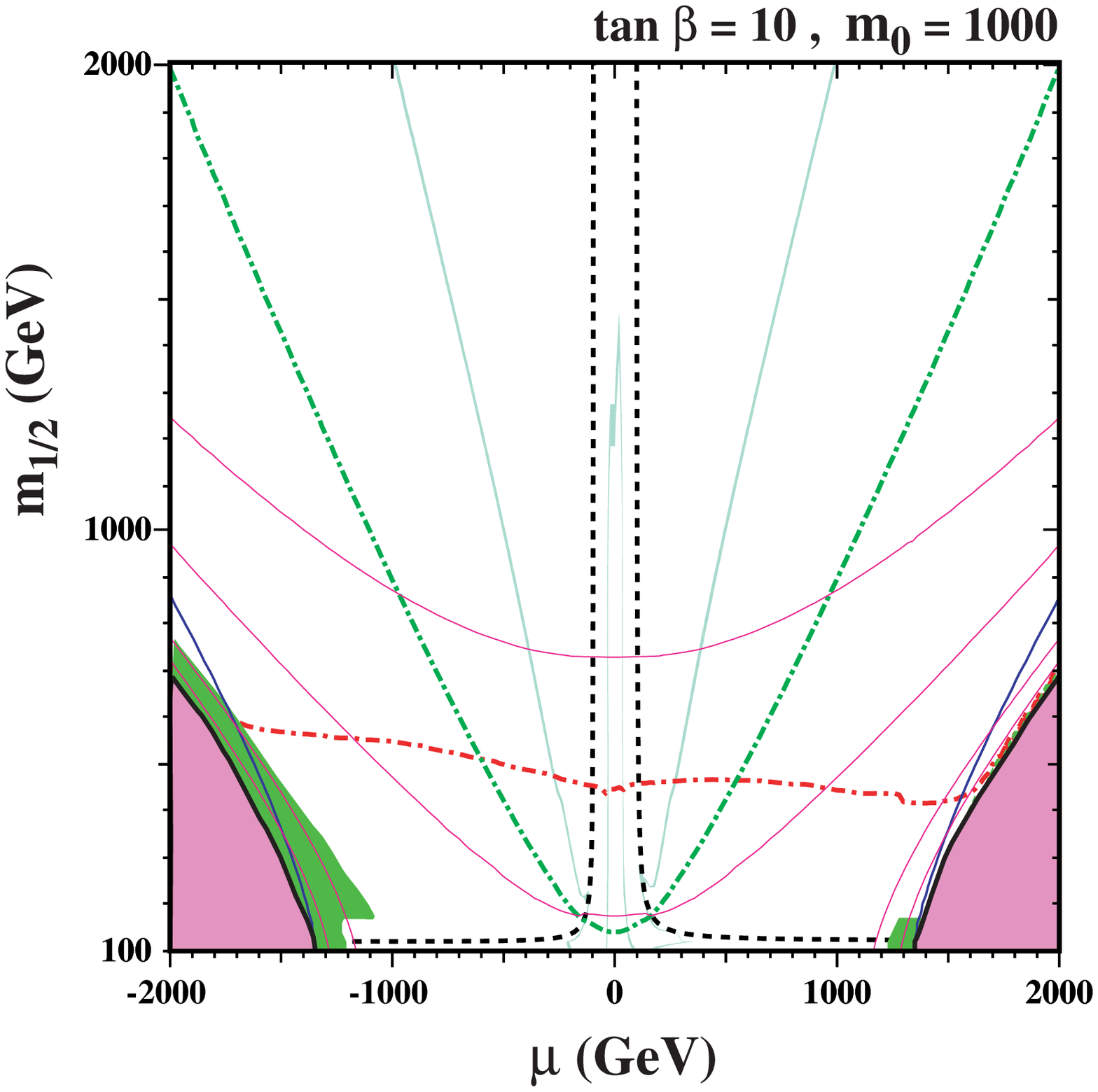,height=7cm}}
\mbox{\epsfig{file=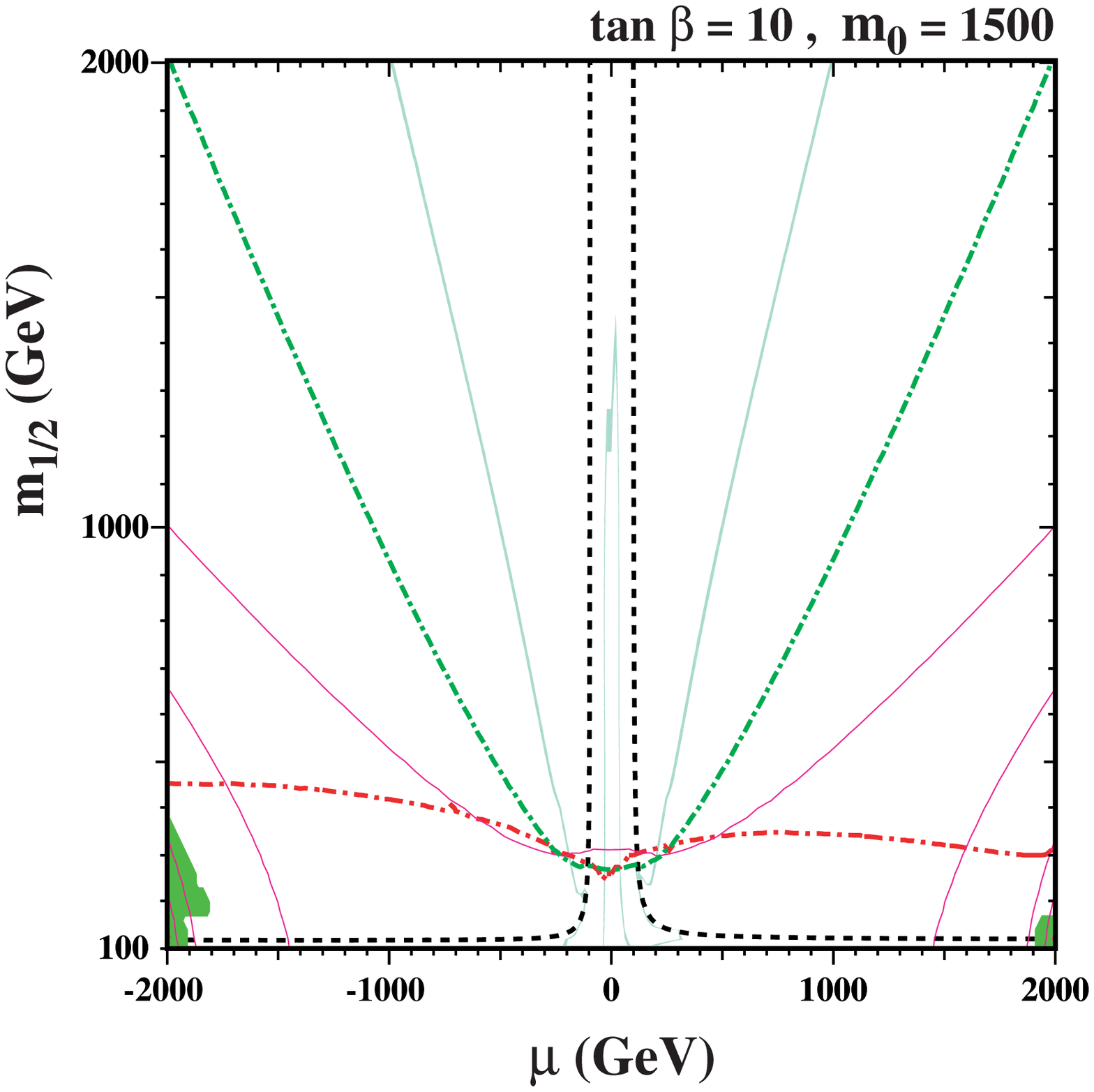,height=7cm}}
\end{center}
\caption{\it Examples of $(\mu,m_{1/2})$ planes with $\tanb = 10$,
$A_0=0$, and $m_0=300$, 500, 1000, and 1500 GeV in Panels
(a), (b), (c), and (d), respectively.  Constraints are displayed as in Figure~\ref{fig:m12m0fixmA}.
}
\label{fig:mum12fixm0}
\end{figure}

Analogously to the discussion in Section~\ref{sec:mA}, alternative ways to view the parameter
space are to fix either $m_{1/2}$ or $m_0$ and scan over $\mu$.  In
Fig.~\ref{fig:mum12fixm0}, we show examples of $(\mu,m_{1/2})$
planes for fixed $m_0 = 300$, 500, 1000, and 1500~GeV in panels (a), (b), (c),
and (d), respectively. For the first time, we display here both positive and
negative values of $\mu$. The unphysical regions excluded by not having 
electroweak symmetry breaking or by having a charged LSP cover a
large part of the plane for $m_0 = 300$~GeV and recede out of the 
visible part of the plane as $m_0$ increases. 
Triangular regions in the lower right and left corners are forbidden
because the pseudoscalar Higgs mass-squared is negative.  For fixed
$\mu$ and $m_{1/2}$, as $m_0$ increases, $m_A$ increases slightly. As a result, the
regions at small $m_{1/2}$ that had been excluded due to
unphysical negative $m_A^2$ recede to larger $|\mu|$, dragging along the
contours of constant $m_A = 300$, 500, 1000, and 1500~GeV. 
For $m_0 = 300$ GeV, the upper right and left portions of the plane are
forbidden because the stau is the LSP, though these regions move
quickly to large $m_{1/2}$ as $m_0$ is increased, almost disappearing
for $m_0 = 1000$~GeV and becoming invisible for larger $m_0$. Bordering
these regions of the plane (but away from the CMSSM contours) the
selectron is lighter than the stau, forming a second region
forbidden by the presence of a charged LSP. 

In each panel, there is a strip at low $|\mu|$ that is excluded by the LEP
chargino constraint. Additionally, at low $m_{1/2}$ (slightly dependent on $m_0$), 
there is a region where the light Higgs mass falls below the LEP limit.  Since $m_h$
increases with $m_{1/2}$, the region below the Higgs mass contour is excluded, a
constraint that is slightly stronger for $\mu < 0$.  The branching ratio of $b
\goto s \gamma$ constrains significantly more strongly the $\mu < 0$ half of
the plane, with the green area being excluded. However, the half-planes with
$\mu < 0$ are not all excluded in the NUHM1. The region
favored by $g_\mu - 2$ is found at small positive $\mu$ and low $m_{1/2}$.
However, it lies below the Higgs mass contour even at $m_0 =
300$ GeV, and shrinks and then evaporates as $m_0$ is increased. 

There are two cosmologically preferred regions in each plane~\footnote{In
addition, at low $|\mu|$ there are regions disallowed by the LEP chargino constraint.}. 
Crossover regions form a long, narrow
`Vee' at relatively small $\mu$, roughly proportional to $m_{1/2}$.
The relic density of neutralinos is below the WMAP
range inside the crossover `Vee', and
above the WMAP range at larger $\mu$. In addition, rapid-annihilation 
funnels occur along diagonals that form a broader `Vee'
with slightly curved walls. These are very thin
cosmologically preferred strips on either side of the blue lines where
$2 m_{\chi} = m_A$, and the relic density
is again below the WMAP range between the two strips of each rapid-annihilation funnel.
We see that there are allowed regions of both the crossover strips and the 
rapid-annihilation funnels when $\mu < 0$, as well as in the conventionally favoured
case $\mu > 0$. However, the latter also include lower values of $m_{1/2}$ where
(in panel (a) for $m_0 = 300$~GeV and panel (b) for $m_0 = 500$~GeV) the
preferred range for $g_\mu - 2$ may also be obtained.

Comparison with the CMSSM case shown in
panel~(a) of Fig.~\ref{fig:m12m0fixmu} yields insight into the
appearance of the CMSSM contours in the NUHM1 planes of Fig.~\ref{fig:mum12fixm0}. 
Following a contour of constant
$m_0$, at low $m_{1/2}$ we begin in either the bulk region excluded by
the LEP Higgs and/or chargino bounds and in the unphysical $\mu<0$
region.  As we move to larger $m_{1/2}$, the
sparticle masses and relic density generally increase, until one reaches the forbidden
$\stau$-LSP region at very large $m_{1/2}$. Thus, the CMSSM contours
in Fig.~\ref{fig:mum12fixm0} begin at $\mu = 0$ in a portion of the
plane excluded by LEP, rising up to larger $m_{1/2}$ and $\mu$. In the
CMSSM, for $m_{1/2} = 2000$~GeV, $|\mu| \approx 2000$~GeV and is sensitive
to $m_0$ only at the level of $\sim 2\%$ for 300~GeV $\leq m_0 \leq$
1500~GeV. It is well-known that in the CMSSM there is no
rapid-annihilation funnel for $\tanb=10$, so we do not expect the
funnel regions in the NUHM1 to cross the CMSSM contours, as
seen in all the panels of Fig.~\ref{fig:mum12fixm0}.  At large
$m_0$, however, the CMSSM crossover WMAP
strip appears at very low $\mu$, so there is a crossing between each
crossover WMAP strip 
and the CMSSM contour for $m_{1/2} \gtrsim 1400$~GeV.

According to previous studies~\cite{lhc,cmstdr}, 
the range of $m_{1/2}$ accessible to
the LHC depends on the value of $m_0$ chosen, being roughly $900 (900)
(800) (700)$~GeV for the choices $m_0 = 300 (500) (1000) (1500)$~GeV
shown in Fig.~\ref{fig:mum12fixm0}. This implies that there are increasing
portions of the crossover and rapid-annihilation strips that are likely to be
inaccessible as $m_0$ increases from panels (a) and (b) to panels (c) and (d).

\subsubsection{Fixed $m_{1/2}$}

\begin{figure}[ht!]
\begin{center}
\mbox{\epsfig{file=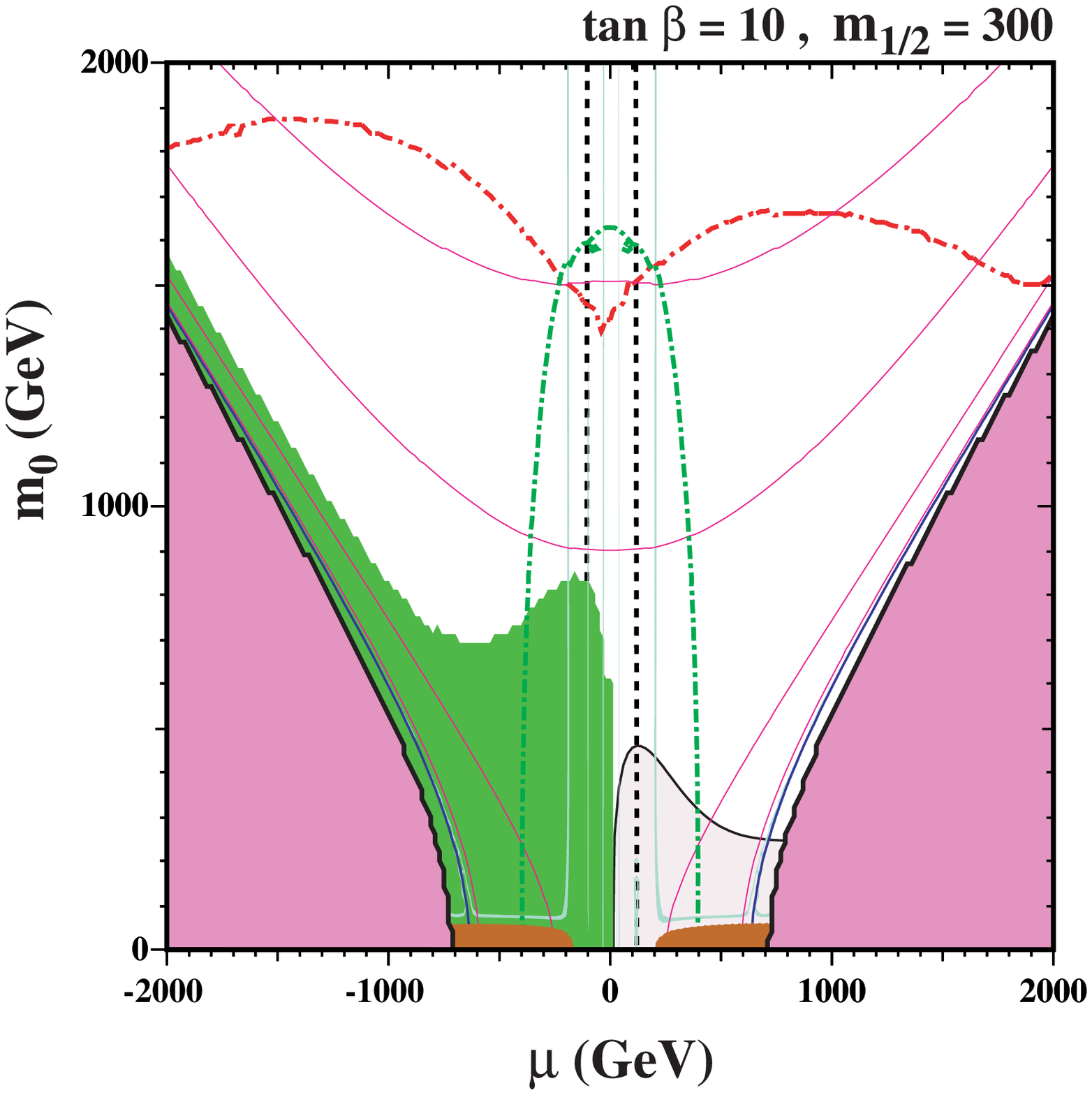,height=7cm}}
\mbox{\epsfig{file=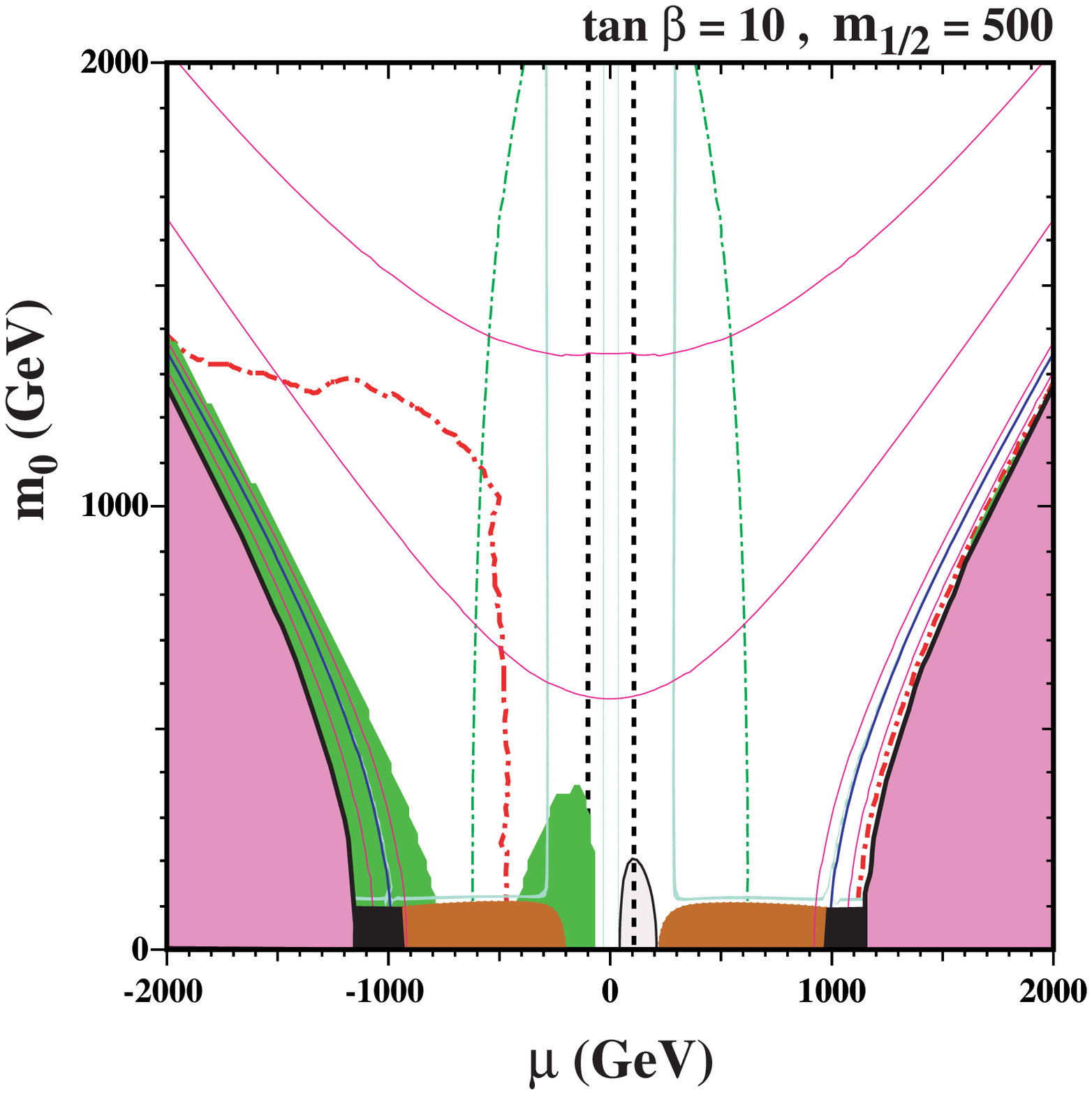,height=7cm}}
\end{center}
\begin{center}
\mbox{\epsfig{file=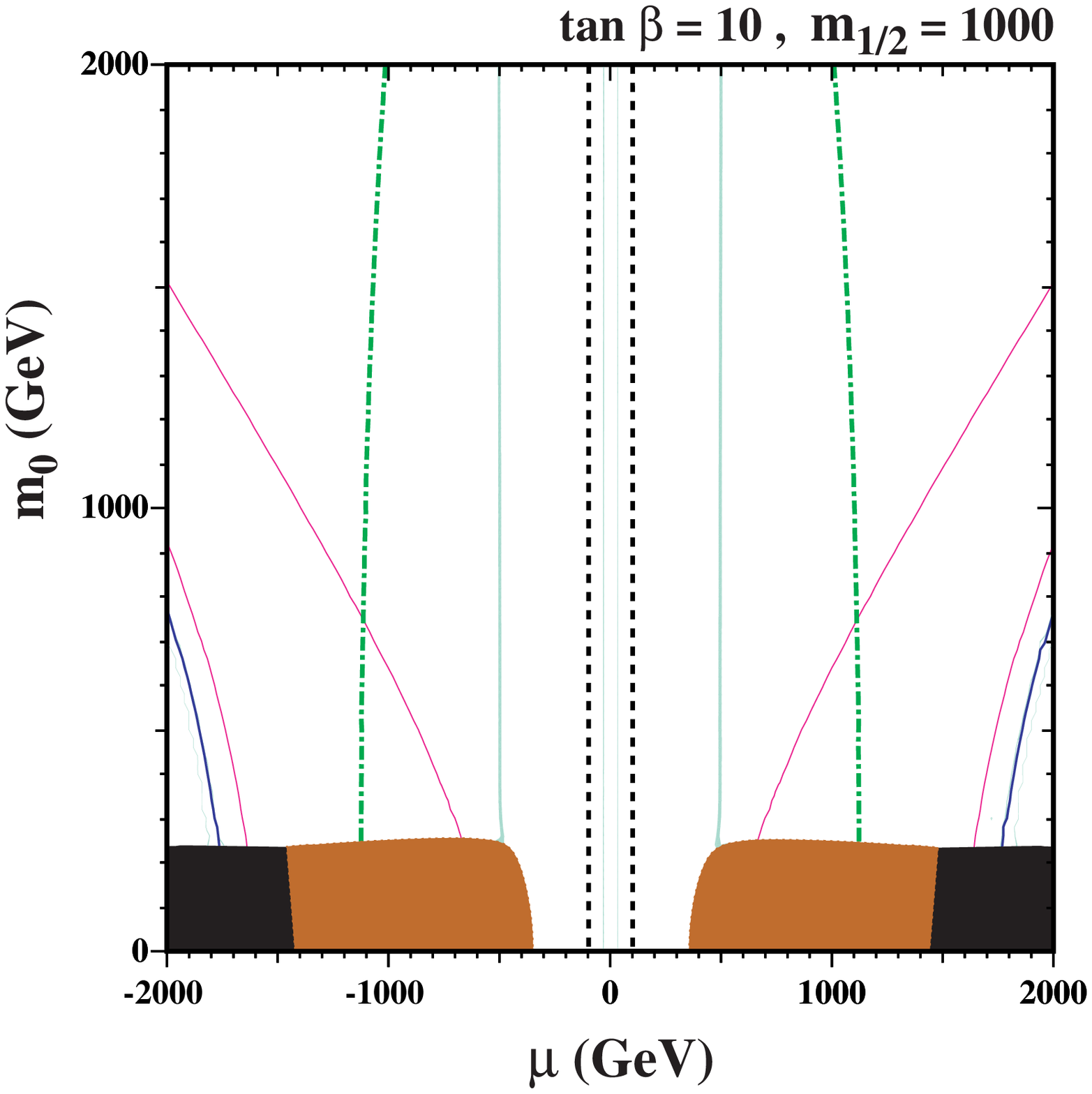,height=7cm}}
\mbox{\epsfig{file=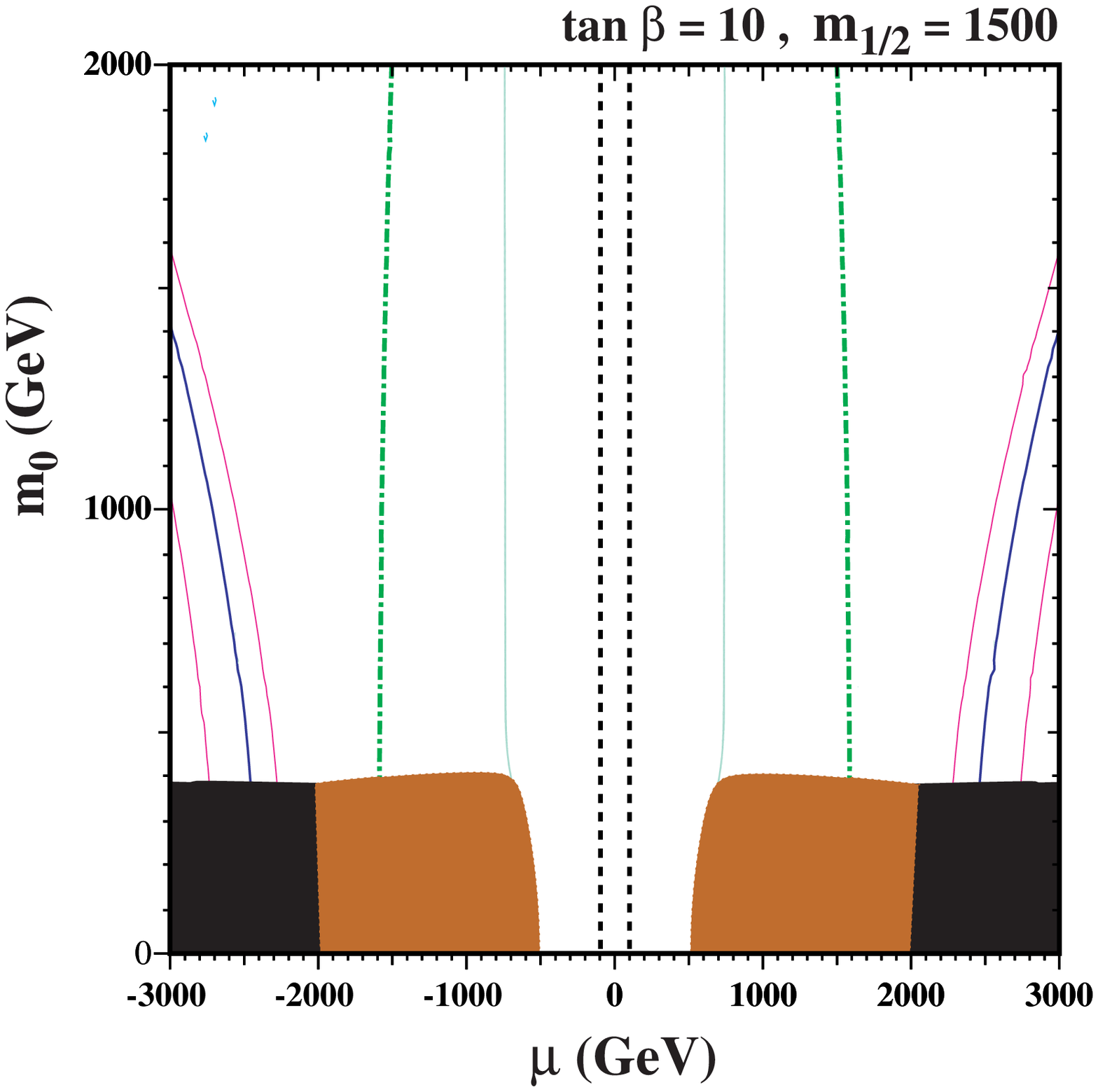,height=7cm}}
\end{center}
\caption{\it Examples of $(\mu,m_0)$ planes with $\tanb = 10$,
$A_0=0$, and $m_{1/2}=300$, 500, 1000, and 1500 GeV in Panels
(a), (b), (c), and (d), respectively.  Constraints are displayed as in Figure~\ref{fig:m12m0fixmA}.}
\label{fig:mum0fixm12}
\end{figure}

Fig.~\ref{fig:mum0fixm12} shows NUHM1 $(\mu,m_0)$ planes with
$m_{1/2}$ fixed to be 300, 500, 1000, and 1500~GeV in panels~(a), (b),
(c), and (d), respectively. Again, regions at large $|\mu|$ excluded
because there is no electroweak symmetry breaking (since $m_A^2 < 0$) 
are bordered by contours of constant $m_A$ and
parallel rapid-annihilation funnels. These regions recede and disappear
for $m_{1/2} \ge 1000$~GeV. There are also excluded charged LSP regions 
at  small $m_0$, which expand as $m_{1/2}$ increases.

For $m_{1/2} = 300$~GeV, shown in panel (a) of Fig.~\ref{fig:mum0fixm12}, the LEP
constraint on the Higgs mass excludes all of the plane below the
contour at $m_0 \sim 1500$~GeV. The branching ratio of $b \goto s \gamma$ also
excludes a region with $\mu < 0$ at
lower $m_0$. The chargino mass bound from LEP appears as vertical
black dot-dashed lines at small $|\mu|$, and a region favored by
$g_\mu - 2$ is visible at small positive $\mu$.
For $m_{1/2} = 500$~GeV, shown in panel (b), the Higgs constraint is
weakened for $\mu < 0$ and disappears for $\mu > 0$, and the region favoured by
$g_\mu - 2$ for $\mu > 0$ contracts. The Higgs and $b \goto s \gamma$
constraints disappear completely when $m_{1/2} \ge 1000$~GeV.

The relic density of neutralinos may fall in the range favoured by WMAP
in three regions of each $(\mu, m_0)$ plane: along the rapid-annihilation funnels
that straddle the blue lines where $m_A = 2 m_\chi$, in the thin
crossover strips that run outside and roughly parallel to the LEP
chargino limits, and, at small $m_0$, along coannihilation strips
close to the excluded $\stau$- and $\sel$-LSP regions. 

The CMSSM contours appear in these planes as parabolas, symmetric about
$\mu = 0$, with a peak height that increases dramatically with
$m_{1/2}$. Since $m_{1/2}$ is constant in each of the planes, 
each half of each parabola may be regarded as tracing a line of
constant $m_{1/2}$ in the standard CMSSM $(m_{1/2}, m_0)$ plane. 
When $m_{1/2} = 300$~GeV, at low $m_0$ one encounters the bulk region
that is excluded by the Higgs constraint and (for $\mu < 0$) the $b \goto s \gamma$
constraint. The only points compatible the dark matter and all other constraints
are at $|\mu| \sim 100$~GeV and $m_0 \sim 1550$~GeV, barely satisfying the
Higgs constraint. As $m_0$
increases, these CMSSM WMAP-compatible points move up to very large
$m_0 > 2000$~GeV, a relic of the focus-point region in the familiar
CMSSM $(m_{1/2}, m_0)$ plane. However, for $500$~GeV~$\le m_{1/2} \le 900$~GeV
we also encounter WMAP-compatible $\stau$-coannihilation points at the bottoms
of the parabolae, which are compatible with all the other constraints (except the
Higgs when $m_{1/2} = 500$~GeV and $\mu < 0$). The CMSSM contours never
cross the rapid-annihilation funnels for this value of $\tbt = 10$. 

According to previous studies~\cite{lhc,cmstdr}, 
the range of $m_{0}$ accessible to
the LHC depends on the value of $m_{1/2}$ chosen, being above $2000$~GeV 
for the choices $m_{1/2} = 300, 500$~GeV shown in panels (a) and (b)
of Fig.~\ref{fig:mum0fixm12}. In the CMSSM, we do not expect to be
able to probe supersymmetry with $m_{1/2} > 1000$~GeV, however in the
NUHM1, there are regions of parameter space with heavy gauginos and
much lighter scalars that may be accessible, specifically the lower
portions of the crossover strips shown in panels (c) and (d).

\subsubsection{Varying $\tan \beta$}

\begin{figure}[ht!]
\begin{center}
\mbox{\epsfig{file=nuhm1_muvM_10_500s.eps,height=7cm}}
\mbox{\epsfig{file=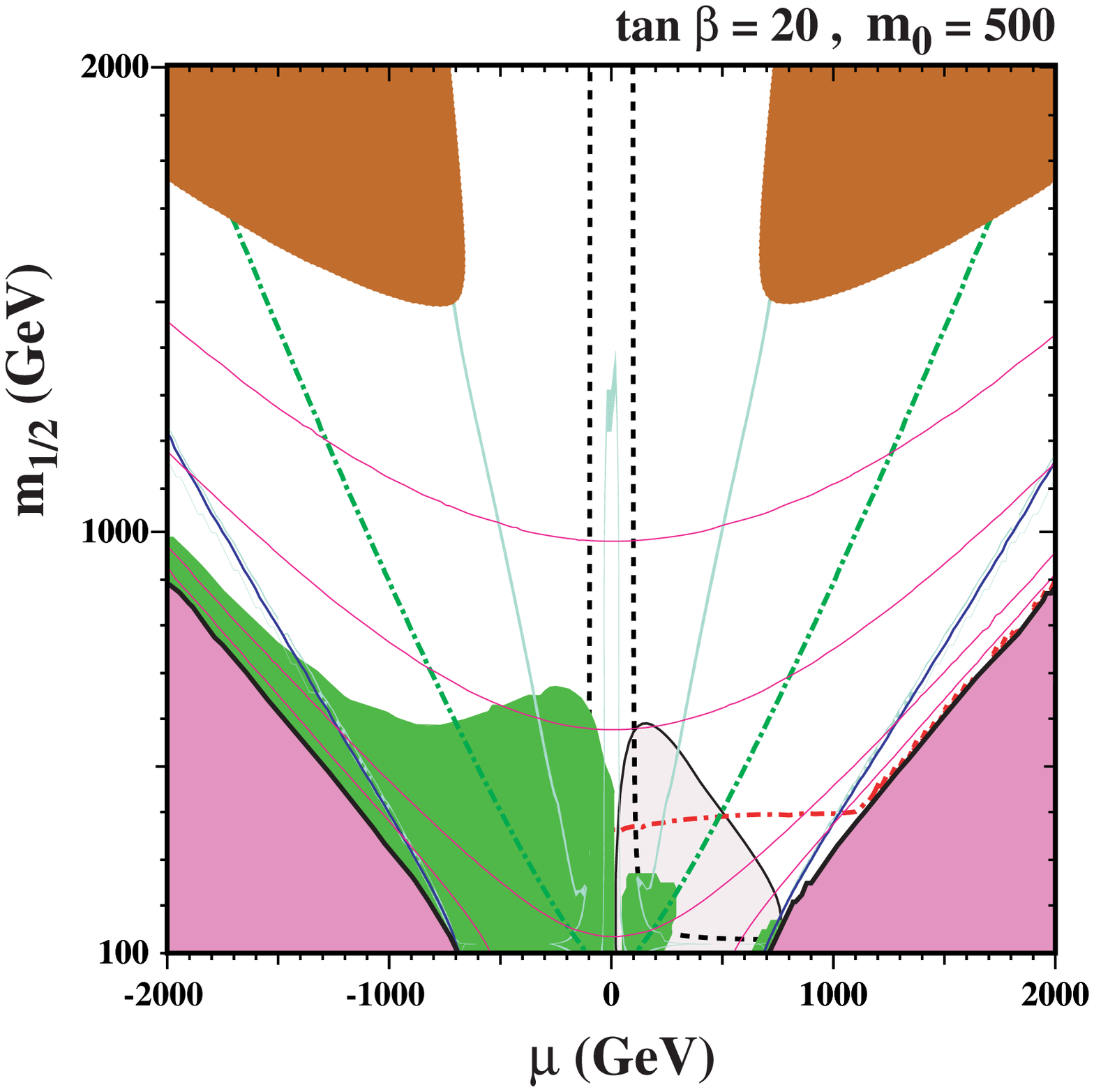,height=7cm}}
\end{center}
\begin{center}
\mbox{\epsfig{file=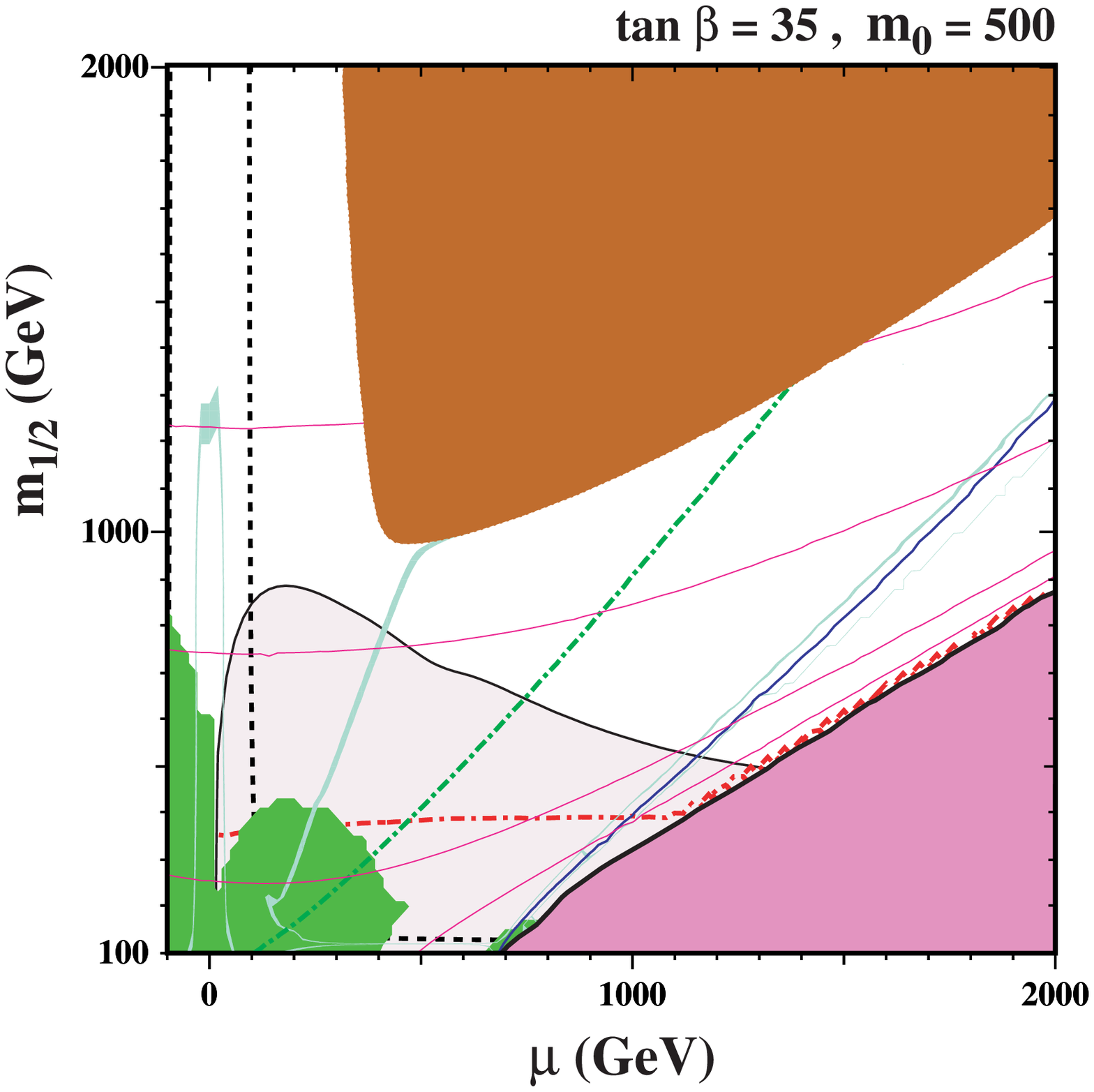,height=7cm}}
\mbox{\epsfig{file=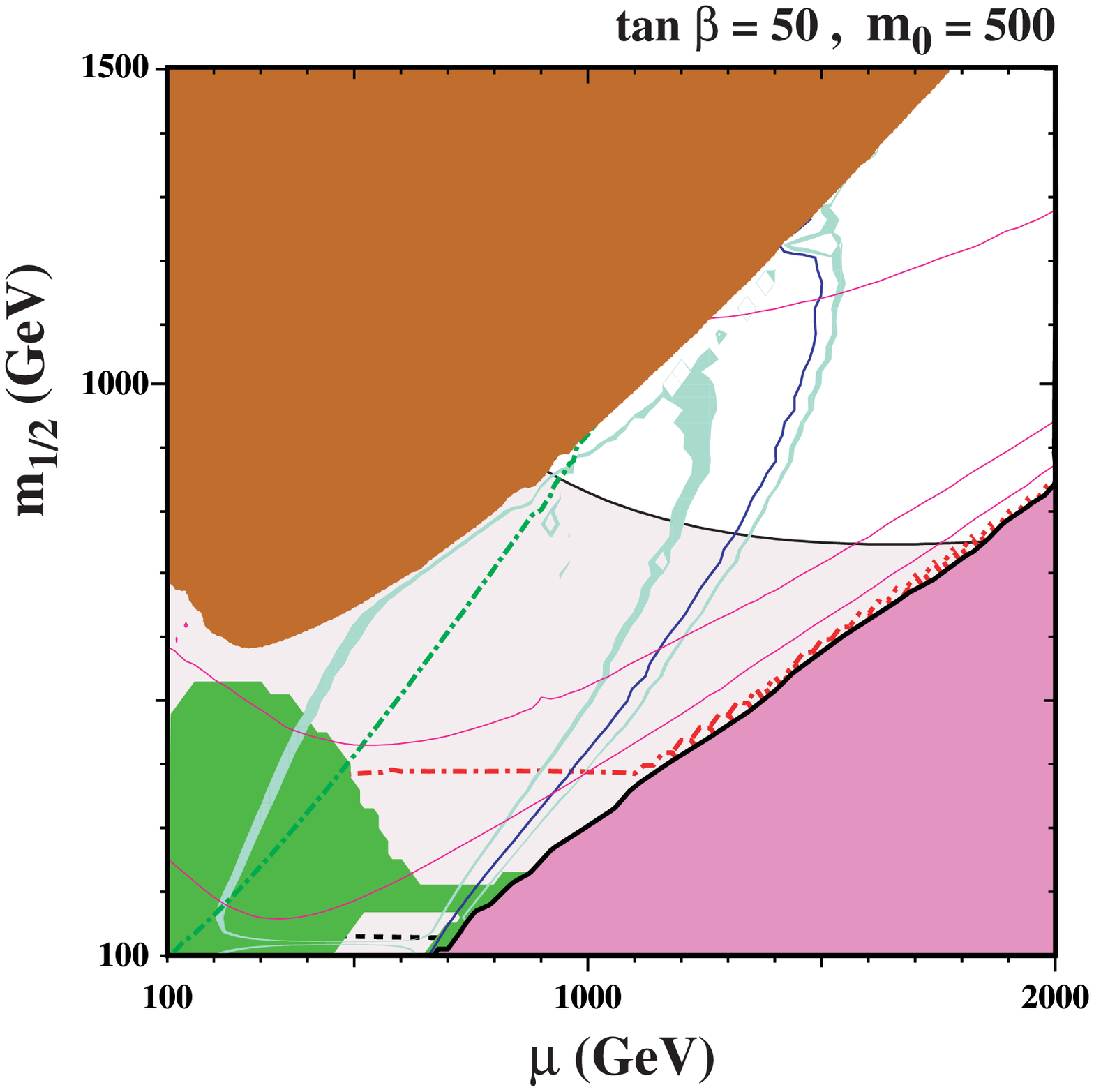,height=7cm}}
\end{center}
\caption{\it Examples of $(\mu,m_{1/2})$ planes with $m_0 = 500$ GeV,
$A_0=0$, and $\tanb=10$, 20, 35, and 50 in Panels
(a), (b), (c), and (d), respectively.  Constraints are displayed as in Figure~\ref{fig:m12m0fixmA}.
}
\label{fig:mum12tanb}
\end{figure}

We now consider the effect of varying $\tan \beta$, initially at fixed
$m_0 = 500$~GeV. Panels (a), (b), (c), and (d) of Figure~\ref{fig:mum12tanb} show NUHM1
$(\mu,m_{1/2})$ planes for $\tanb=10$, 20, 35, and 50,
respectively. In all panels, the requirement of electroweak symmetry
breaking appears identically as a triangular excluded region at large
$|\mu|$ and low $m_{1/2}$. The $\stau$-LSP regions, while remaining 
similar in shape, become more prominent at large $\tanb$, as in the
CMSSM. Focusing on $\mu > 0$, we see the constraint due to the branching
ratio of $b \goto s \gamma$ grows with $\tanb$, while the LEP Higgs
constraint, for fixed $m_0$, has little dependence on $\tanb$.  In panels
(c) and (d), we display only
the $\mu > 0$ half of the plane for $\tbt = 35$ and 50, since solutions are not reliably found
for large $\tan \beta$ with $\mu < 0$.

In all panels, the crossover strips
and the rapid-annihilation funnels are viable cosmologically preferred
regions, both appearing as diagonals forming `Vee'
shapes in the planes. The CMSSM contours lie
between the two `Vees,' intersecting WMAP strips only in regions excluded by
collider constraints in panels (a) through (c).  As $\tanb$ increases, the CMSSM contours shift
to smaller $|\mu|$, while the rapid-annihilation funnel becomes more
prominent and is deformed to lower $|\mu|$. At $\tanb=50$, where a
rapid-annihilation funnel is natural in the CMSSM, the coannihilation
strip connects the crossover strip with the enlarged funnel
region. For this fixed value of $m_0 = 500$~GeV, the CMSSM contour
does not intersect the rapid-annihilation funnel, however an
intersection would occur for larger $m_0$. At $\tanb = 50$, the region
favoured by $g_{\mu}-2$ has expanded to encompass large regions of the plane where collider constraints are
evaded and the dark matter density is in agreement with astrophysical
measurements.

We recall that in the CMSSM, none of the regions of parameter space
with $m_{1/2}
\gtrsim 1000$~GeV may be within the 10~fb$^{-1}$ reach of the
LHC~\cite{lhc,cmstdr} regardless of the value of
$\tanb$. Extrapolating to the NUHM1, it is clear from Fig.~\ref{fig:mum12tanb} that portions of the crossover and
rapid-annihilation strips, and possibly part of the $\stau$
coannihilation strip at $\tanb\sim50$, will be beyond the reach of the LHC. For comparison, in the CMSSM the corresponding
$\stau$ coannihilation strips would be accessible, but not portions of the
focus-point and rapid-annihilation funnels.

\begin{figure}[ht!]
\begin{center}
\mbox{\epsfig{file=nuhm1_muvm0_10_500s.eps,height=7cm}}
\mbox{\epsfig{file=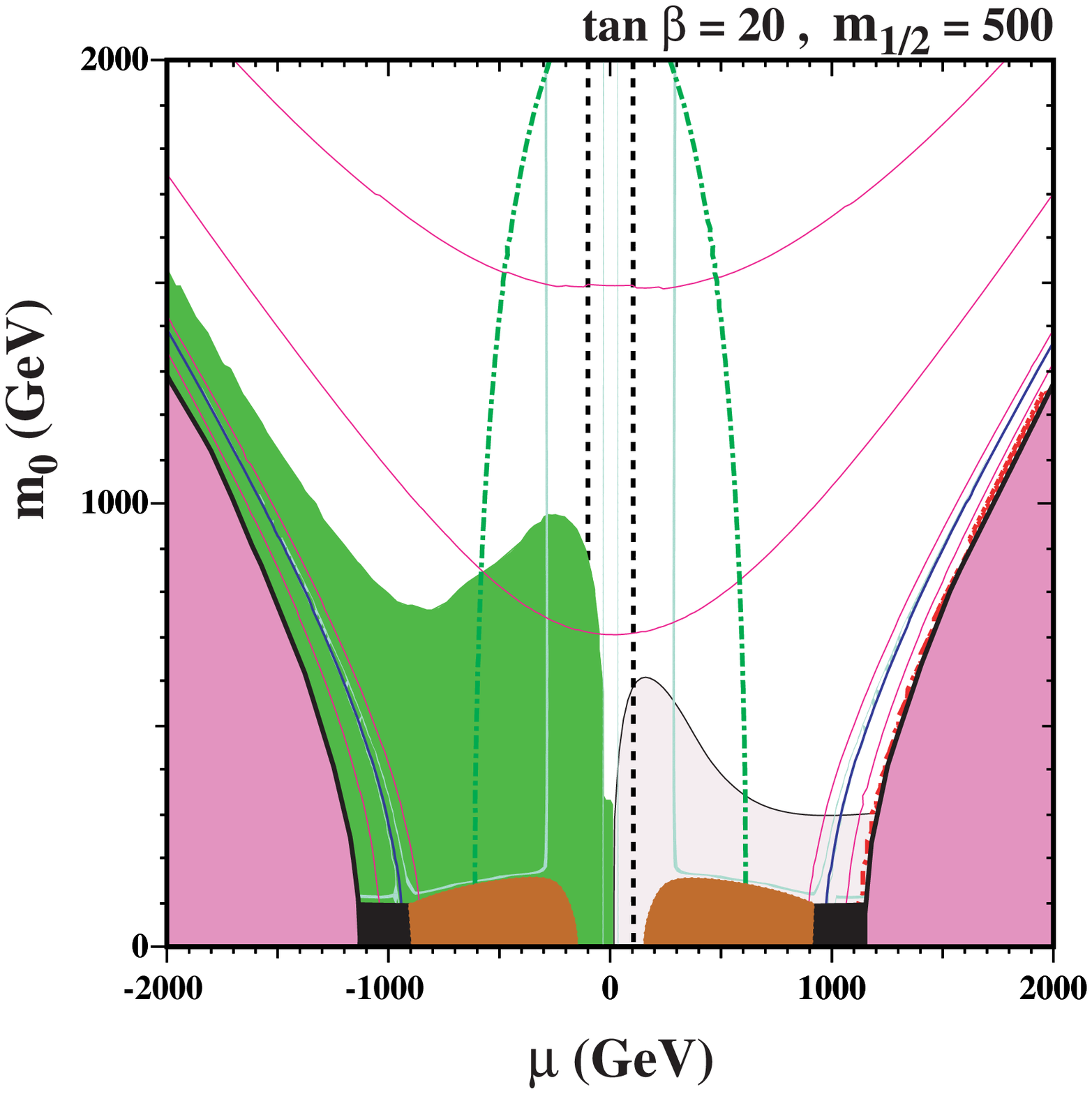,height=7cm}}
\end{center}
\begin{center}
\mbox{\epsfig{file=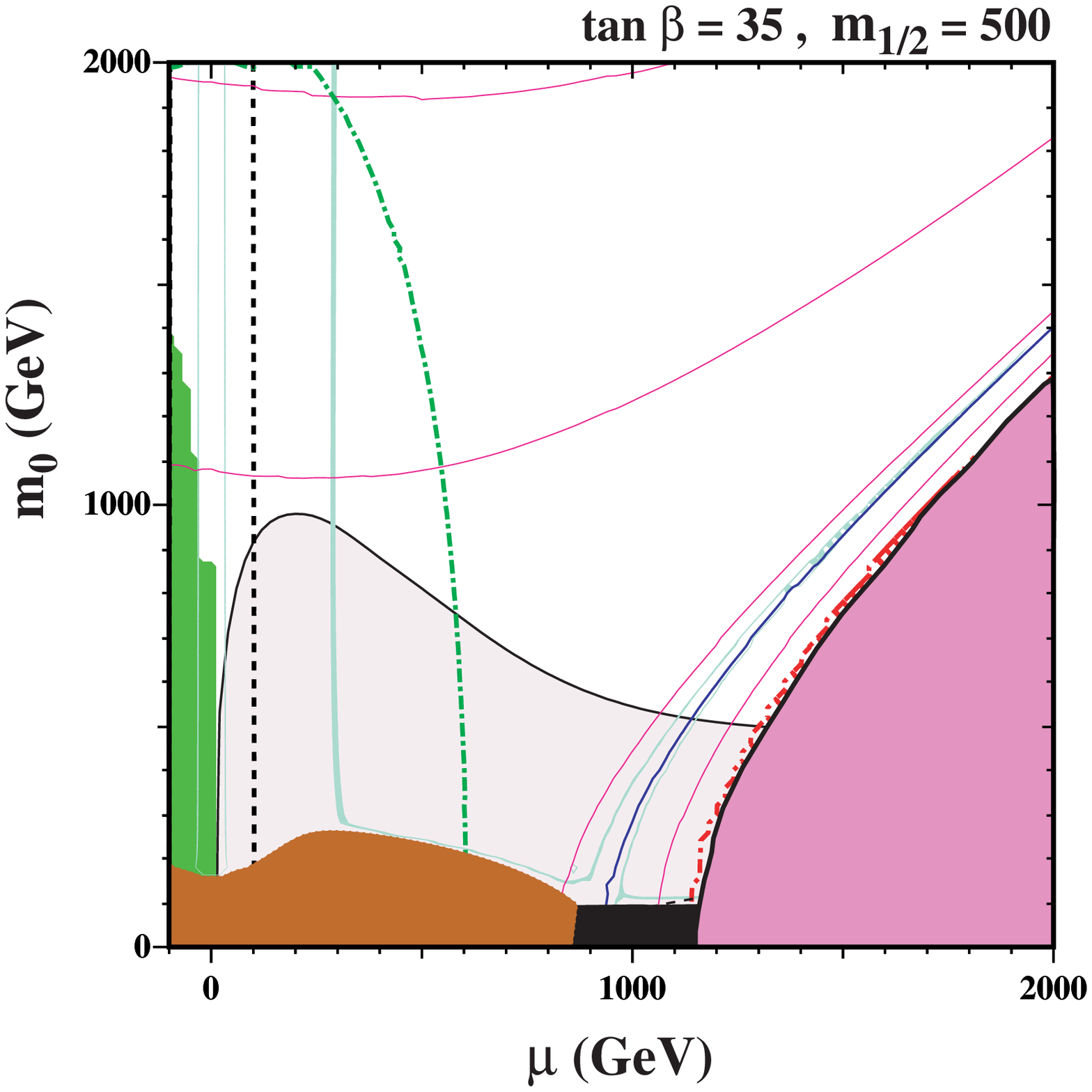,height=7cm}}
\mbox{\epsfig{file=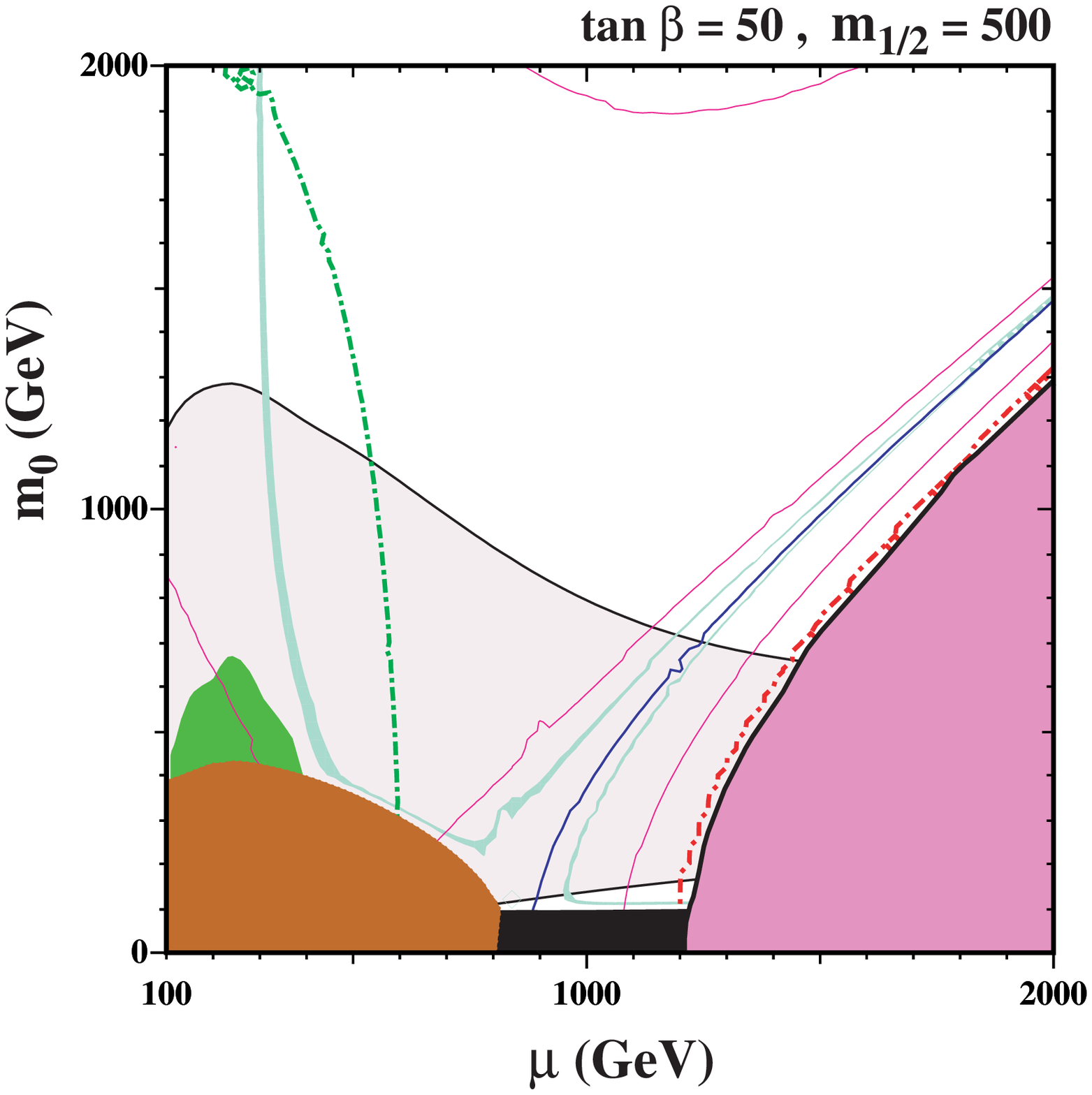,height=7cm}}
\end{center}
\caption{\it Examples of $(\mu,m_0)$ planes with $m_{1/2} = 500$ GeV,
$A_0=0$, and $\tanb=10$, 20, 35, and 50 in panels
(a), (b), (c), and (d), respectively.  Constraints are displayed as in Fig.~\ref{fig:m12m0fixmA}.
}
\label{fig:mum0tanb}
\end{figure}

Figure~\ref{fig:mum0tanb} shows examples of the $(\mu,m_0)$ plane at
fixed $m_{1/2}=500$~GeV for four choices of $\tanb$. 
Progressing from $\tbt = 10$ shown in panel (a), which is the same as panel (b)
of Fig.~\ref{fig:mum0fixm12}, we see that, as $\tbt$ increases to 20 in panel (b), the 
regions excluded by the absence of electroweak symmetry breaking and the
presence of a $\stau$ or $\sel$ LSP are little changed~\footnote{Regions with tachyonic
sfermions are found within the $\stau$-LSP regions shown above for $\tbt \ge 35$.}. However, the Higgs constraint essentially disappears,
whereas the $b \goto s \gamma$ constraint is much more aggressive at $\mu < 0$
and a larger region is favoured by $g_\mu - 2$ at $\mu > 0$. Again, in panels
(c) and (d), we display only
the $\mu > 0$ half of the plane.

The regions favoured by the dark matter density are crossover strips
at $|\mu | \sim 300$~GeV,
rapid-annihilation funnels arching up close to the region excluded by the absence of 
electroweak symmetry breaking, and coannihilation strips close to the
charged LSP regions. For $\tanb \ge 20$, separate $\chi-\stau$ and
$\chi-\sel$ coannihilation strips are easily discerned,
separated by the rapid-annihilation funnel.

The CMSSM lines in the $(\mu, m_0)$ planes remain essentially unchanged as
$\tbt$ increases. They always have intersections with the crossover strips at large
$m_0 \sim 2000$~GeV, for both signs of $\mu$, and also intersect the $\stau$ coannihilation 
strip for $\mu > 0$. This intersection is in the region favoured by $g_\mu - 2$, whereas the
corresponding intersection for $\mu < 0$ is excluded either by the LEP Higgs limit (for
$\tbt = 10$) or $b \goto s \gamma$ (for $\tbt \ge 20$). There are no intersections with the
rapid-annihilation funnels or the ${\tilde e}$-coannihilation regions.

For the choice of $m_{1/2} = 500$~GeV made in Fig.~\ref{fig:mum0tanb},
all the range of $m_0 \le 2000$~GeV should be accessible to the LHC~\cite{lhc,cmstdr}. 
However, fewer of the heavier neutralinos, charginos and Higgs bosons would be
detectable at larger values of $\mu$ (horizontal axis) and $m_A$ (pink contours).

\section{From the NUHM1 to the NUHM2}

Having situated the NUHM1 relative to the CMSSM, we now discuss
the extension to the NUHM2, in which the soft supersymmetry-breaking
contributions to both the Higgs scalar masses $m_{1,2}$ are regarded as free parameters.
These two extra parameters imply that each point in a CMSSM
$(m_{1/2}, m_0)$ plane can be `blown up' into a $(\mu, m_A)$ plane,
as displayed in Figs.~\ref{fig:traditional3_1}, \ref{fig:traditional5_3} and \ref{fig:traditional5_10}.
Alternatively, one may display the NUHM2 parameter space directly in $(m_1, m_2)$
planes, as we do in Figs.~\ref{fig:new3_1}, \ref{fig:new5_3} and \ref{fig:new5_10}.
In the following, we use these `blow-ups' to relate the NUHM2 to the NUHM1 and the
CMSSM, noting that, in each plane, the NUHM1 subspace may be represented as a line, 
and the CMSSM as one or two points on this line.

\subsection{NUHM2 $(\mu, m_A)$ Planes}

\begin{figure}[ht!]
\begin{center}
\mbox{\epsfig{file=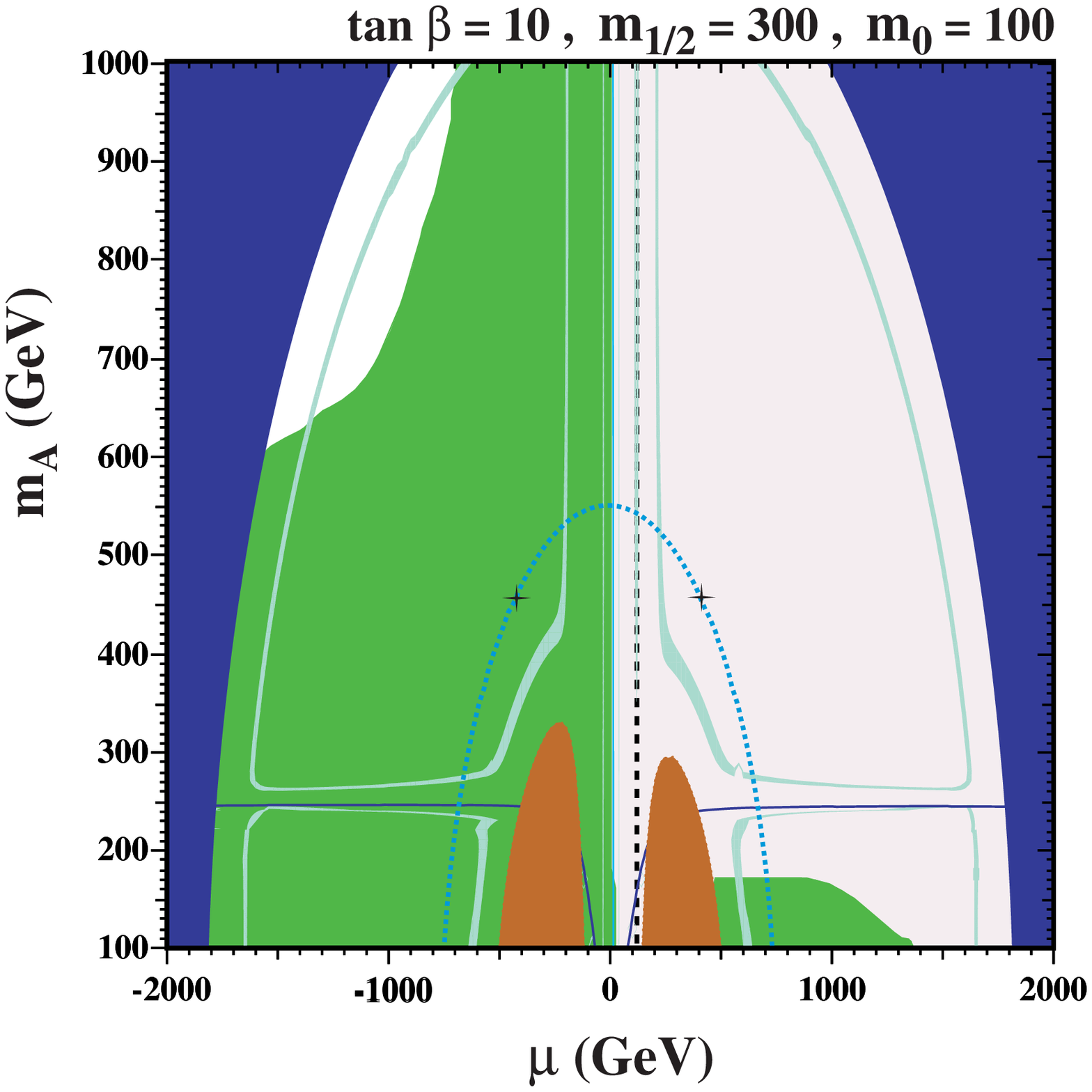,height=7cm}}
\mbox{\epsfig{file=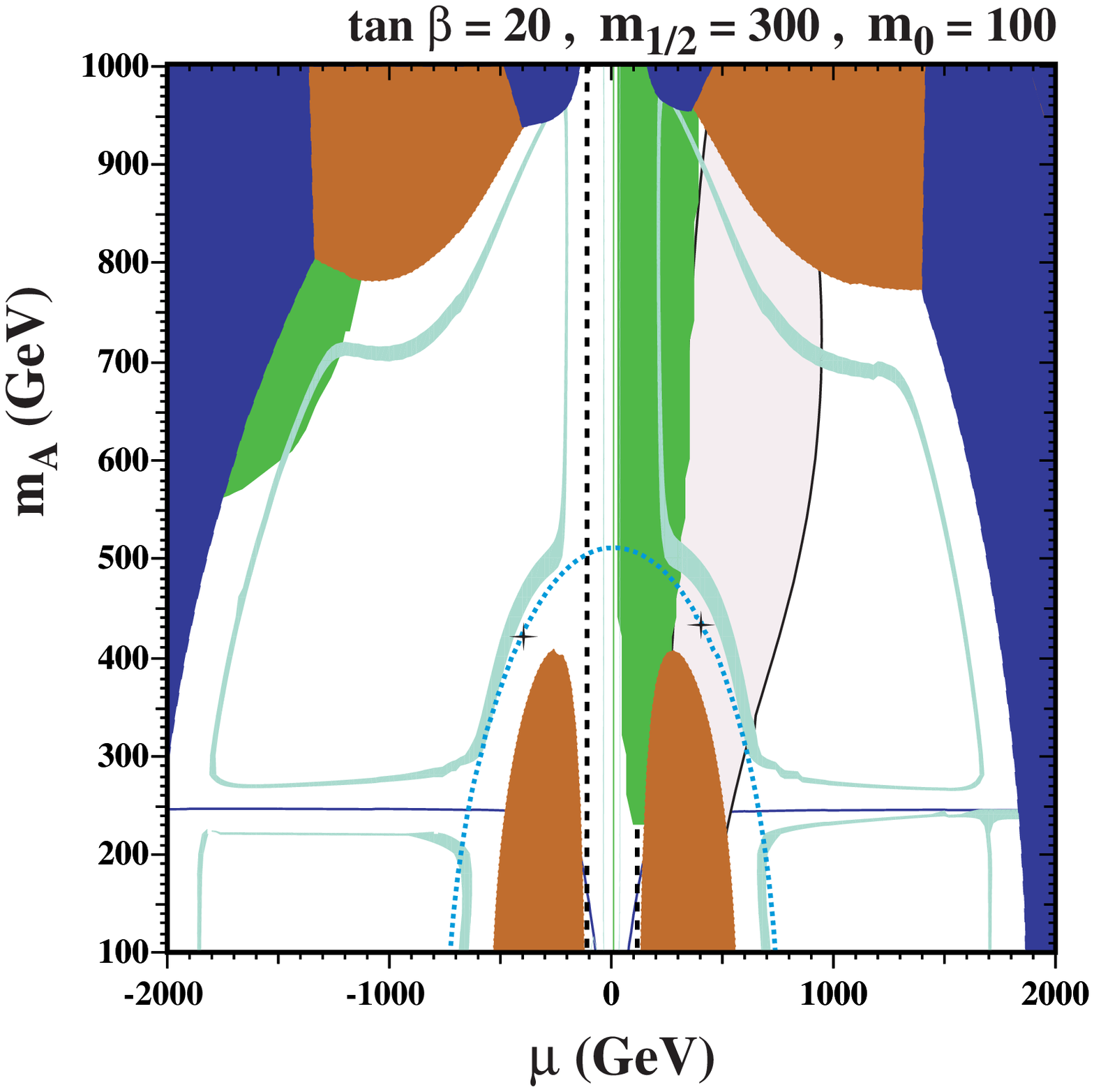,height=7cm}}
\end{center}
\begin{center}
\mbox{\epsfig{file=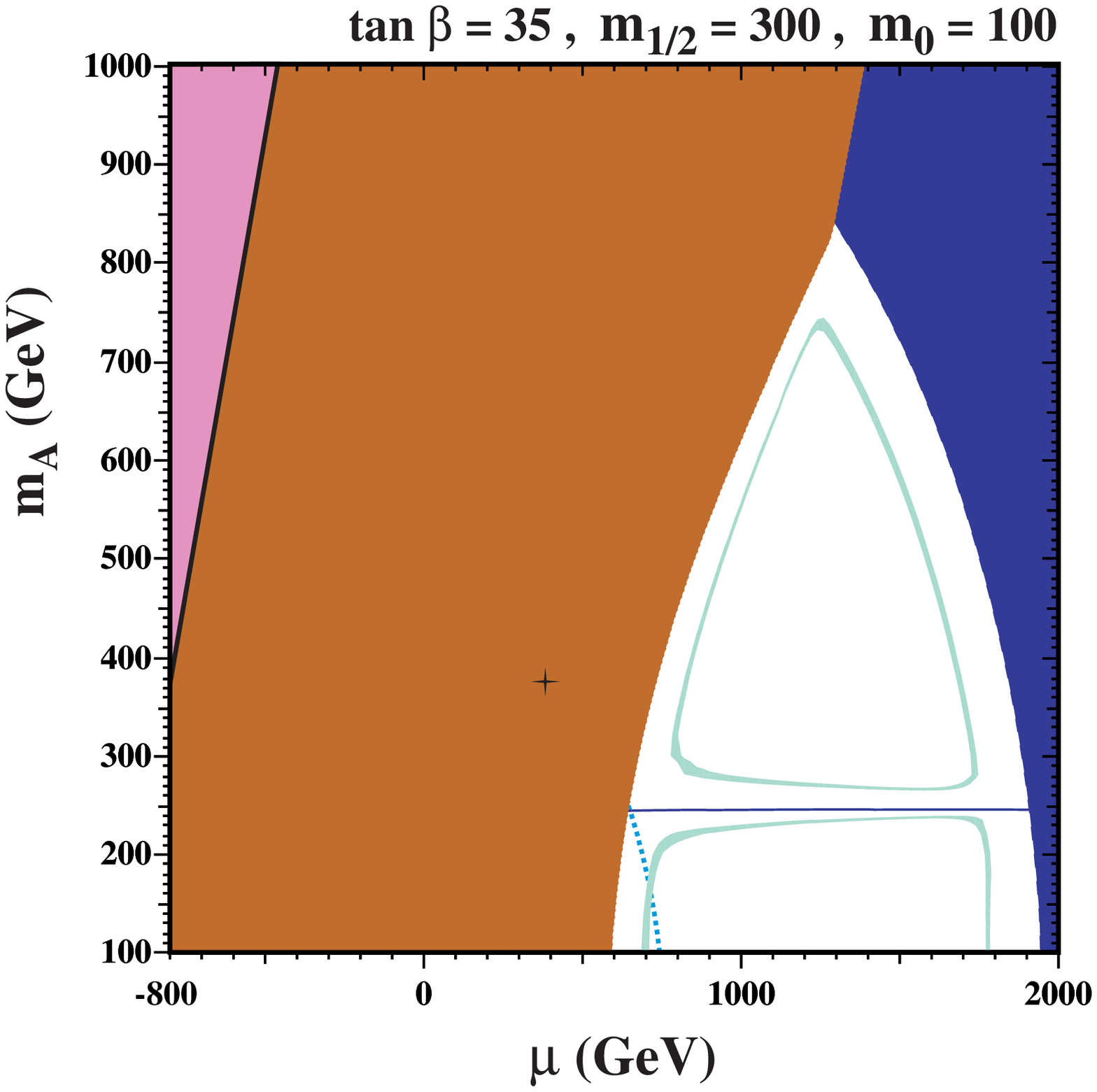,height=7cm}}
\mbox{\epsfig{file=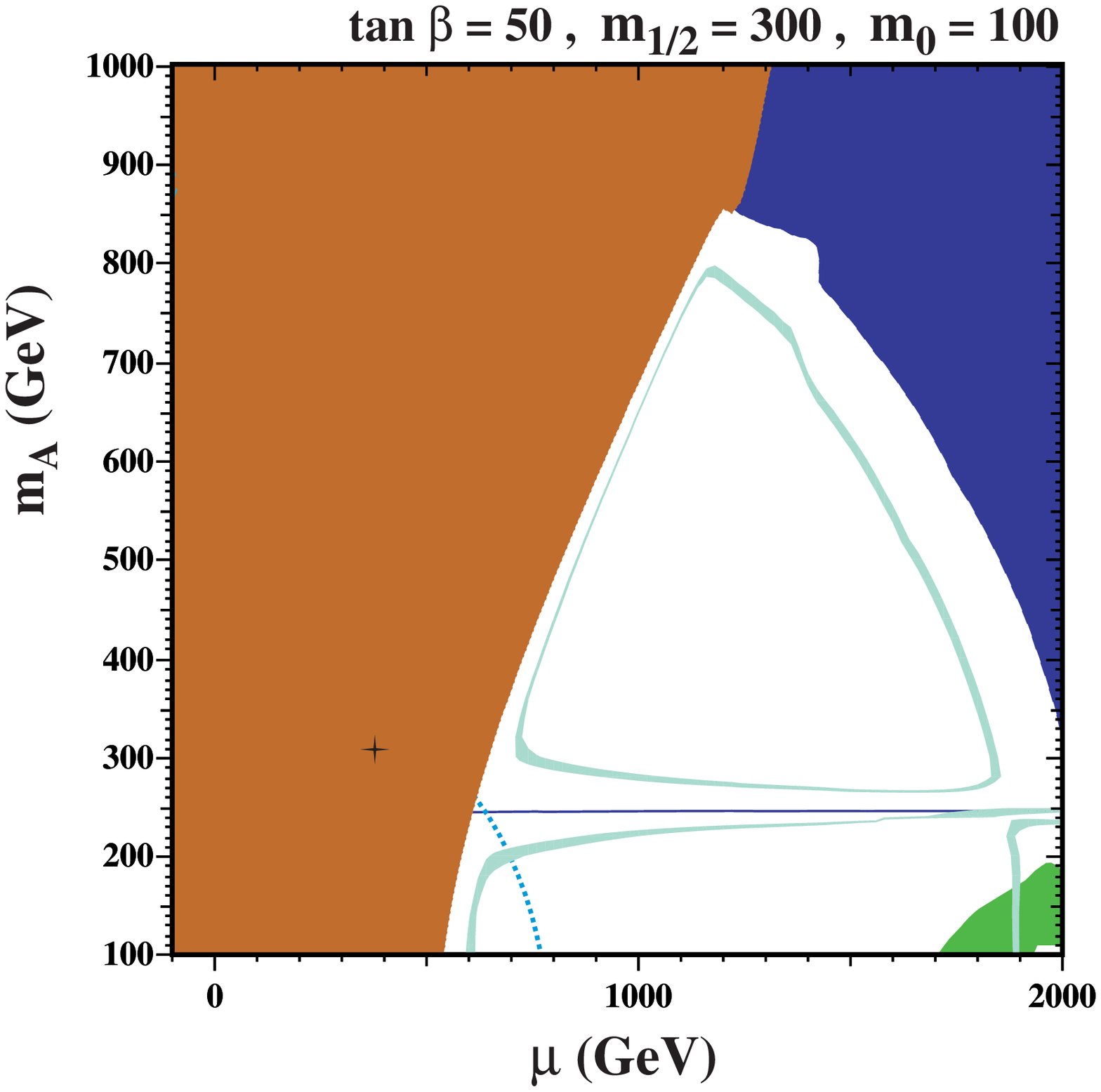,height=7cm}}
\end{center}
\caption{\it Examples of NUHM2 $(\mu,m_A)$ planes with
$m_{1/2}=300$~GeV, $m_0=100$~GeV, $A_0=0$, and $\tanb = 10$, 20, 35, and 50 in
panels (a), (b), (c), and (d), respectively. Constraints are displayed as in Fig.~\ref{fig:m12m0fixmA}.
}
\label{fig:traditional3_1}
\end{figure}

We start by considering the $(\mu, m_A)$ `blow-ups' of points with the relatively small
values $(m_{1/2}, m_0) = (300, 100)$~GeV, shown in Fig.~\ref{fig:traditional3_1}. Panel (a)
is for $\tbt = 10$. We see (brown) regions excluded because of a $\stau$ LSP at small values of
$|\mu|$ and $m_A$, and other regions at large values of
$|\mu|$ and $m_A$ excluded because either the $\stau$ (brown) or sneutrino
(dark blue) is the LSP. Most of the half-plane with $\mu < 0$ is excluded by $b \goto s \gamma$,
and also a small region with small $m_A$ and $\mu > 0$.
The Higgs mass is slightly below the LEP constraint over the entire plane in all four panels of 
Fig.~\ref{fig:traditional3_1}. In panel (a) compatibility with $g_{\mu}-2$ is found for $\mu > 0$.
The dark matter density favoured by WMAP {\it et al.} is attained in narrow strips that stretch
around the non-excluded regions. They feature a gaugino-Higgsino crossover at small $|\mu|$
and large $m_A$, sneutrino coannihilation at large $|\mu|$ and $m_A$, rapid-annihilation
funnels at $m_A \sim 250$~GeV, and $\stau$ coannihilation at small $|\mu|$ and $m_A$.

The NUHM1 line is a symmetric parabola passing through $(|\mu|, m_A)
= (\sim 700, 0)$~GeV and $(0, \sim 550)$~GeV. For $\mu>0$, this passes through the WMAP strip
in three locations, once in the crossover strip at $m_A \sim 520$~GeV, and once on
either side of the rapid-annihilation funnel at $\mu \sim
650$~GeV. These NUHM1 WMAP-preferred region crossings are visible in
the NUHM1 planes, as well.  For example, in panel (a)
of Figure~\ref{fig:mAm0fixm12}, where $m_{1/2}=300$~GeV, by following
$m_0 = 100$~GeV, one encounters precisely these three WMAP preferred
strips, one at $m_A=520$~GeV near the boundary of the region where
electroweak symmetry breaking is not obtained, plus both walls of
the rapid-annihilation funnel at lower $m_A$. The same crossings can
be observed in the $(\mu,m_0)$ plane when $m_{1/2}$ is fixed to be
300~GeV, by examining the $m_0=100$~GeV contour in a similar manner. On the other hand,
the NUHM1 line in the NUHM2 plane completely misses the sneutrino coannihilation region at
large $\mu$ and $m_A$, which is a new feature for the NUHM2. In this
case, the CMSSM point (marked by a + sign) is in a region interior to
the WMAP strip, where the relic LSPs are overdense. 

Turning now to the corresponding $(|\mu|, m_A)$ plane for 
$m_{1/2} = 300$~GeV, $m_0 = 100$~GeV and $\tbt = 20$, shown in
panel (b) of Fig.~\ref{fig:traditional3_1}, we see that the $\stau$-LSP regions at low
$|\mu|$ and $m_A$ have expanded somewhat, and the $\snu$-LSP regions at
large $|\mu|$ have changed little, whereas the $\stau$-LSP region has concentrated at
large $m_A$. The $b \goto s \gamma$ constraint is of reduced importance compared
to panel (a), and $g_\mu - 2$ now favours a region of small $\mu > 0$. The WMAP
strip is qualitatively similar to that in panel (a), except that there are now separate
$\stau$ and $\snu$ coannihilation regions at large $m_A$.

The NUHM1 line follows closely the $\stau$ coannihilation strip at low $|\mu|$
and $m_A$ missing, in this case, both the crossover strip and the $\snu$
coannihilation strip. In particular, the CMSSM points for both positive and
negative $\mu$ would, with only minor adjustment, satisfy the WMAP
constraint as well as the phenomenological constraints including $b \goto s \gamma$.
The CMSSM point with $\mu > 0$ also lies in the region favoured by $g_\mu - 2$,
as does a portion of the NUHM1 strip extending from $m_A \sim 300$ to 500~GeV.

For larger values of $\tbt$, as seen in panels (c) and (d) of Fig.~\ref{fig:traditional3_1},
the half-plane with $\mu < 0$ and a large part of the half-plane with $\mu > 0$ are
excluded because the $\stau$ is the LSP. The $\snu$-LSP region at
large $\mu > 0$ has also expanded, leaving only a (curved) triangle of allowed
parameters at $\mu > 0$. The WMAP strip now consists of a $\stau$
coannihilation strip and a $\snu$ coannihilation strip, linked by a
rapid-annihilation funnel. 
Since the values of $m_{1/2}$ and $m_0$ chosen for 
Fig.~\ref{fig:traditional3_1} are not large, all the WMAP-compatible points are
accessible to the LHC~\cite{lhc,cmstdr}, and several types of sfermions should be detectable.
Some neutralinos, charginos and heavy Higgs bosons should also be
detectable in the $\stau$ coannihilation strip and the rapid-annihilation funnel,
but this would be more difficult in the $\snu$ coannihilation strip.

In both panels (c) and (d), only a small portion of the NUHM1 line is allowed. It
intersects the WMAP strip close to a junction between the $\stau$ coannihilation
strip and the rapid-annihilation funnel. The CMSSM points in both panels are
well within the excluded $\stau$-LSP region, as could have been
anticipated from the well-known fact that this region extends to higher $m_0$
(at fixed $m_{1/2}$) as $\tbt$ increases.

\begin{figure}
\begin{center}
\mbox{\epsfig{file=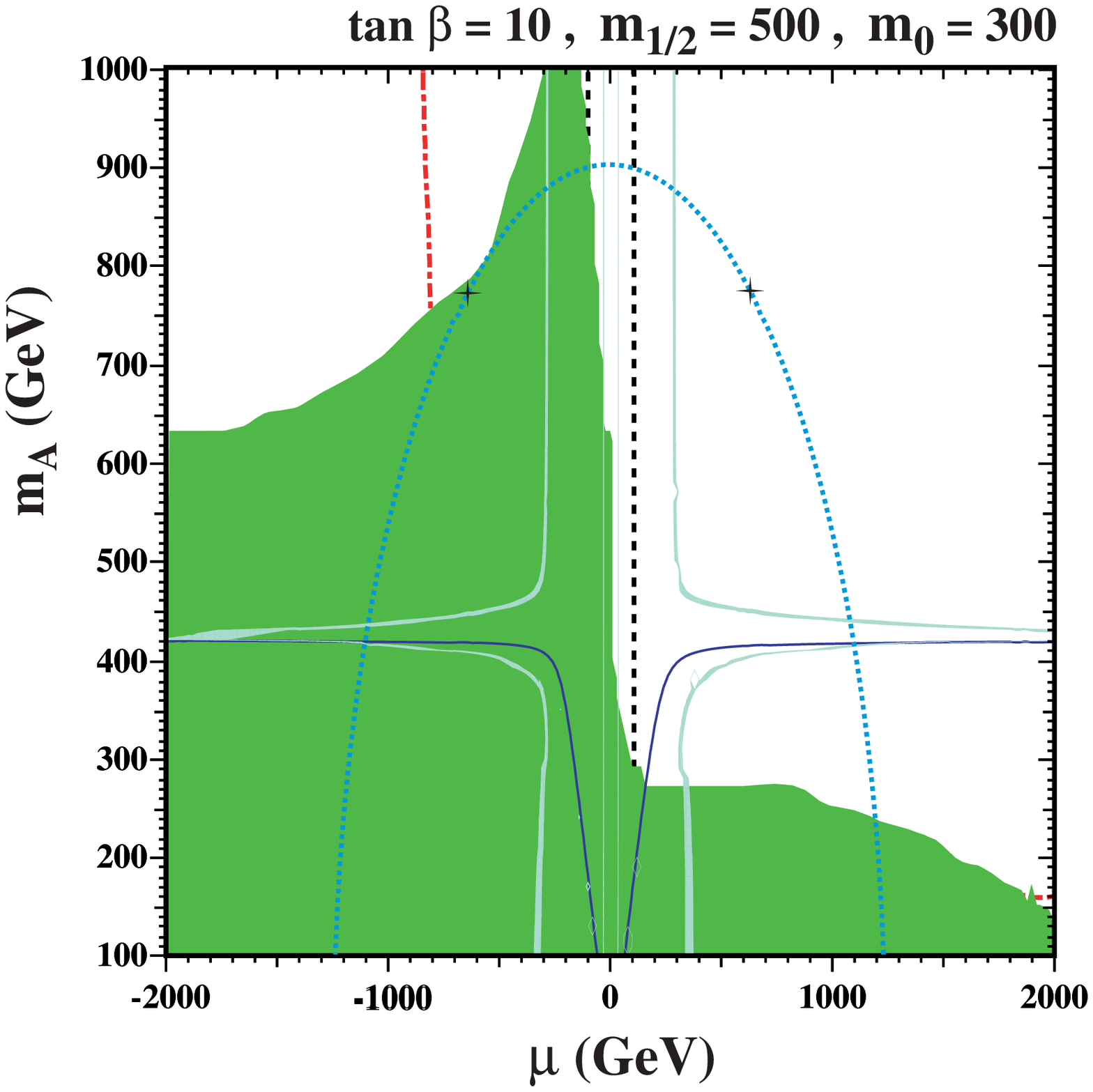,height=7cm}}
\mbox{\epsfig{file=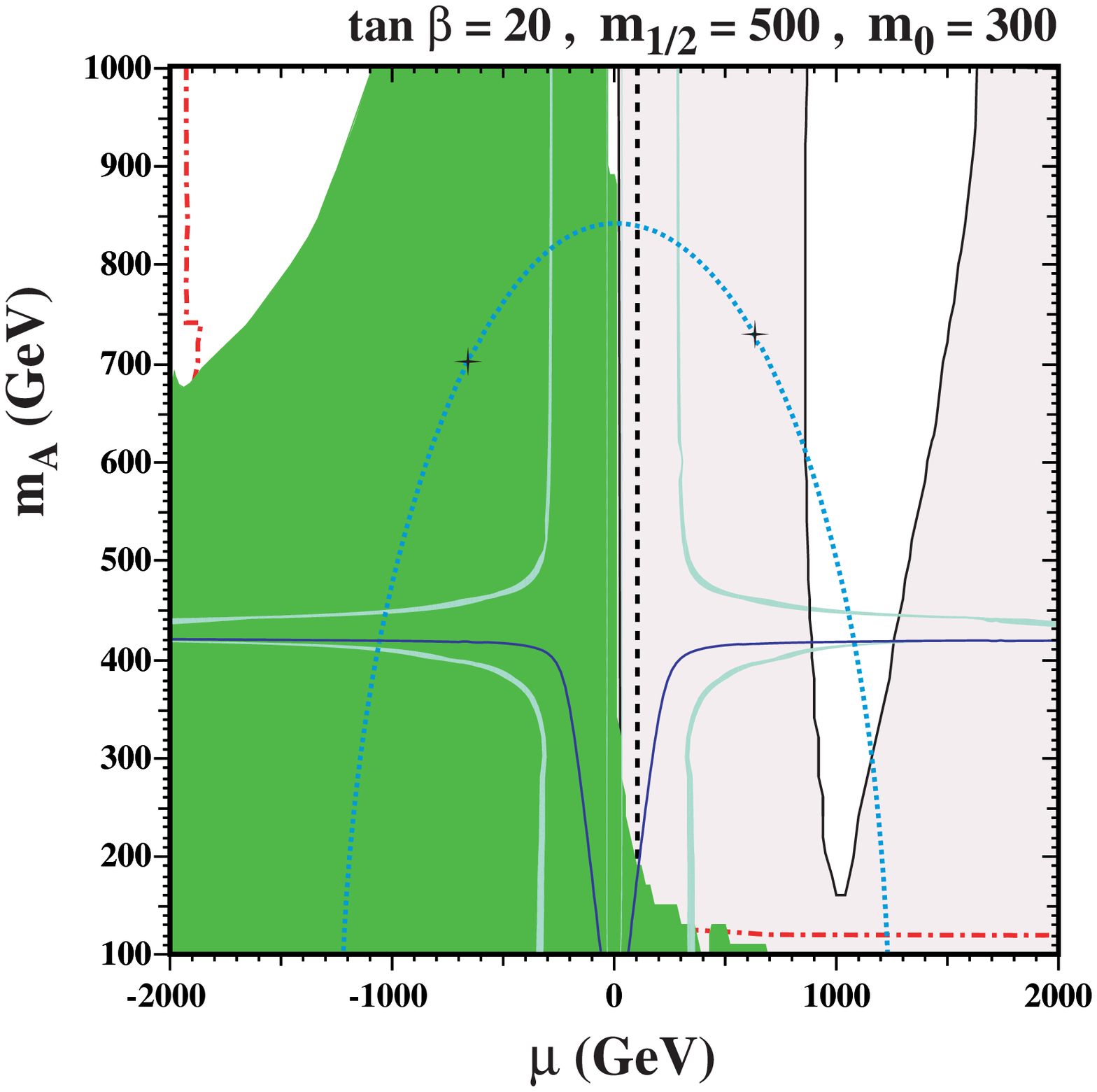,height=7cm}}
\end{center}
\begin{center}
\mbox{\epsfig{file=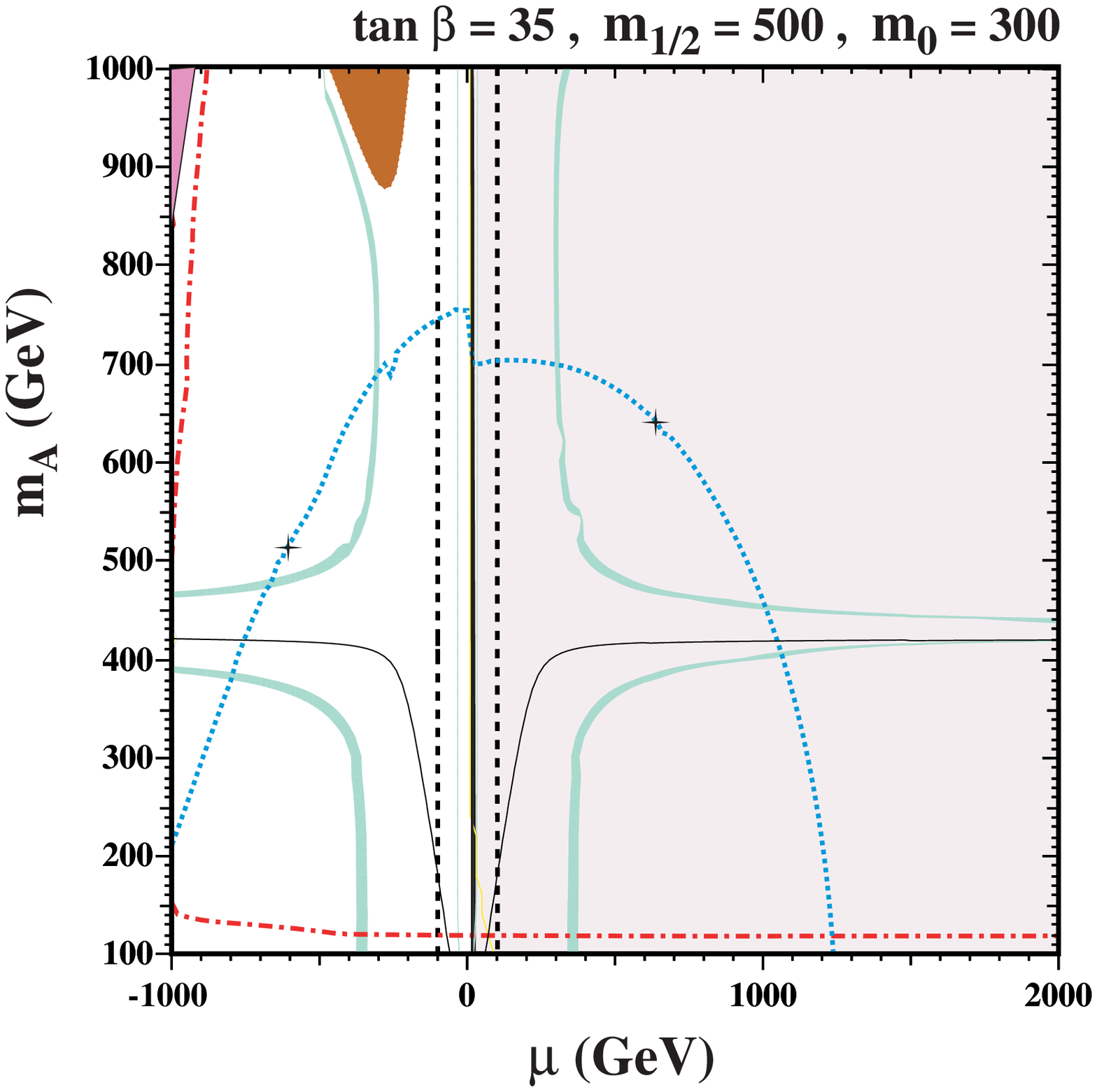,height=7cm}}
\mbox{\epsfig{file=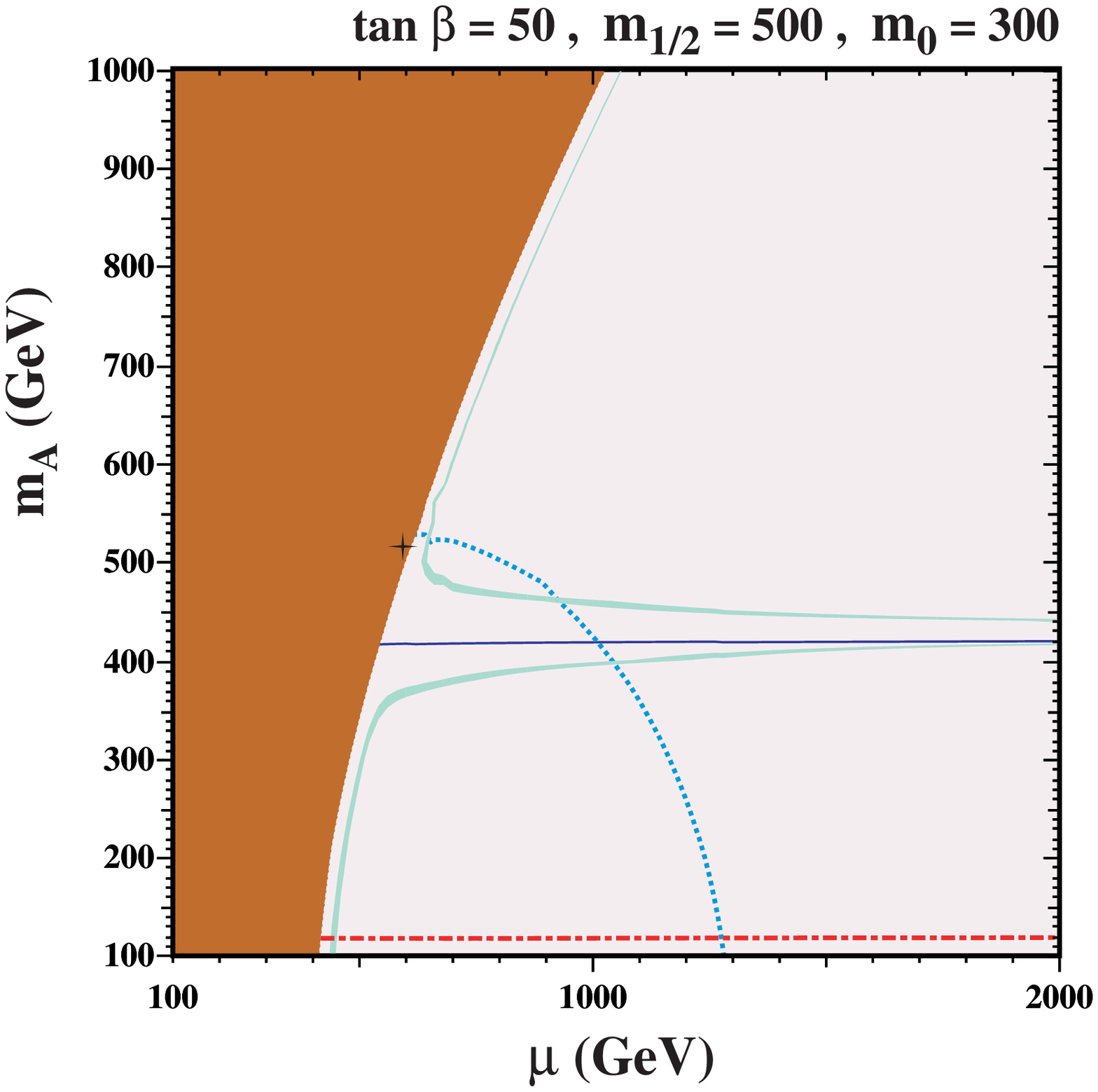,height=7cm}}
\end{center}
\caption{\it  Examples of NUHM2 $(\mu,m_A)$ planes with
$m_{1/2}=500$~GeV, $m_0=300$~GeV, $A_0=0$, and $\tanb = 10$, 20, 35, and 50 in
panels (a), (b), (c), and (d), respectively. Constraints are displayed as in Fig.~\ref{fig:m12m0fixmA}.}
\label{fig:traditional5_3}
\end{figure}

The configurations of the $(\mu, m_A)$ planes change significantly for
$(m_{1/2}, m_0) =  (500, 300)$ GeV, as seen in Fig.~\ref{fig:traditional5_3}.
The $\stau$- and $\snu$-LSP regions disappear completely in panels (a) and (b) for
$\tbt = 10$ and 20, respectively. There is only a small excluded region
in panel (c) for $\tbt = 35$, which grows finally in panel (d) for $\tbt = 50$.
Much of the $\mu < 0$ half-plane is excluded by $b \goto s \gamma$
for $\tbt = 10$ and 20, but this constraint disappears for larger
$\tbt$.
The LEP Higgs constraint is not important in the regions allowed by
$b \goto s \gamma$. The $\mu > 0$ half-planes are favoured by $g_\mu - 2$
for $\tbt \gtrsim 18$.
The regions favoured by WMAP are rapid-annihilation funnels for all values of $\tbt$,
crossover strips for $\tbt = 10, 20$ and $35$, and $\stau$ coannihilation strips for
$\tbt = 50$ and (fleetingly) for $\tbt = 35$.

The NUHM1 lines are again (approximate) parabolae in all four panels.
They intersect the WMAP strips in crossover and rapid-annihilation regions
in panels (a, b) and (c), for $\tbt \le 35$, and in the rapid-annihilation and
$\stau$ coannihilation regions for $\tbt = 50$ in panel (d). We note that
in panel (d) the approximate NUHM1 parabola has shifted such that for
some values of $m_A$ there is no unique solution for $\mu$~\footnote{For this
reason, the boundary of the region where there are no consistent
solutions to the electroweak vacuum conditions appears augmented in the
the NUHM1, $\tanb=50$ planes of Section~\ref{sec:mA}.}
The CMSSM points are in strongly overdense regions in panels
(a, b) and (c), but in the forbidden $\stau$-LSP region of panel (d). However,
this point is close to an allowed region where the relic density would be
within the favoured range. Therefore, there are nearby CMSSM points with 
similar values of $m_{1/2}, m_0, \tbt, m_A$ and $\mu$ that are consistent
with all the constraints.
All these planes in Fig.~\ref{fig:traditional5_3}
should be accessible to the LHC~\cite{lhc,cmstdr}, because of the moderate values chosen for
$m_{1/2}$ and $m_0$, but some heavier neutralinos, charginos and Higgs bosons
would only be accessible for relatively small values of $\mu$ and $m_A$.

\begin{figure}
\begin{center}
\mbox{\epsfig{file=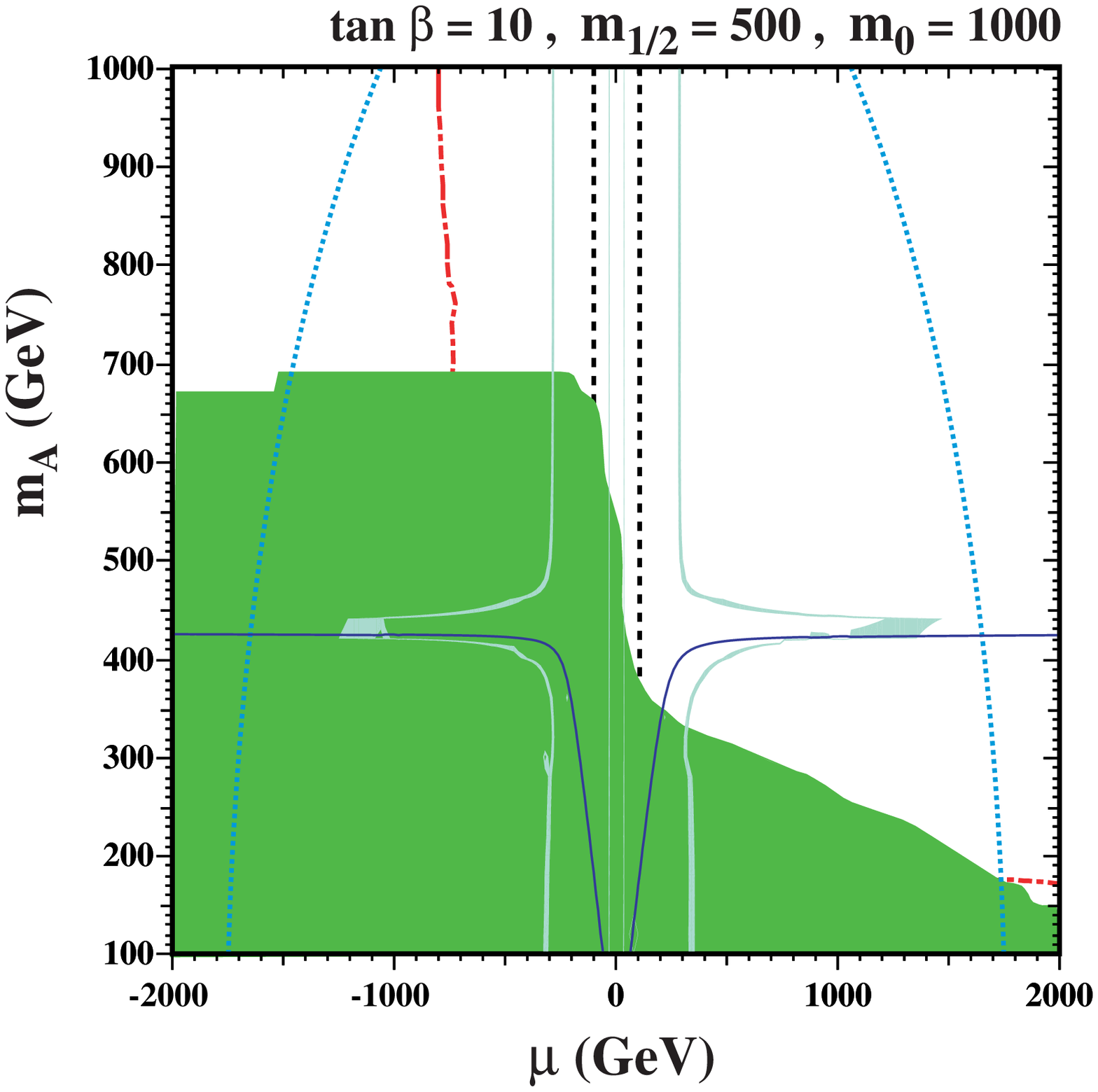,height=7cm}}
\mbox{\epsfig{file=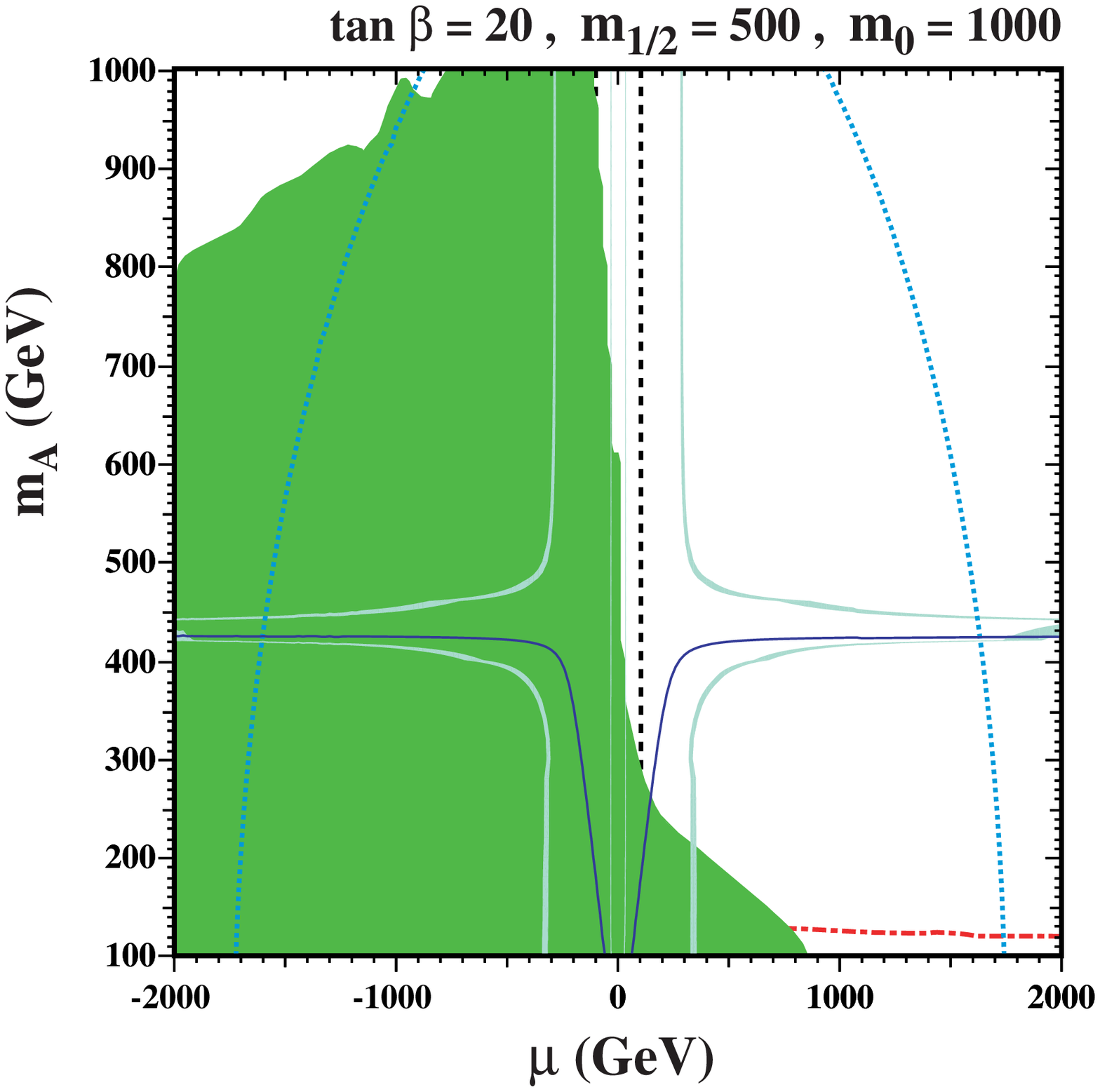,height=7cm}}
\end{center}
\begin{center}
\mbox{\epsfig{file=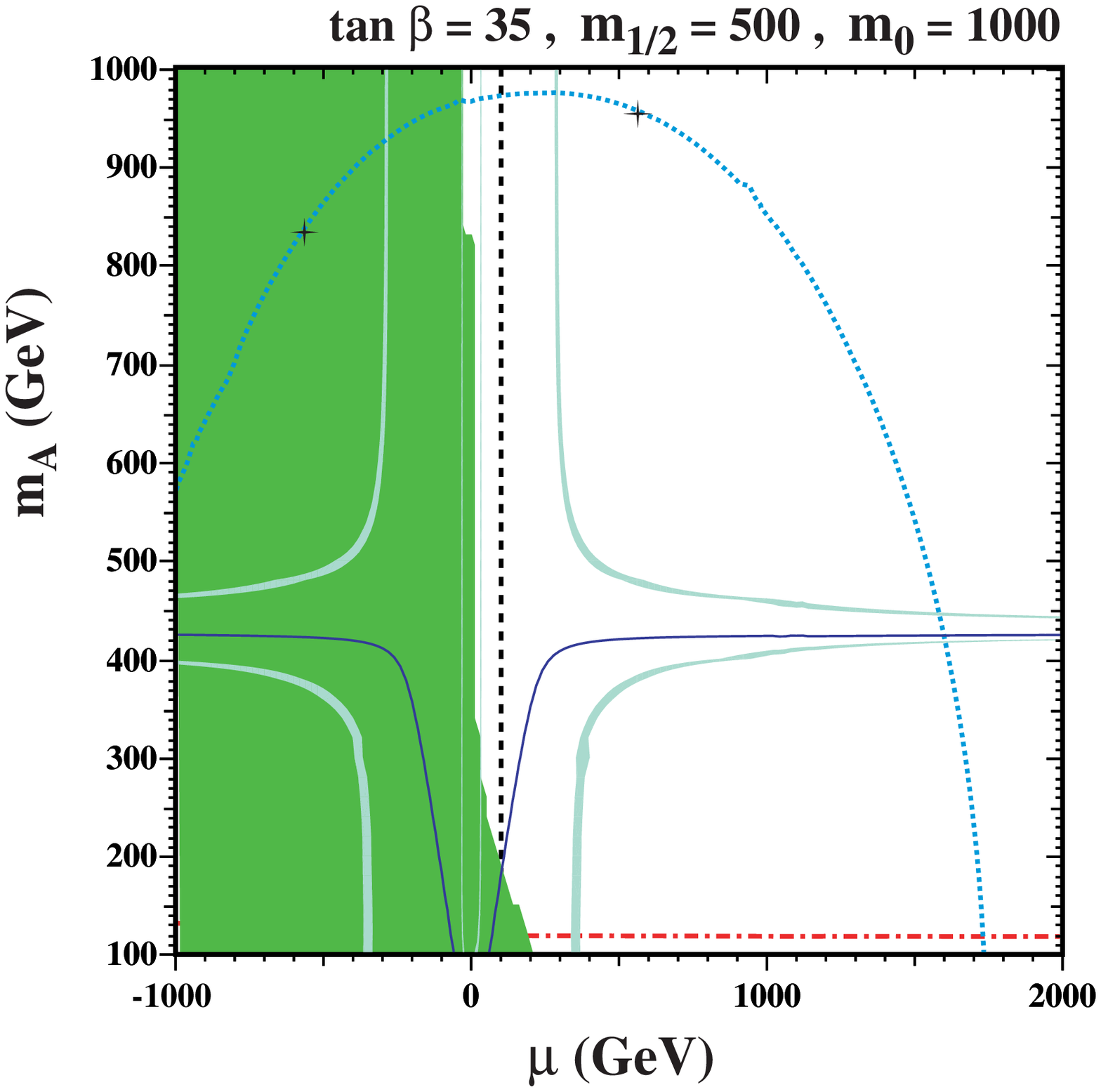,height=7cm}}
\mbox{\epsfig{file=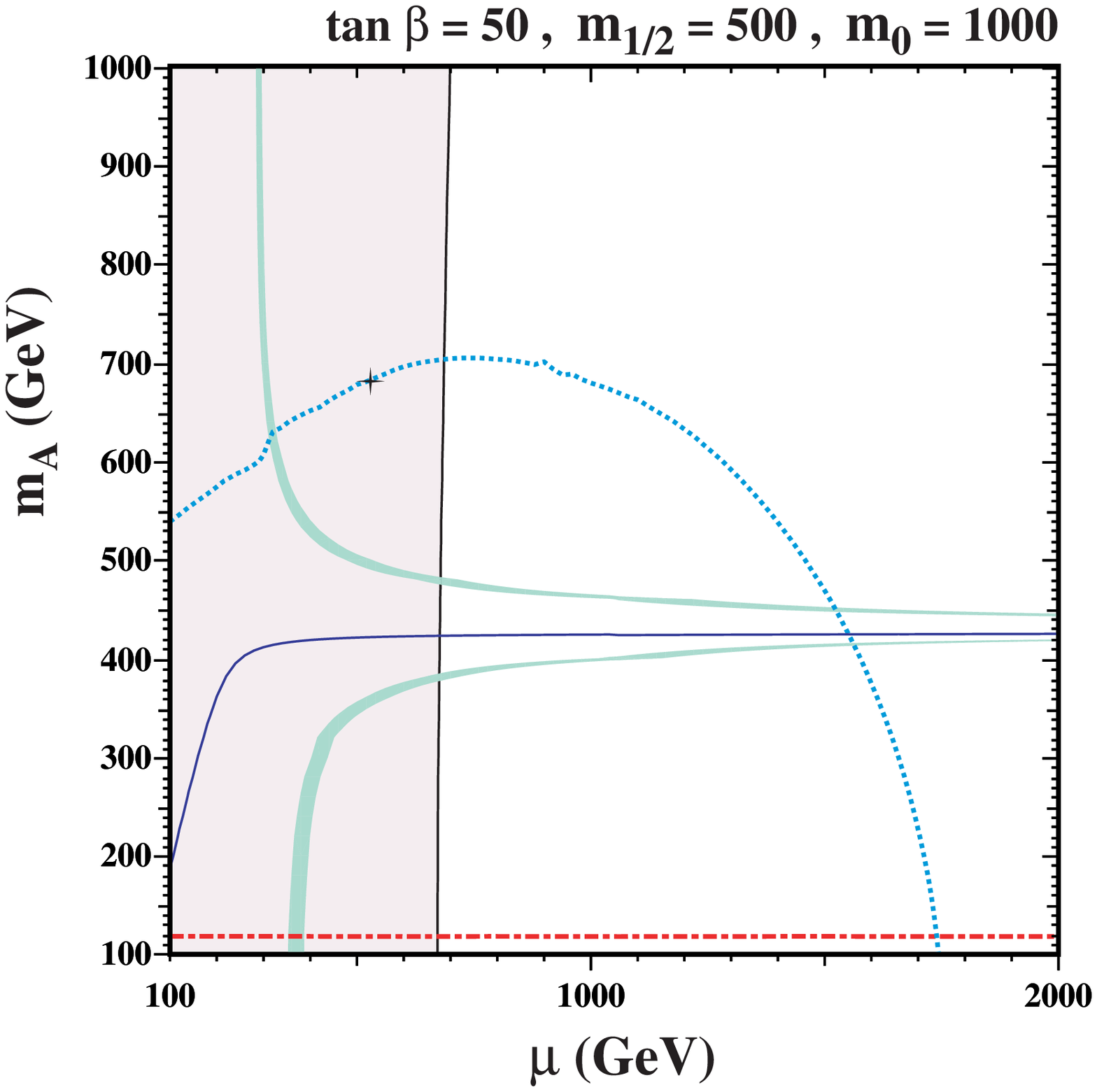,height=7cm}}
\end{center}
\caption{\it  Examples of NUHM2 $(\mu,m_A)$ planes with
$m_{1/2}=500$~GeV, $m_0=1000$~GeV, $A_0=0$, and $\tanb = 10$, 20, 35, and 50 in
panels (a), (b), (c), and (d), respectively. Constraints are displayed as in Fig.~\ref{fig:m12m0fixmA}.}
\label{fig:traditional5_10}
\end{figure}

Finally, we present in Fig.~\ref{fig:traditional5_10} some $(\mu, m_A)$ planes for
the choices $(m_{1/2}, m_0) = (500, 1000)$ GeV. Unlike the previous cases,
these choices are in a region of the $(m_{1/2}, m_0)$ plane that is far from the
coannihilation strip in the CMSSM. No parts of any of the planes are excluded
by the absence of electroweak symmetry breaking or the presence of a charged LSP.
We see explicitly in the panels (a, b) and (c) for
$\tbt \le 35$ that $b \goto s \gamma$ again excludes most of the half-plane
with $\mu < 0$. For $\tbt = 50$, shown in panel (d), reliable solutions are not found with $\mu < 0$. The LEP Higgs limit does not exclude a significant extra
region of the $(\mu, m_A)$ plane in any of the panels. In panel (d) for $\tbt = 50$ there is a
region at $\mu < 700$~GeV that is favoured by $g_\mu - 2$, but not for the lower values
of $\tbt$. In each of the panels, the region favoured by WMAP consists of a crossover
strip at $\mu \sim 300$~GeV and a rapid-annihilation funnel with
400~GeV$ < m_A <$ 450~GeV.
These planes should also be accessible to the LHC~\cite{lhc,cmstdr}, though more
luminosity would be required than in the previous cases because of the larger
value of $m_0$, in particular. This would also render more difficult the searches
for some heavier neutralinos, charginos and Higgs bosons.

The NUHM1 lines are again parabolae,
reaching values of $m_A$ that decrease from $> 1000$~GeV to $\sim 700$~GeV
as $\tbt$ increases, and becoming increasingly asymmetric in
$\mu$~\footnote{Again, this leads to a lack of unique solutions for
$\mu$ for some choices of $m_A$, $m_0$, and $m_{1/2}$ with $\tanb=50$.}. They intersect the WMAP regions in both the rapid-annihilation 
strips and the crossover strips (the latter at $m_A > 1000$~GeV for $\tbt < 20$).
Thus, the NUHM1 lines do sample both the WMAP possibilities in these NUHM2
planes. On the other hand, the CMSSM points are always in strongly overdense
regions of the $(\mu, m_A)$ planes.

As the GUT-scale values of the gaugino and scalar masses are fixed in
the NUHM2 planes in
Figures~\ref{fig:traditional3_1}-\ref{fig:traditional5_10} (as well as
Figures~\ref{fig:new3_1}-\ref{fig:new5_10} in the next subsection), the sparticle
spectrum does not vary much over any individual panel, the
primary exceptions being the Higgs masses.  What is novel
in the NUHM2 is that there are allowed regions of the NUHM2 parameter
space with very low $(m_{1/2},m_0)$, leading to sparticle masses below
what would be expected in the CMSSM.  
Alternatively, inspection of
$(\mu,m_A)$ planes for large $(m_{1/2},m_0)$ would show that there are
indeed cosmologically preferred strips that evade all collider
constraints and have very heavy sparticles.

\subsection{NUHM2 $(m_1, m_2)$ Planes}

We now present a novel analysis of the NUHM2, based directly on the input
non-universal soft supersymmetry-breaking parameters, $m_1$ and $m_2$,
for the same choices of $m_{1/2}$ and $m_0$ as were used in the previous
subsection.

\begin{figure}
\begin{center}
\mbox{\epsfig{file=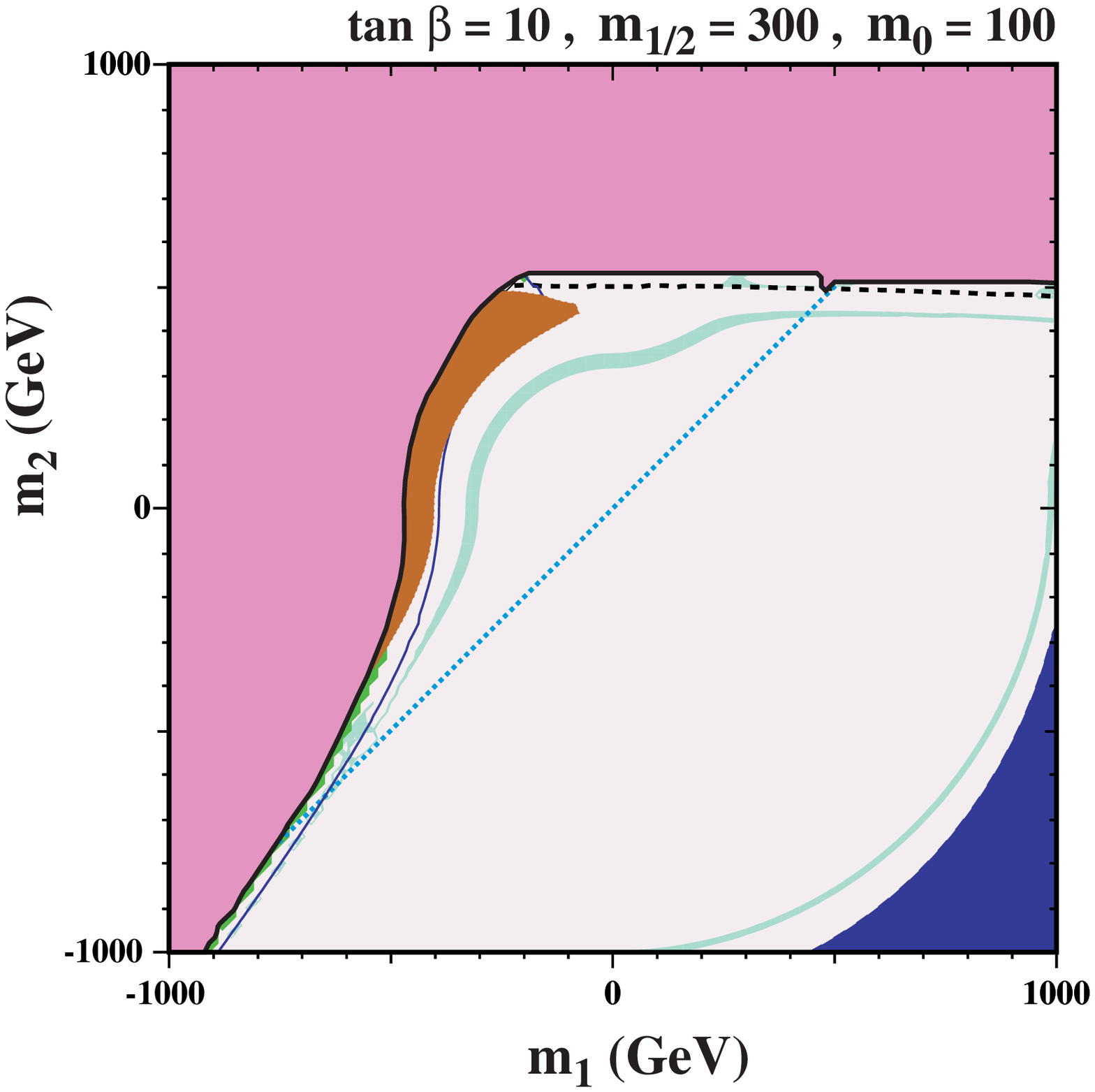,height=7cm}}
\mbox{\epsfig{file=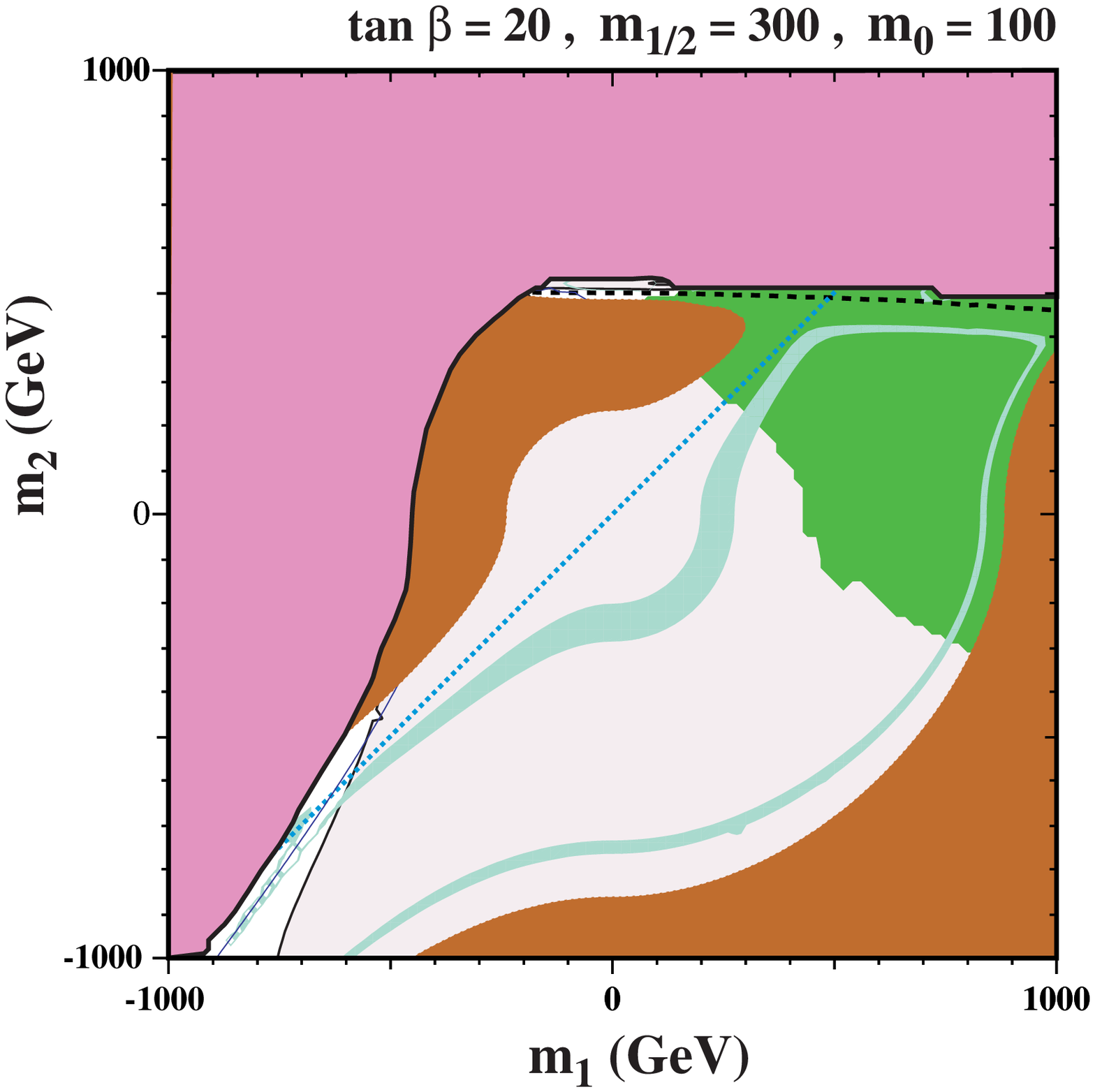,height=7cm}}
\end{center}
\begin{center}
\mbox{\epsfig{file=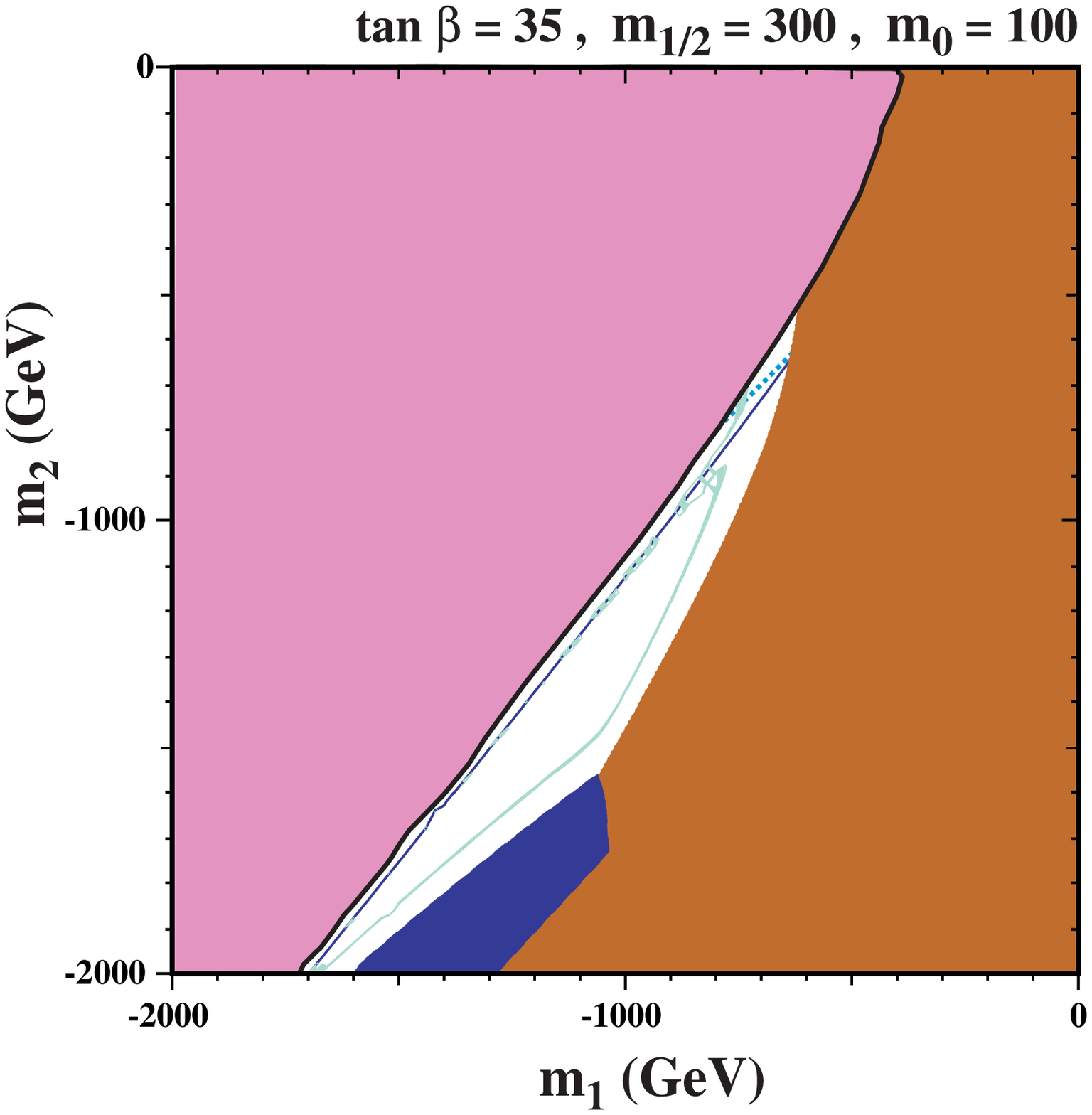,height=7cm}}
\mbox{\epsfig{file=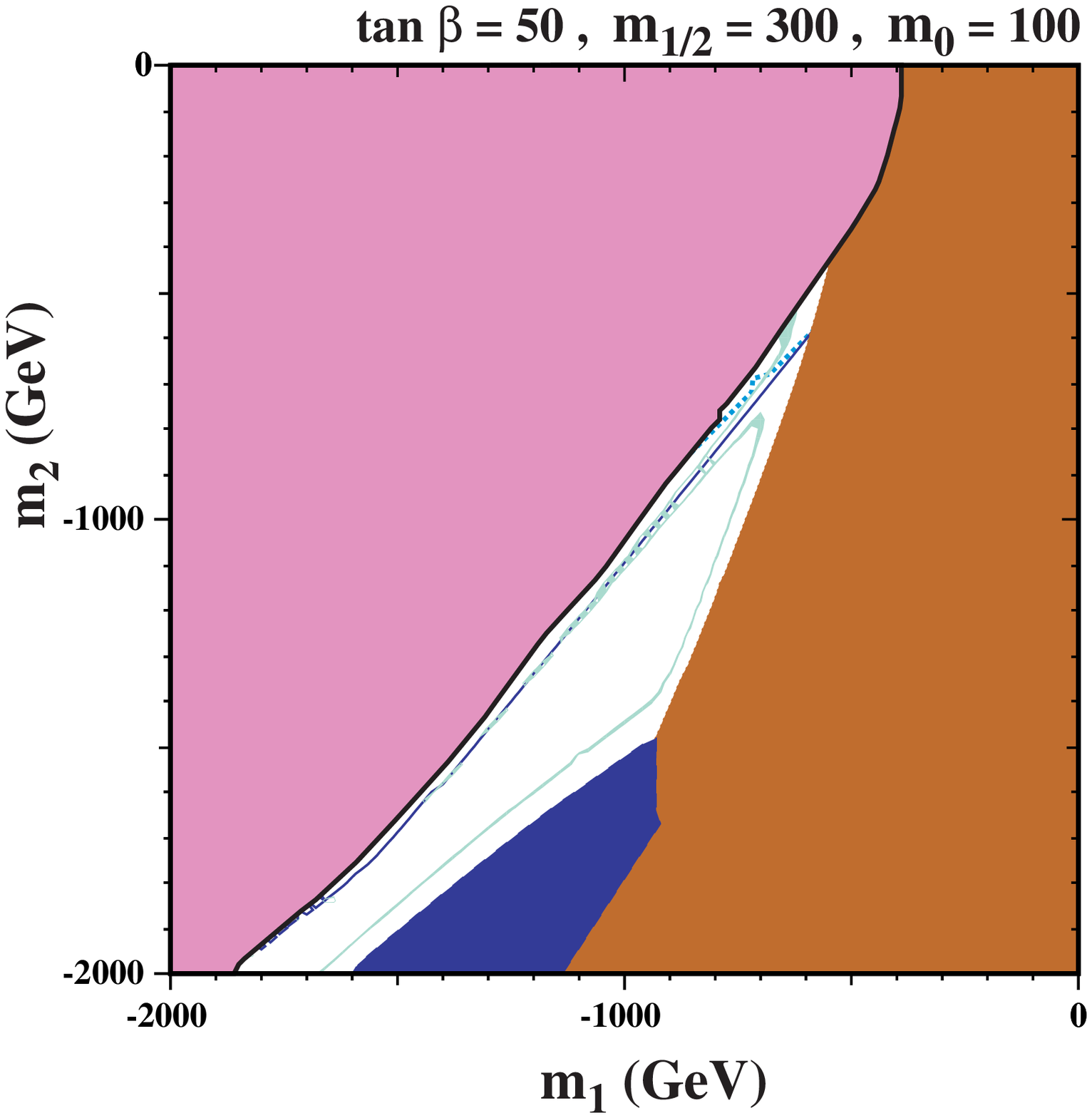,height=7cm}}
\end{center}
\caption{\it  Examples of NUHM2 $(m_1,m_2)$ planes with
$m_{1/2}=300$~GeV, $m_0=100$~GeV, $A_0=0$, and $\tanb = 10$, 20, 35, and 50 in
panels (a), (b), (c), and (d), respectively. The sign in the axes labels refer to the
sign of $m_{1,2}^2$. Constraints are displayed as in Fig.~\ref{fig:m12m0fixmA}.}
\label{fig:new3_1}
\end{figure}

Fig.~\ref{fig:new3_1} shows a selection of $(m_1, m_2)$ planes for the same
values $(m_{1/2}, m_0) = (300, 100)$ GeV as in Fig.\ref{fig:traditional3_1}.
We notice immediately that large negative values of $m_1$ and positive values
of $m_2$ are excluded by the electroweak symmetry-breaking requirement,
and regions of positive $m_1$ and negative $m_2$ are excluded because the
$\stau$ or $\snu$ is the LSP. There are also $\stau$-LSP
excluded regions in the second quadrant of panels (a) and (b), for $\tbt = 10$ and 20.
The slepton-LSP constraints
become much stronger as $\tbt$ increases, with the effect that the allowed
region of parameter space is pushed to values of $m_{1,2}^2 \ll 0$, far away from
values where $m_{1,2}/m_0 = {\cal O}(1)$. The dashed blue diagonal lines 
in panels (a) and (b) are
the NUHM1 lines where $m_1 = m_2$, and the CMSSM points are found at $m_1 = m_2 = m_0$. 

As in Fig.~\ref{fig:traditional3_1}, the Higgs mass is slightly below the LEP constraint over the entire plane in all four panels of 
Fig.~\ref{fig:new3_1}.
The LEP chargino constraint runs close to the upper boundaries of the
allowed regions in panels (a) and (b) of Fig.~\ref{fig:new3_1}.
The $b \goto s \gamma$ constraint is visible only in panel (b), for
$\tbt = 20$, where it excludes a large part of the first quadrant.
Likewise, the region favoured for $g_\mu - 2$ is also visible only
in panels (a) and (b), where it covers most of the allowed part of the $(m_1, m_2)$
plane.

It is a common feature of all the panels that the WMAP strip skirts
the boundaries of the allowed region. In panels (a) for $\tbt = 10$
and (b) for $\tbt = 20$, it comprises a crossover strip at the top and, 
combined with a rapid-annihilation funnel
in the bottom left corner, a $\stau$ coannihilation strip 
on the left side, and a $\snu$ coannihilation strip (in (a)) and a $\stau$ coannihilation
strip (in (b)) on the right side.
In panels (c) for $\tbt = 35$ and (d) for $\tbt = 50$, it comprises a crossover
strip/rapid-annihilation funnel on the left side and a coannihilation strip on the right side.

In panel (a), the NUHM1 line intersects the WMAP strip in the crossover
strip in the first quadrant and in the crossover/rapid-annihilation strip in the
third quadrant. We note that both these regions have the common value
of $|m_{1,2}| \gg m_0$. Most of the rest of the NUHM1 line has excessive
relic density. On the other hand, in panel (b), the relic density lies below the WMAP
range along all the NUHM1 line, except in the third quadrant. The CMSSM points
in these two panels have relic densities that are too large, in panel (a) for $\tbt = 10$,
or too small, in panel (b) for $\tbt = 20$. The NUHM1 lines lie mostly in regions 
which are excluded and the CMSSM
points are in the disallowed regions of panels (c) and (d).

\begin{figure}
\begin{center}
\mbox{\epsfig{file=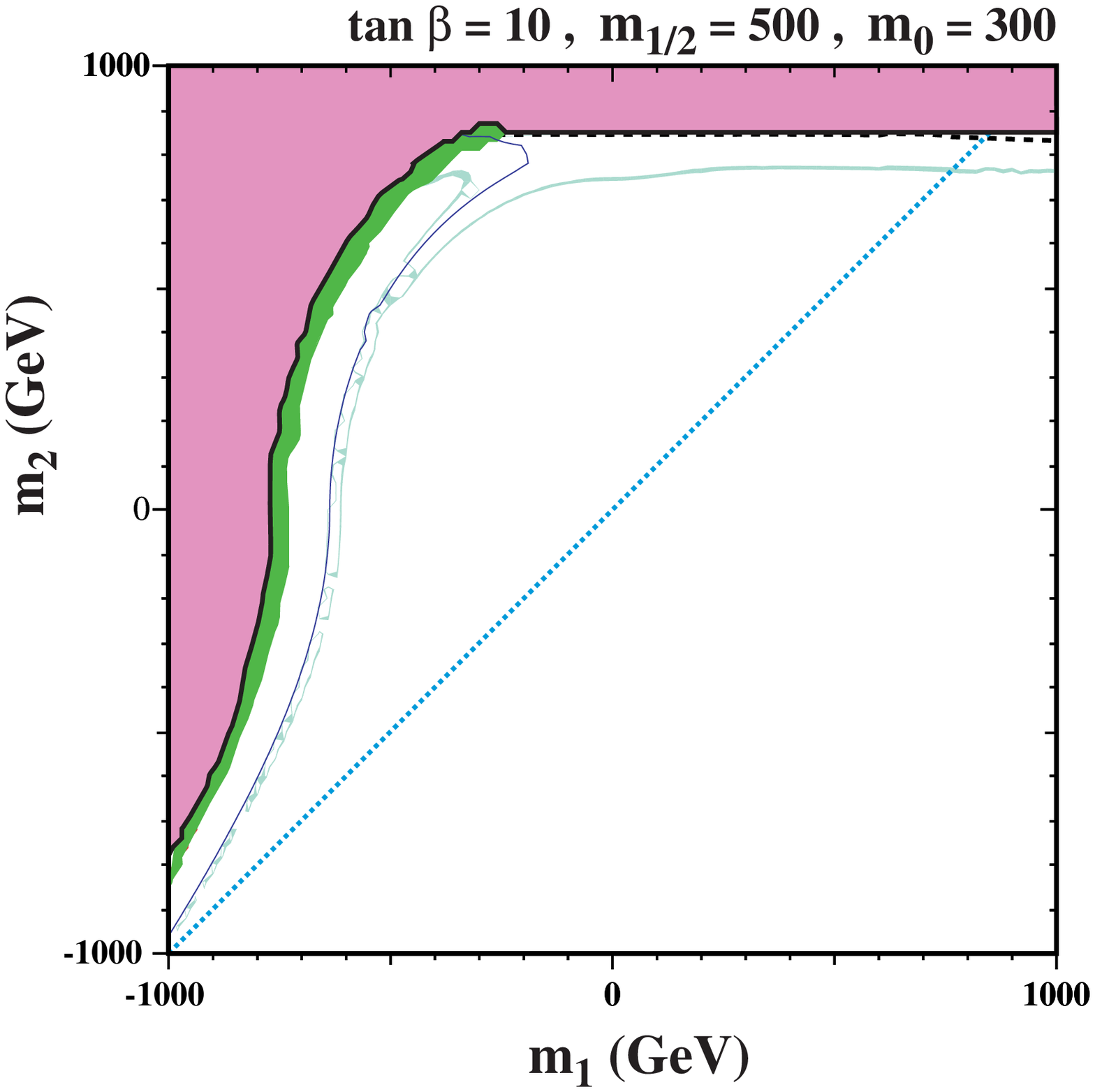,height=7cm}}
\mbox{\epsfig{file=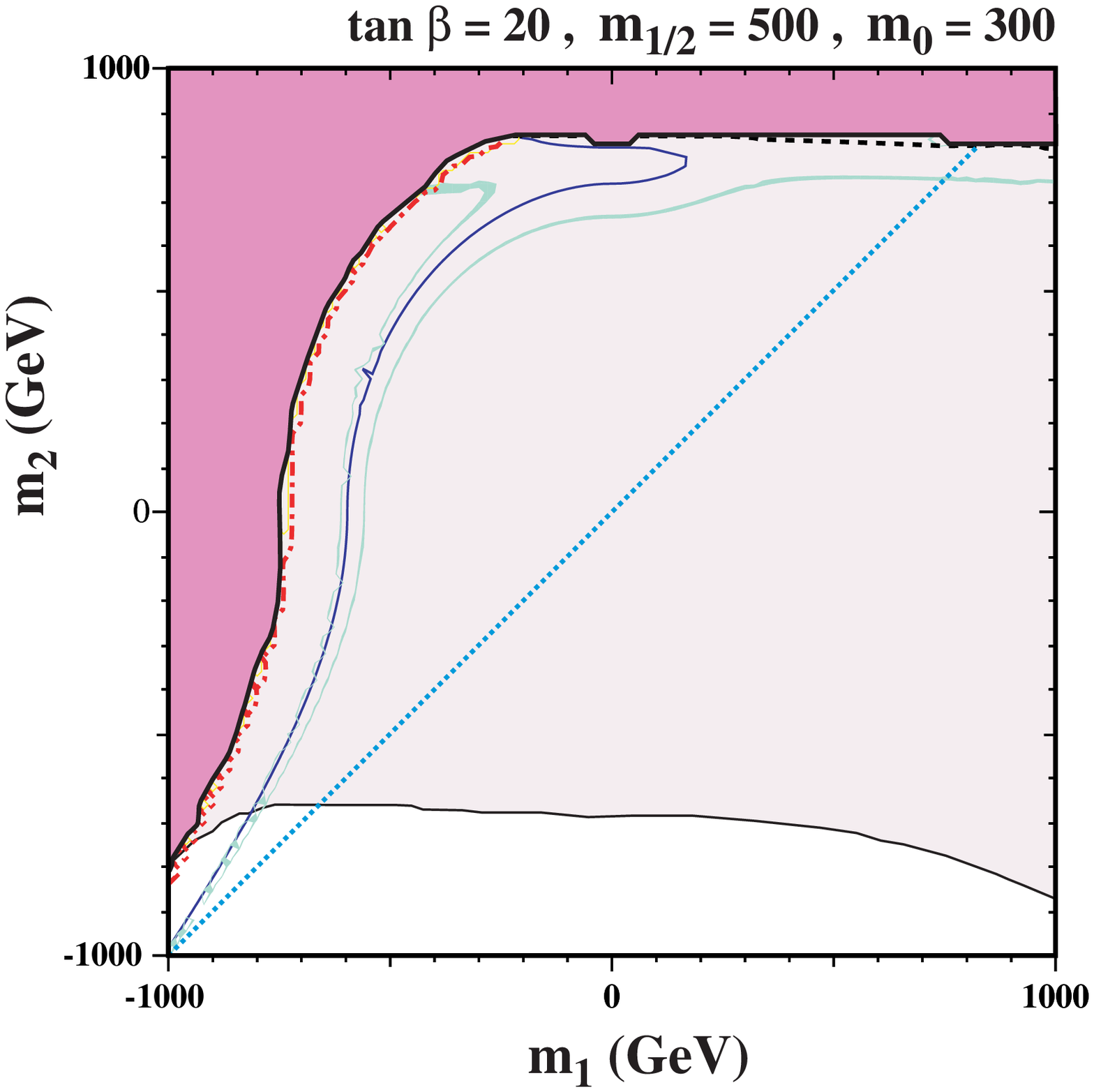,height=7cm}}
\end{center}
\begin{center}
\mbox{\epsfig{file=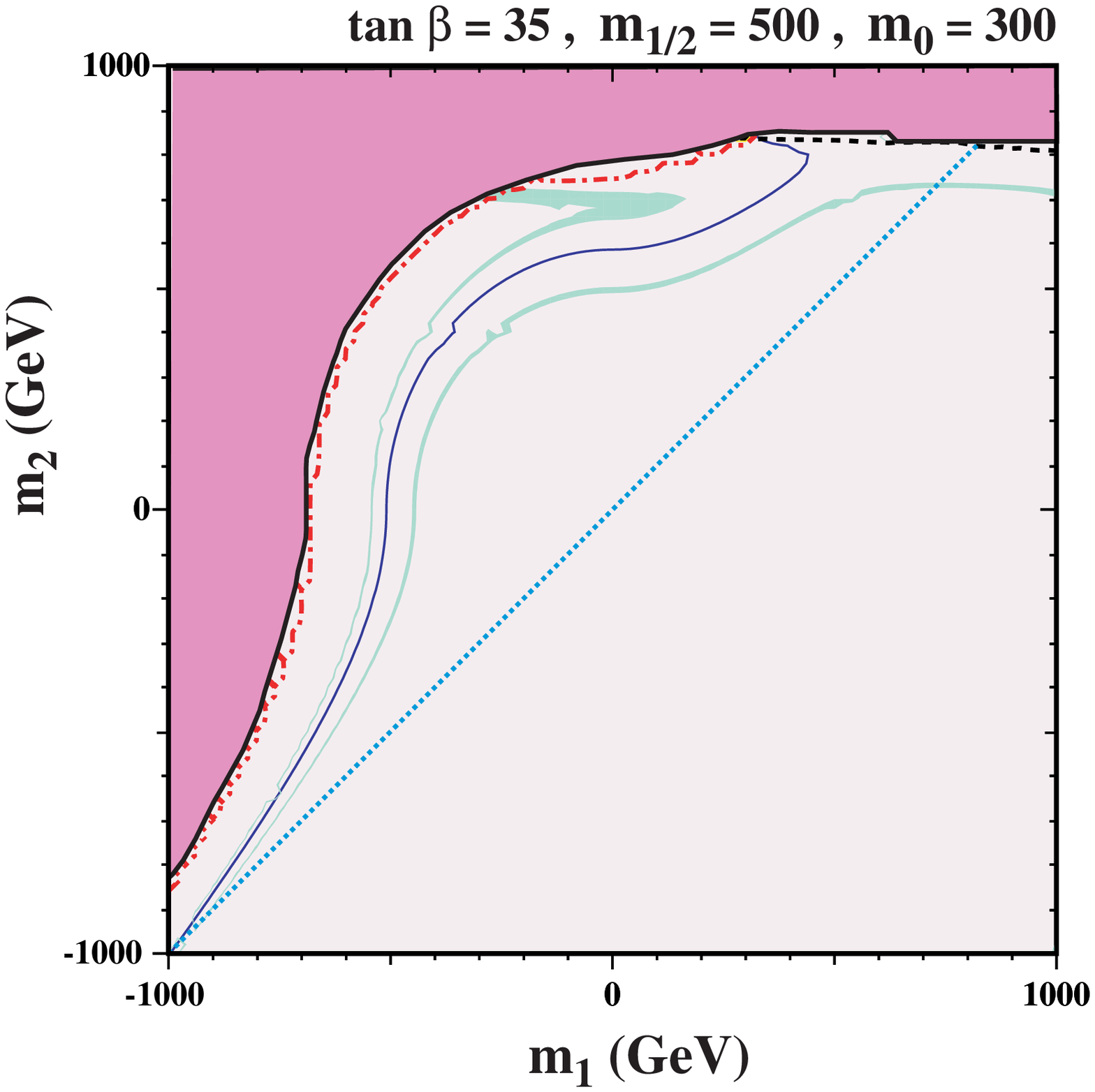,height=7cm}}
\mbox{\epsfig{file=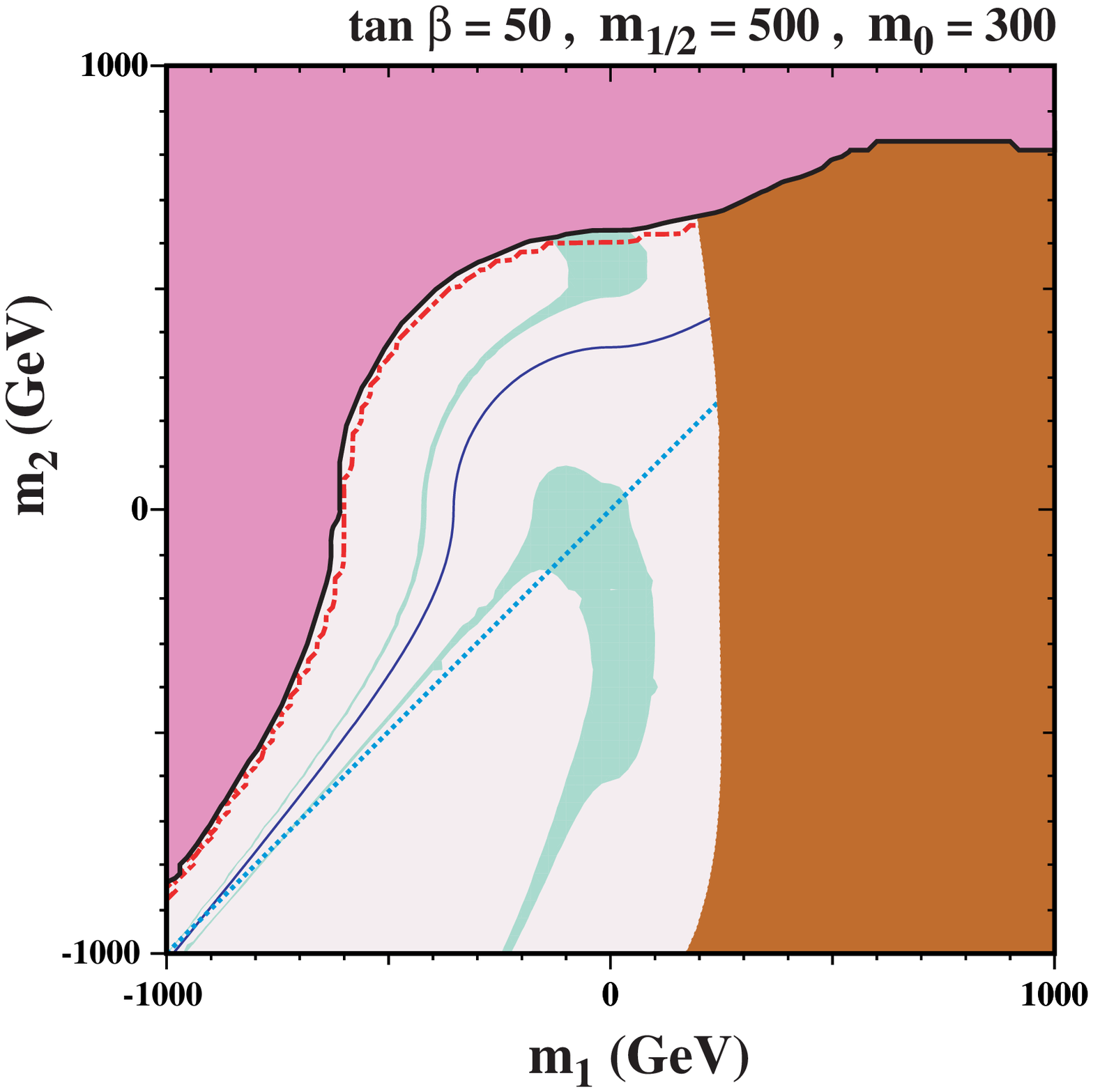,height=7cm}}
\end{center}
\caption{\it Examples of NUHM2 $(m_1,m_2)$ planes with
$m_{1/2}=500$~GeV, $m_0=300$~GeV, $A_0=0$, and $\tanb = 10$, 20, 35, and 50 in
panels (a), (b), (c), and (d), respectively. Constraints are displayed as in Fig.~\ref{fig:m12m0fixmA}.}
\label{fig:new5_3}
\end{figure}

Analogous $(m_1, m_2)$ planes for the choices $(m_{1/2}, m_0) = (500, 300)$~GeV 
are shown in Fig.~\ref{fig:new5_3}. In this case, the electroweak symmetry-breaking
condition excludes strips at large positive $m_2$ and large negative $m_1$. The
condition for the absence of a $\stau$ LSP forbids a region with large $m_1 > 0$
for $\tbt = 50$, as shown in panel (d). The LEP chargino constraint excludes a
narrow strip close to the boundary in the first and second quadrants in panels
(a, b) and (c), i.e., for $\tbt \le 35$, and the Higgs constraint excludes a narrow
strip along the boundary in the second and third quadrants in all panels. The
$b \goto s \gamma$ constraint is absent except for $\tbt = 10$, 
whilst there are large regions favoured
by $g_\mu - 2$ for $\tbt = 20$, 35, and 50, but not for $\tbt = 10$.

The rapid-annihilation funnel evolves in an interesting way as $\tbt$ increases.
After starting close to the left boundary for $\tbt = 10$, it moves out into the
allowed region as $\tbt$ increases, and becomes increasingly serpentine.
The two sides of the funnel run almost parallel for $\tbt \le 35$, with the right
side extended by a crossover strip along the top boundary for $\tbt \le 35$.
On the other hand, for $\tbt = 50$, the right boundary of the rapid-annihilation 
funnel expands and evolves into a $\stau$ coannihilation strip.

In panels (a, b) and (c) for $\tbt \le 35$,
the NUHM1 lines intersect the WMAP region in the crossover strip at large
positive $m_1$ and $m_2$, and in the rapid-annihilation funnel at large
negative $m_1$ and $m_2$. These intersections lie far from the CMSSM
point, which is in an overdense region. On the other hand, for $\tbt = 50$
in panel (d), the CMSSM point lies very close to a WMAP strip, in an
underdense region.

\begin{figure}
\begin{center}
\mbox{\epsfig{file=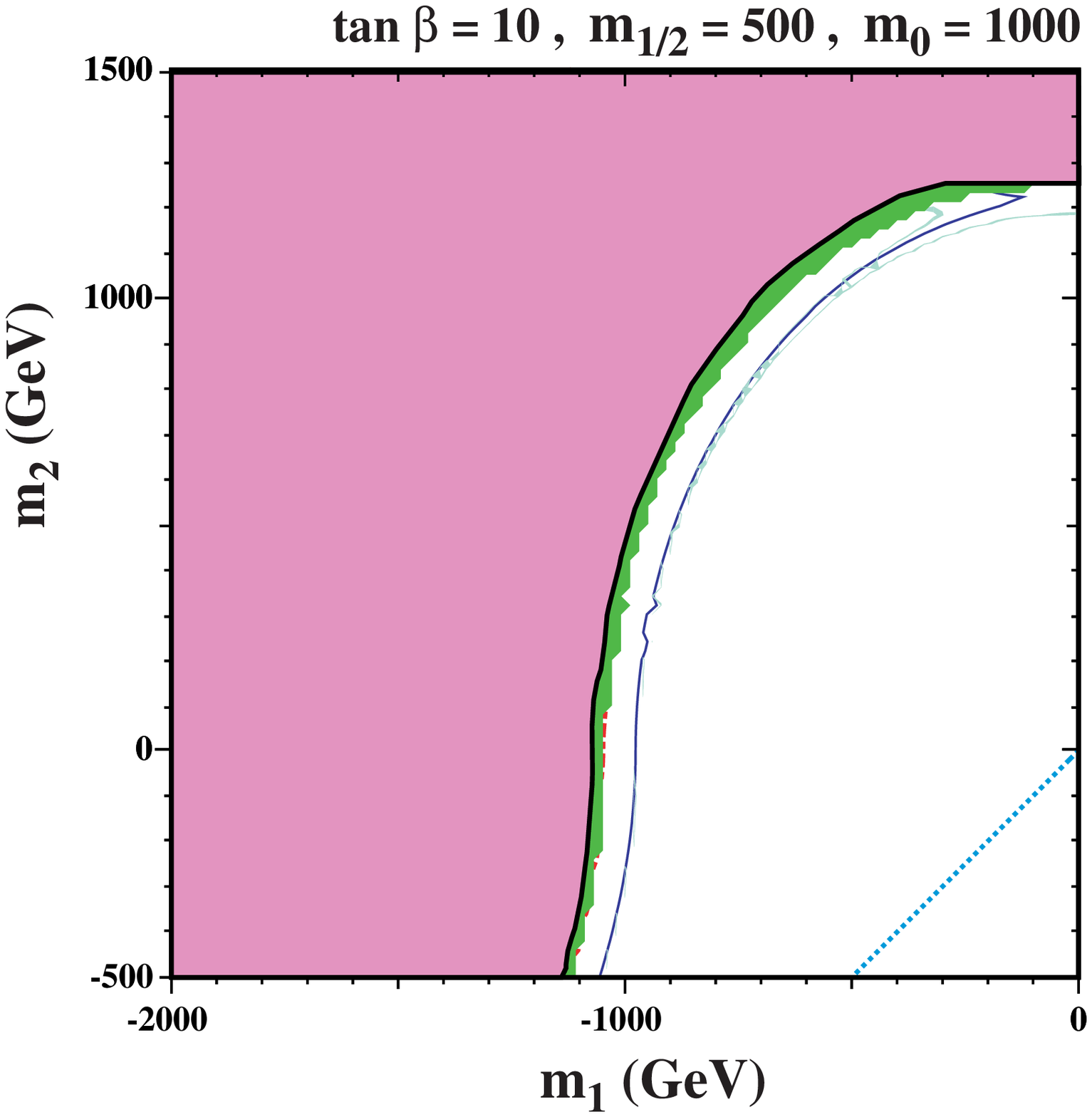,height=7cm}}
\mbox{\epsfig{file=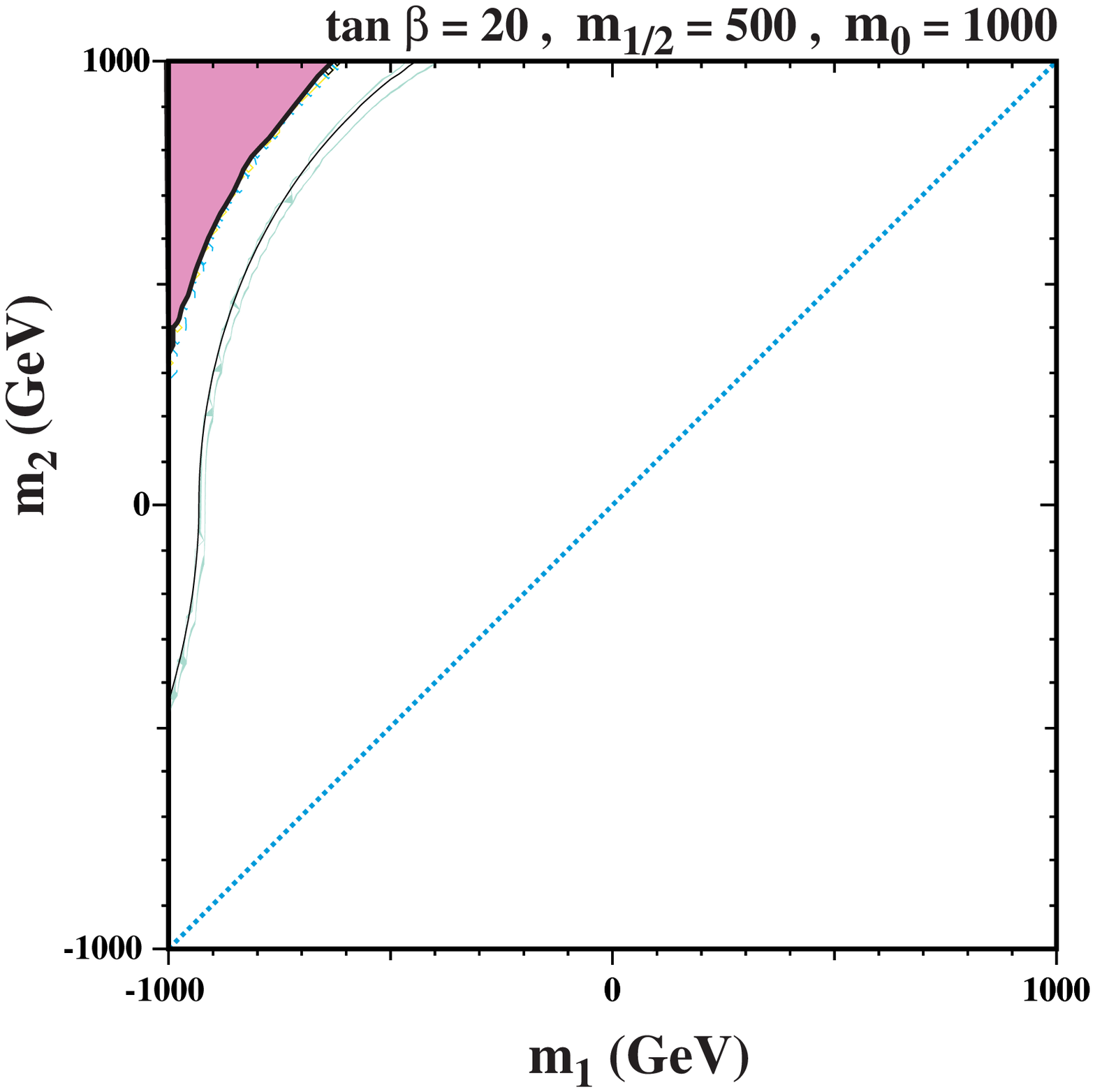,height=7cm}}
\end{center}
\begin{center}
\mbox{\epsfig{file=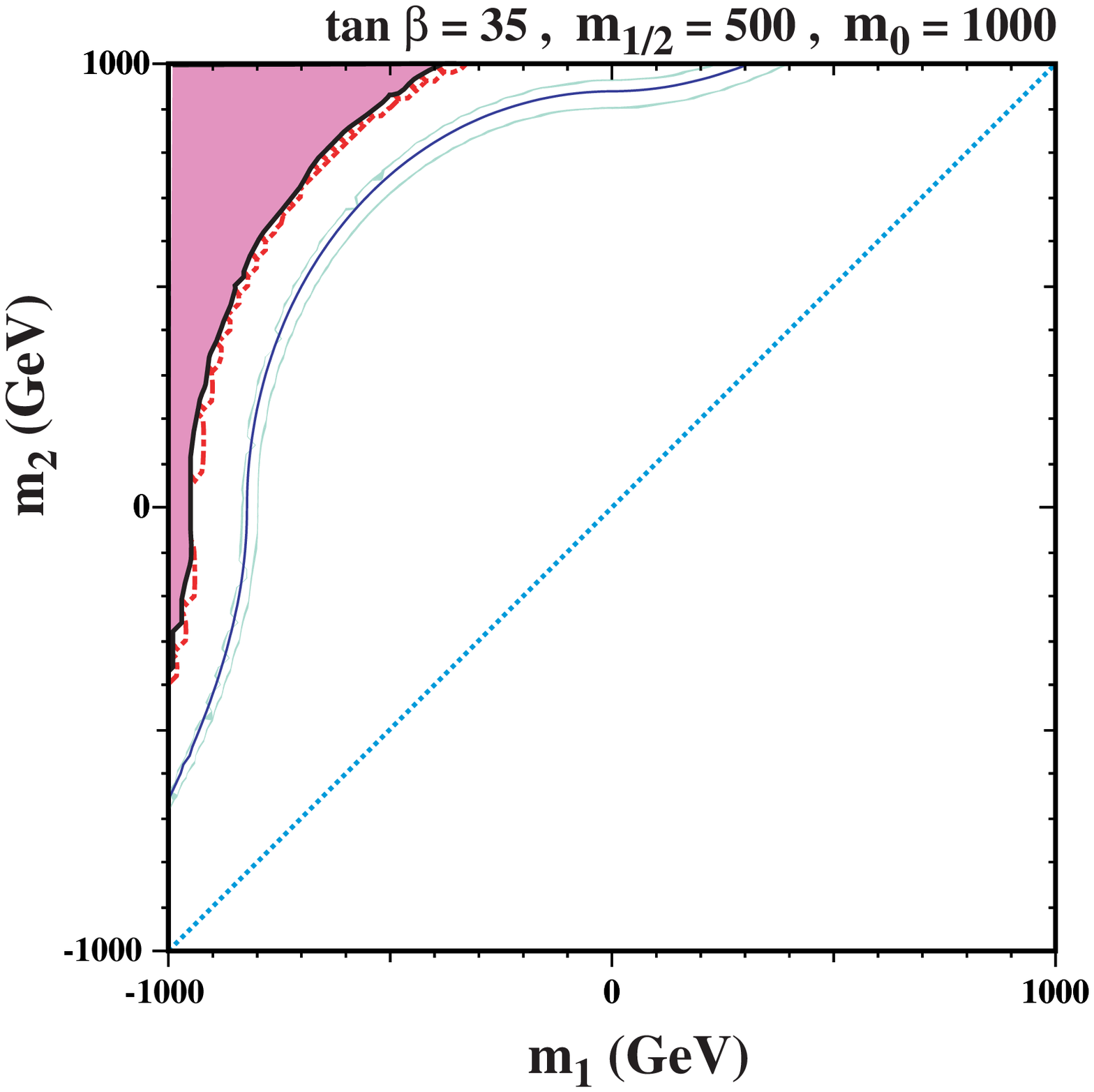,height=7cm}}
\mbox{\epsfig{file=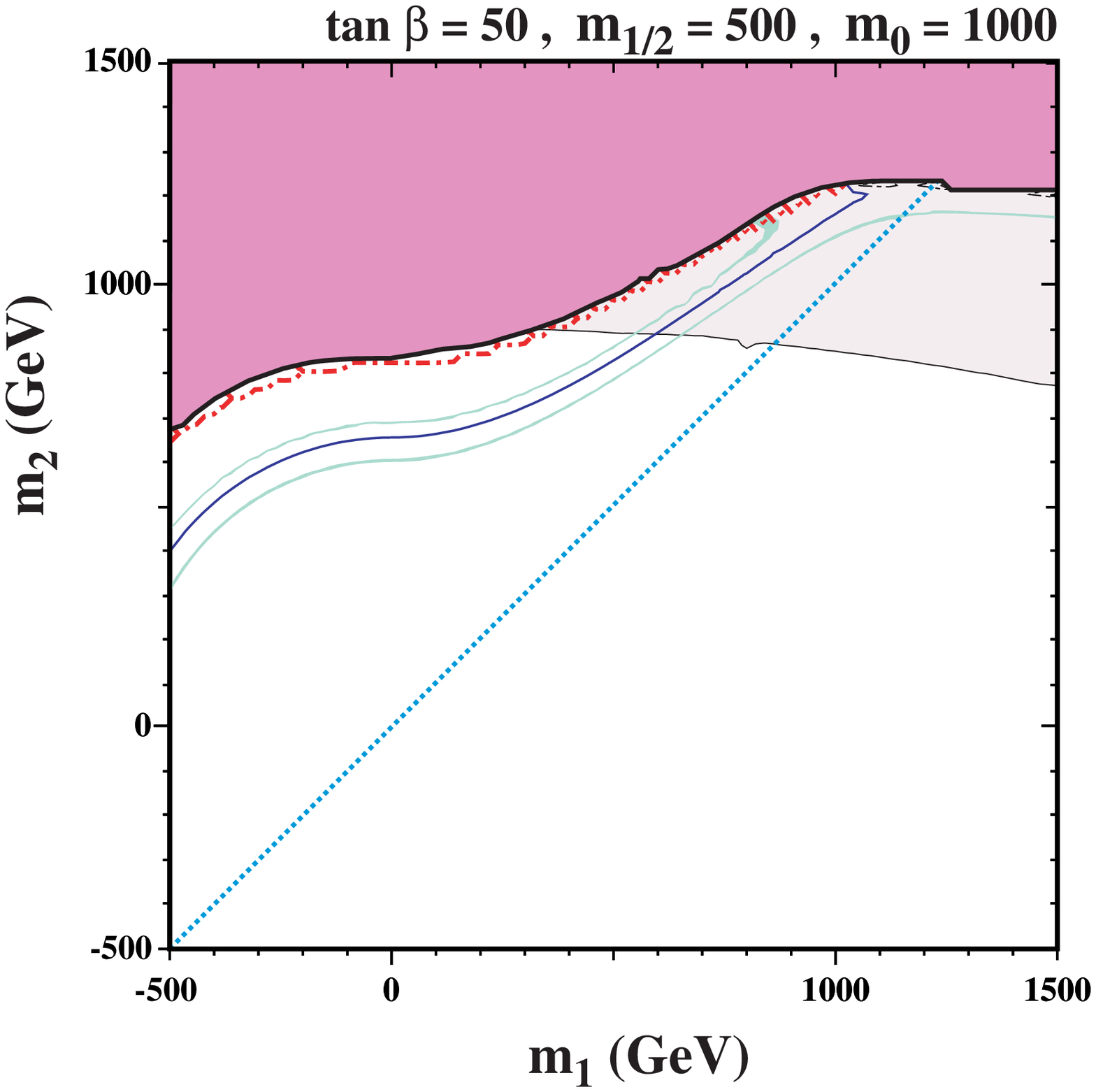,height=7cm}}
\end{center}
\caption{\it Examples of NUHM2 $(m_1,m_2)$ planes with
$m_{1/2}=500$~GeV, $m_0=1000$~GeV, $A_0=0$, and $\tanb = 10$, 20, 35, and 50 in
panels (a), (b), (c), and (d), respectively. Constraints are displayed as in Fig.~\ref{fig:m12m0fixmA}.}
\label{fig:new5_10}
\end{figure}

In the case $(m_{1/2}, m_0) = (500, 1000)$~GeV shown in Fig.~\ref{fig:new5_10},
the electroweak symmetry-breaking condition again excludes large
portions of the plane that expand somewhat as $\tanb$ is
increased~\footnote{Note that the ranges of $m_1$ and $m_2$ displayed in Fig.~\ref{fig:new5_10}
differ from panel to panel.}. 
The requirement that the LSP be neutral does not
constrain the parameter space. The $b \goto s \gamma$ constraint excludes
only a narrow strip along the left boundary of panel (a), and the LEP Higgs
constraint also excludes only narrow boundary regions in all four panels.

The only WMAP regions are confined to rapid-annihilation funnels,
supplemented in panel (d) for $\tbt = 50$ by an extension to a crossover
strip. Imitating its behaviour in Fig.~\ref{fig:new5_3}, the funnel is again
quite serpentine. The NUHM1 line does not intersect this funnel, but
does cross the crossover strip in panel (d). Thus, the NUHM1 lines lie
almost entirely in overdense regions, and the CMSSM points are all
overdense.

\section{Conclusions}

We have studied in this paper how the CMSSM parameter space may be
embedded successively in the larger NUHM1 and NUHM2 parameter spaces.
We find several qualitatively new features in making these generalizations.

One new feature of the NUHM1 is that the allowed domain  is restricted in places
by the requirement that the LSP not be a selectron, a possibility that does not
arise within the CMSSM. Another feature of the NUHM1 is that there may be 
funnels of parameter space where rapid annihilation
through direct-channel Higgs poles extends the WMAP-compatible part of
parameter space to large $m_0$ and/or $m_{1/2}$ even for $\tan \beta = 10$,
whereas this feature appears only at much larger $\tan \beta$ in the CMSSM.
This is because $m_A$ can be regarded as a free parameter within the NUHM1,
whereas it is calculable in terms of $m_{1/2}, m_0, A_0$ and $\tan \beta$ in the
CMSSM. Other features of the dark matter density in the NUHM1 include the
possibility that neutralino-selectron coannihilation may be important close to
the forbidden selectron-LSP region, and the possibility that the relic density may
be suppressed into the WMAP-compatible range in regions where the
neutralino composition crosses over from being mainly bino to mainly Higgsino.

Additional new features appear in the further generalization to the NUHM2.
For example, the allowed region of parameter space is partly restricted by
the requirement that the LSP not be a sneutrino. Near this boundary,
neutralino-sneutrino coannihilation can be important for bringing the
relic neutralino density into the WMAP-compatible range.

One of the novel features of this study has been the presentation of constraints in
the $(m_1, m_2)$ plane for certain fixed values of the other parameters. It is striking
that the relic density requirement, in particular, often favours values of the parameters
where both $|m_{1,2}| \gg m_0$, and they are far from being equal to each other.
There is no hint that the NUHM1 subspace is favoured within the larger NUHM2
space, and still less suggestion that the smaller CMSSM subspace is favoured in
any way. 

One of the prime motivations for this study has been to understand to what
extent the good coverage of the WMAP-compatible CMSSM region by the LHC can
be generalized to the NUHM1 and NUHM2. In the CMSSM, the LHC covers the
stau coannihilation region, but not completely the focus-point region (which can be
regarded as an example of a bino/Higgsino crossover), nor the rapid-annihilation
funnel that appears at large $\tan \beta$. In the NUHM1,  the appearance of selectron
coannihilation does not add to the woes of the LHC. However, the
rapid-annihilation funnels extending to large $m_0$ and/or $m_{1/2}$ may be
problematic for the LHC, as may the crossover strips that may also appear
at relatively large $m_{1/2}$ and extend to large $m_0$.
However, it remains the case that the LHC can cover a large fraction of the
NUHM1 and NUHM2 parameter spaces. If the LHC does indeed discover
supersymmetry, a key check whether the scalar masses are universal,
in addition to sfermion mass measurements, will
be to determine the values of $m_A$ and $\mu$, and to explore whether
they are compatible with the values required by the electroweak vacuum
conditions within the CMSSM. This would be possible, e.g., by measuring
the masses of heavier Higgs bosons, neutralinos and charginos. This should be
possible if $m_{1/2}$ and $m_0$ are not too large, but such a study lies
beyond the scope of this paper.

\section*{Acknowledgments}
\noindent 
The work of K.A.O. and P.S. was supported in part
by DOE grant DE--FG02--94ER--40823. P.S. would like to acknowledge
support from the Doctoral Dissertation Fellowship Program at the
University of Minnesota.

\end{document}